\documentclass[aps,prx,reprint,superscriptaddress]{revtex4-2}
\usepackage{graphicx}

\usepackage{amsmath}
\usepackage{ulem}

\usepackage{amssymb}
\usepackage{bm}
\usepackage{physics}
\usepackage[unicode, colorlinks=true, linkcolor=blue, citecolor=blue, urlcolor=blue]{hyperref}
\usepackage{amsthm}
\theoremstyle{definition}
\newtheorem{proposition}{Proposition}
\usepackage{siunitx}
\usepackage{enumitem}

\newcommand{\bop}[1]{\mathbf{#1}}
\newcommand{\ah}{\bop{a}}

\newcommand{\Oh}{\bop{O}}
\newcommand{\ahdag}{\bop{a}^\dagger}
\newcommand{\sigmah}{\bm{\sigma}}
\newcommand{\mrin}{\mathrm{in}}
\newcommand{\mrout}{\mathrm{out}}
\newcommand{\mrinout}{\mrin/\mrout}

\newcommand{\ain}{\ah_{\mrin}}
\newcommand{\aout}{\ah_{\mrout}}

\newcommand{\Sh}{\bop{S}}
\newcommand{\Ph}{\bop{P}}
\newcommand{\Pht}{\tilde{\Ph}}

\newcommand{\Pin}{\Ph_{\mrin}}
\newcommand{\Pout}{\Ph_{\mrout}}
\newcommand{\Pinout}{\Ph_{\mrinout}}

\newcommand{\Ptin}{\Pht_{\mrin}}
\newcommand{\Ptout}{\Pht_{\mrout}}

\newcommand{\Dh}{\bop{D}}
\newcommand{\Dht}{\tilde{\Dh}}
\newcommand{\rmin}{r_\mathrm{min}}
\newcommand{\xh}{\bop{x}}
\newcommand{\ph}{\bop{p}}
\newcommand{\Hh}{\bop{H}}
\newcommand{\Hcav}{\bop{H}_{\mathrm{cav}}}
\newcommand{\Hspin}{\bop{H}_{\mathrm{det}}}
\newcommand{\Hint}{\bop{H}_{\mathrm{int}}}
\newcommand{\sigmag}{\sigma_{g}}
\newcommand{\rhog}{\rho(g)}
\newcommand{\nomega}{n(\omega)}
\newcommand{\Omeff}{\Omega_{\mathrm{eff}}}

\newcommand{\Nh}{\bop{N}}

\newcommand{\mrase}{\mathrm{A}}
\newcommand{\mrrase}{\mathrm{R}}
\newcommand{\aase}{\ah_\mrase}
\newcommand{\arase}{\ah_\mrrase}
\newcommand{\tdelay}{t_\mathrm{d}}

\newcommand{\pipulse}{\(\pi\)-pulse}

\usepackage{xcolor}
\newcommand{\nxixst}[1]{{\color{black}#1}}

\begin{document}
\title{Inhomogeneous~Light-Matter~Coupling~as~a~Resource~for~Noiseless~Quantum~Memories}

\author{Fumiya Hanamura}
\affiliation{Centre for Quantum Technologies, Singapore}

\author{Sicheng Bao}
\affiliation{Centre for Quantum Technologies, Singapore}

\author{Jie Lerk Yoo}
\affiliation{National University of Singapore, Department of Physics, Faculty of Science, Singapore}

\author{Alexia Auff\`eves}
\affiliation{Centre for Quantum Technologies, Singapore}
\affiliation{MajuLab, CNRS-UCA-SU-NUS-NTU International Joint Research Laboratory}

\author{Steven Touzard}
\affiliation{Centre for Quantum Technologies, Singapore}
\affiliation{National University of Singapore, Department of Physics, Faculty of Science, Singapore}
\affiliation{MajuLab, CNRS-UCA-SU-NUS-NTU International Joint Research Laboratory}
\date{\today}
\begin{abstract}
Inhomogeneous ensembles of two-level systems are central to both fundamental light--matter physics and quantum-network applications. Understanding and optimizing ensemble-based quantum memories and entanglement protocols requires a unified framework that describes how to store quantum states of light as collective matter excitations and retrieve them on demand. Here we develop such a framework, the \textit{waveguide model}, by mapping the dark collective modes of the ensemble onto an effective waveguide with well-defined input–output relations, valid in both the weak-excitation regime and near population inversion. This model reveals that inhomogeneous coupling---often regarded as a limitation---is instead the physical origin of noisy-echo suppression by adiabatic pulses, a key ingredient for realizing noiseless quantum memories. Direct comparison with experimental data from O'Sullivan et al. [Phys. Rev. X 12, 041014 (2022)] shows quantitative agreement between the model predictions and experimental observations. For entanglement generation, the same mechanism exposes a previously unexplored shortcoming of robust control pulses and leads to a new composite-pulse protocol that overcomes it. These results show that this waveguide model translates the collective physics of inhomogeneous ensembles into design rules for quantum-network components, recasting coupling from an obstacle into a control knob for collective emission.
\end{abstract}
\maketitle

\section{Introduction}
Ensembles of two-level quantum systems (TLSs) are central to both fundamental collective light--matter physics and quantum technologies.
They exhibit collectively enhanced light emission and absorption, known as superradiance~\cite{kaluznyObservationSelfInduced}, and give rise to many-body quantum effects such as enhanced charging in quantum batteries~\cite{kurmanPoweringQuantum}.
At the same time, they provide a promising platform for quantum memories and entanglement sources, which are key building blocks for quantum networks~\cite{kimbleQuantumInterneta}.
These capabilities have been explored in a wide range of systems, including rare-earth-doped crystals~\cite{maOnehourCoherent,duttaAtomicFrequency,hedgesEfficientQuantum}, cold atoms~\cite{hsiaoHighlyEfficient}, nitrogen-vacancy centers~\cite{grezesMultimodeStorage}, and quantum dots~\cite{dyteStoringQuantum}.

In realistic systems, such ensembles are inherently inhomogeneous, with variations in both transition frequencies and light-matter coupling.
Controlling atomic transitions in the presence of coupling inhomogeneity is a long-standing problem in driven two-level systems, with roots in the Landau--Zener paradigm~\cite{LandauEnergy,zenerNonadiabaticCrossing,rosenDoubleSternGerlach}.
Robust-control techniques such as adiabatic rapid passage (ARP) have therefore been developed to mitigate its effects~\cite{degraafAdiabaticRf,abragam1961principles,malinovskyGeneralTheory}.
However, coupling inhomogeneity is usually treated as an imperfection \cite{julsgaardQuantumMemory}, and its impact on collective dynamics remains underexplored.

In contrast, frequency inhomogeneity has a well-established physical consequence that underlies ensemble quantum memories.
It dephases the collective superradiant state that initially resides in the \textit{bright mode} of the ensemble, suppressing emission on a timescale set by the inhomogeneous linewidth.
The corresponding steady-state frequency response, including effects such as cavity protection, has been extensively studied~\cite{dinizStronglyCoupling,kuruczSpectroscopicPropertiesa,julsgaardDynamicalEvolution}.
The same process has a complementary time-domain interpretation: a quantum state initially stored in the bright mode is transferred into long-lived \textit{dark modes} that are decoupled from the radiation field.
Since the seminal works on Hahn echoes~\cite{hahnSpinEchoes}, it has been known that this information can be rephased and retrieved on demand.
For such echo-based quantum memories, the dark modes provide storage space that enables broadband and multimode operation~\cite{damonRevivalSilenced,mcauslanPhotonechoQuantum,grezesMultimodeStorage,osullivanRandomAccessQuantum,kamelMultimodeRandomAccess}.
Existing memory models describe these modes using effective degrees of freedom, mean-field variables, or protocol-specific constructions~\cite{afzeliusImpedancematchedCavity,wesenbergDynamicsCollective,greggioOptimalAbsorption,awilliamsonCavityEnhanced}, often in the broad-inhomogeneous-broadening limit.
However, these approaches do not provide a unified time-domain input--output description of how the bright mode couples to the dark-mode manifold, and they typically neglect coupling inhomogeneity.
A general framework that explicitly connects the radiative bright mode to long-lived dark modes, in the relevant regimes of weakly-excited and nearly-inverted ensemble, therefore remains lacking.

In this work, we develop an analytical \textit{waveguide model} for ensembles of TLSs with arbitrary frequency and coupling distributions.
The central idea is to map the dark modes to an effective waveguide coupled to the bright mode.
This mapping makes the input--output relation between the radiative collective excitation and the long-lived storage modes explicit.
The resulting description reduces the light-ensemble dynamics to an intuitive linear system of two coupled resonators.
This model retains a fully quantum treatment applicable to both cases of weakly-excited and nearly-inverted ensemble.
It therefore enables analytical predictions for the time-domain dynamics of a broad class of quantum-memory and entanglement protocols without requiring full many-body simulations.

Our model resolves the mechanism behind a striking recent observation by identifying coupling inhomogeneity as a valuable control resource that can switch superradiant emission on and off.
In a modified Hahn-echo experiment, using an ARP pulse for ensemble inversion was found to strongly suppress the echo \cite{osullivanRandomAccessQuantum,kamelMultimodeRandomAccess}. 
Existing analyses attribute this suppression to the frequency-dependent phase imprinted by ARP.
We show, however, that this phase alone would not suppress the echo; instead, it would only chirp it, producing an amplified chirped echo (ACE). Importantly, this ACE is not compatible with a low-noise quantum memory.
We find that true suppression instead requires coupling inhomogeneity, which imprints an additional coupling-dependent phase.
We compare the model directly with the microwave spin-ensemble experiment of Ref.~\cite{osullivanRandomAccessQuantum}, finding that including coupling inhomogeneity reproduces the observed echo suppression, whereas the homogeneous-coupling assumption retains a finite ACE.
Furthermore, we show that the same mechanism has a counterpart when the inverted ensemble is used for entanglement generation \cite{awilliamsonCavityEnhanced}: ARP pulses combined with coupling inhomogeneity also suppress the generated entanglement.
Guided by the model, we propose an alternative scheme that generates this entanglement with the robustness of ARP pulses while bypassing this entanglement suppression.
Together, these results could strongly impact the development of practical quantum memories by identifying how coupling inhomogeneity suppresses noisy echo and how its adverse effect on entanglement can be reversed.
The waveguide model thus serves two purposes: it explains existing observations,  the echo suppression of Refs.~\cite{osullivanRandomAccessQuantum,kamelMultimodeRandomAccess}, and it produces a new protocol, the composite-pulse sequence that restores entanglement.
More broadly, our results show that the waveguide model connects collective light--matter physics to the design of protocols for quantum-network components based on inhomogeneous ensembles, by expressing the noise and entanglement of these protocols in terms of the frequency and coupling distributions of the ensemble and the control pulses applied to it.

\begin{figure*}[htb]
    \centering
    \includegraphics[scale=1]{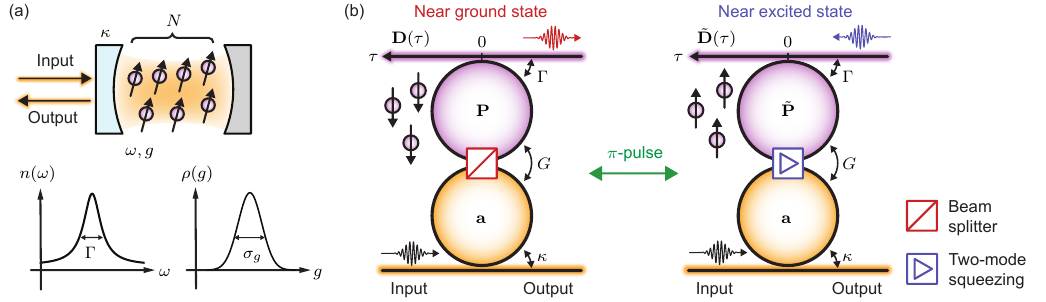}
 \caption{Waveguide model.
        (a) Physical system: an ensemble of \(N\) TLSs interacts with a single resonator mode, with detunings \(\omega\) and couplings \(g\).
        The detunings follow a Lorentzian distribution of width \(\Gamma\) (see Appendix \ref{sec:non_lorentzian} for the non-Lorentzian case), while the couplings follow an arbitrary distribution (Gaussian with standard deviation \(\sigmag\) is used in numerical simulations unless otherwise specified).
        (b) Equivalent waveguide model near the ground state: the ensemble is mapped to a resonator-like bright mode \(\Ph\) and a dark waveguide, formed by a continuum of dark modes \(\Dh(g,\tau)\) with transverse coordinate \(g\).
        The mode \(\Ph\) couples to \(\ah\) with strength \(G\), and to the dark waveguide at \(\tau=0\) with rate \(\Gamma\).
        \(P\) couples only to the transverse mode with profile \(u(g)\), whose field \(\Dh(\tau)=\int\dd{g}\,u(g)\,\Dh(g,\tau)\) is drawn here.
        Near the fully excited state, the corresponding modes are denoted by \(\Pht\) and \(\Dht(g,\tau)\).
        An ideal \pipulse{} switches the interaction between beam-splitter and two-mode-squeezing forms, while also reversing the propagation direction in the dark waveguide.}
    \label{fig:waveguide}
\end{figure*}

\section{Echo-based quantum memories}
To fully capture the progress enabled by our model, we briefly recall the basic operations of a quantum memory, the main protocols developed by the community and their limitations.

A quantum state of light is input into a resonator coupled to an ensemble of TLSs, where it is mapped onto a state of the ensemble's bright mode.
The number of excitations is taken small compared to the number of TLSs in the ensemble, which is typically \(10^6\)–\(10^9\), depending on the system.
This defines the weakly excited regime of the ensemble. 
Because of frequency inhomogeneity, the state of the bright mode rapidly leaks into the long-lived dark modes of the ensemble, where it is stored.
A quantum memory is operated by retrieving this input state at a desired time, without attenuating it or corrupting it with noise.
This retrieval requires reversing the dephasing process so that the stored quantum state refocuses into an echo.
There are several ways to realize this reversal, including atomic frequency combs (AFC)~\cite{duttaAtomicFrequency}, controlled reversible inhomogeneous broadening (CRIB)~\cite{krausQuantumMemory}, and gradient echo memory (GEM)~\cite{hetetMultimodalProperties}.

Alternatively, the most direct reversal is to invert the ensemble's population with a \pipulse{}.
Such a direct \pipulse{}-based echo memory faces a major issue: population inversion produces a noisy echo through amplified spontaneous emission (ASE).
This process not only generates a continuous noisy background of photons, but also entangles the ensemble with the output field, thereby destroying the stored quantum state.
The revival-of-silenced-echo (ROSE) protocol addresses this problem by suppressing the noisy first echo, applying a second \pipulse{}, and using the second echo for retrieval~\cite{damonRevivalSilenced,afzeliusProposalCoherent,julsgaardQuantumMemory,julsgaardFundamentalLimitations,fuCavityassistedRevival}.
This second echo is noiseless because it forms after the population has been returned close to the ground state.
Protocols such as AFC, CRIB and GEM,  which do not rely on a \pipulse{}, avoid the problem altogether since they do not invert the population and therefore ASE is not generated.

Meanwhile, the very same process that produces noisy ASE can also be used as a resource for entanglement generation between the memory and the light.
In the rephased amplified spontaneous emission (RASE) protocol \cite{ledinghamNonclassicalPhoton,awilliamsonCavityEnhanced}, a \pipulse{} is applied after ASE emission. It rephases the collective excitation of the memory and produces a RASE field that is entangled with the initial ASE field.
This generates time-separated entanglement, a useful resource for quantum networking.

Both ROSE and RASE rely on high-fidelity inversion of the ensemble population, which is limited by coupling inhomogeneity when using simple \pipulse{}s.
For example, in ROSE, imperfect second inversion leaves residual population that adds noise during state retrieval, often requiring additional experimental overhead such as auxiliary energy levels~\cite{maEliminationNoise}.
To overcome this limitation, ARP pulses are a known route to provide robust inversion over a wide range of coupling inhomogeneity.
Hence, recent experiments incorporated ARP pulses to improve inversion within the ROSE protocol~\cite{osullivanRandomAccessQuantum,kamelMultimodeRandomAccess}.
This led to an intriguing additional observation: when ARP pulses were used, the expected noisy echo was strongly suppressed.

Below we show that our waveguide model, by quantitatively accounting for these coupling inhomogeneities, allows us both to explain these recent results and to propose new protocols.
In particular, through direct comparison with the experimental data, we provide analytical and numerical evidence that coupling inhomogeneities account for the echo suppression observed in the experiment of Ref.~\cite{osullivanRandomAccessQuantum}, in contrast to the conventional interpretation based on frequency inhomogeneity.

\section{Waveguide Model}
An inhomogeneous ensemble coupled to a resonator comprises a radiative collective dipole and a continuum of collective excitations that do not radiate into the resonator. We develop an effective waveguide representation of these degrees of freedom that retains the inhomogeneous light--matter couplings across the ensemble. The argument below provides a physical motivation for the model and its input--output relations. The additional details of its microscopic construction are given in Appendix~\ref{sec:waveguide_model_appendix}.
\subsection{Physical system}\label{sec:system}

We consider an ensemble of \(N\) TLSs coupled to a single resonator mode (Fig.~\ref{fig:waveguide}(a)). A TLS is characterized by its coupling \(g\) to the resonator and its detuning \(\omega\) from the center of the ensemble frequency distribution. We assume that the two parameters are independent, with normalized densities \(\rho(g)\) and \(n(\omega)\). In a frame rotating at the central TLS frequency, the resonator has detuning \(\delta\) and couples to an external input--output channel at rate \(\kappa\). The Tavis-Cummings Hamiltonian reads
\begin{subequations}\label{eq:main_hamiltonian}
\begin{align}
    \Hh&=\Hcav+\Hint+\Hspin,\\
    \Hcav&=\delta\ahdag\ah,\\
    \Hint&=N\int\dd{g}\dd{\omega}\rhog\nomega\,g
    \qty[\sigmah^-(g,\omega)\ahdag+\mathrm{h.c.}],\label{eq:main_interaction_hamiltonian}\\
    \Hspin&=N\int\dd{g}\dd{\omega}\,\rhog\nomega\,\frac{\omega}{2}\sigmah^z(g,\omega),\label{eq:main_detuning_hamiltonian}
\end{align}
\end{subequations}
where \(\ah\) (\(\ahdag\)) is the annihilation (creation) operator of the resonator, and \(\sigmah^-(g,\omega)\), \(\sigmah^+(g,\omega)\), and \(\sigmah^z(g,\omega)\) are the coarse-grained Pauli operators of the TLSs with coupling \(g\) and detuning \(\omega\) (Appendix~\ref{sec:waveguide_model_appendix}).

Through \(\Hint\), the field of the resonator couples to a
single superposition of the TLS’ dipoles. The profile of
this distribution defines the bright mode (Sec.~\ref{sec:bright}). Because
of the spread in TLS detunings, \(\Hspin\) transfers excitations
from the bright mode to the other collective
modes of the ensemble, the dark modes (Sec.~\ref{sec:dark}).

\subsection{Bright mode}\label{sec:bright}

In \(\Hint\), each TLS couples to the resonator with a rate set only by its coupling \(g\), independent from its detuning \(\omega\). We therefore group the TLSs into coupling classes, whose collective dipole is 
\begin{align}
    \Ph(g)=\sqrt{N}\int\dd{\omega}\,\sqrt{\rho(g)}\nomega\,
    \sigmah^-(g,\omega),
    \label{eq:main_class_mode}
\end{align}
where the factor \(\sqrt{N\rho(g)}\nomega\) appears for normalization. 
Only one superposition of coupling class dipoles couples to the resonator. We call it the \textit{bright mode} and write its collective dipole
\begin{align}
    \Ph=\int\dd{g}\,u(g)\,\Ph(g),\qquad
    u(g)=\frac{g\sqrt{\rhog}}{\bar{g}},
    \label{eq:main_bright_mode}
\end{align}
\nxixst{where \(\bar{g}=\qty[\int\dd{g}\,g^2\rhog]^{1/2}\) is the root-mean-square coupling. The mode profile \(u(g)\) is normalized, $\int dg\,|u(g)|^2=1$, and specifies the contribution of each coupling class to the bright mode. Consequently,
\(\Hint=G(\Ph \ah^\dagger+\Ph^\dagger \ah)\),
where \(G=\sqrt{N}\,\bar g\) is the collectively-enhanced coupling rate. Therefore, \(\Hint\) takes on a simple bilinear form. }

We focus on dynamics involving only weak signal pulses and strong \pipulse{}s, so that the ensemble remains close to either its fully polarized ground state or its fully excited state. In these regimes, the Holstein--Primakoff approximation \cite{holsteinFieldDependence} yields, near the ground state, 
\begin{equation}\label{eq:main_bright_commutator}
[\Ph(g),\Ph^\dagger(g')]=\delta(g-g'),\qquad [\Ph,\Ph^\dagger]= 1
\quad.
\end{equation}
Thus, \(\Ph(g)\) and \(\Ph\) behave as bosonic lowering operators and \(\Hint\) reduces to a beam-splitter Hamiltonian. Near full inversion, the Holstein--Primakoff approximation yields \(\left[\Ph^\dagger, \Ph\right]=1\). Therefore, \(\Pht\equiv\Ph^\dagger\) behaves as a bosonic lowering operator, and \(\Hint=G\left(\Pht \ah + \Pht^\dagger \ahdag\right)\) reduces to a two-mode squeezing Hamiltonian. The conditions to remain within this approximation are detailed in Appendix \ref{sec:waveguide_applicability_appendix}.

Both near the ground- and excited-state, the dynamics of the bright-mode and the resonator correspond to a simpler quantum optics problem. We summarize the properties of the bright mode below.

\begin{proposition}[Bright mode]
    The bright mode and its coupling to the resonator are described by the following properties:
    \begin{itemize}
        \item[(i)] The resonator couples with collectively enhanced
        strength \(G\) to a single mode of the TLS ensemble, the
        bright mode. Within the Holstein--Primakoff approximation,
        the bright mode behaves as a bosonic mode,
        with lowering operator \(\Ph\) near the ground state and
        \(\Pht\) near the fully excited state.

        \item[(ii)] The interaction between the resonator and the bright mode can be conveniently formulated as a beam splitter near the ground state and a two-mode squeezer near the fully excited state:
        \begin{align*}
          \Hint &= G\qty(\Ph\ahdag + \Ph^\dagger\ah)
            && \text{(ground state)},\\
          \Hint &= G\qty(\Pht^\dagger\ahdag + \Pht\ah)
            && \text{(excited state)}.
        \end{align*}

        \item[(iii)] A \pipulse{} exchanges the roles of \(\Ph\) and \(\Pht\) as lowering operator of the bright mode, and thereby switches the interaction between the beam-splitter and two-mode-squeezing forms.
    \end{itemize}
\end{proposition}

Under \(\Hspin\), the bright mode loses its excitations to the dark modes, which are studied in the next subsection.

\subsection{Dark waveguide}\label{sec:dark}

The dynamics of \(\Hspin\) imprints a different phase onto TLSs with different detunings. Therefore, the dipole of a coupling class \(\Ph(g)\) loses its collective coherence, while keeping its initial excitation in the ensemble. This coherence decay is typically referred to as the free induction decay (FID). The initial excitation of \(\Ph(g)\) is redistributed among other combinations of TLS dipoles. We seek a representation in which this redistribution appears as propagation away from the coupling class, into a continuum of dark modes. In this picture, an inversion pulse reverses the propagation and produces an echo. 

If the detuning distribution \(\nomega\) is a Lorentzian with full width at half maximum \(\Gamma\), the coherence of \(\Ph(g)\) exponentially decays with rate \(\Gamma\). This is the same decay law as a bosonic mode coupled to a frequency-flat continuum of waveguide modes, acting as a loss channel. While we focus on Lorentzian broadening in the main text, the construction can be extended to non-Lorentzian distributions using standard techniques for non-Markovian dynamics (Appendix \ref{sec:non_lorentzian}). The parallel between FID and resonator dynamics suggests rewriting \(\Hspin\) in the Gardiner--Collett representation \cite{gardinerInputOutputb} of input--output theory. We therefore postulate that for each coupling \(g\), the coupling class \(\Ph(g)\) interacts through \(\Hspin\) with a continuum of modes \(\Dh(g, \omega)\), as
\begin{align}\label{eq:main_waveguide_hamiltonian}
    \Hspin={}&\int\dd{g}\dd{\omega}\,\omega\,\Dh^\dagger(g,\omega)\Dh(g,\omega)\nonumber\\
    &+i\sqrt{\frac{\Gamma}{2\pi}}\int\dd{g}\dd{\omega}\nonumber\\
    &\quad\times\qty[\Ph^\dagger(g)\Dh(g,\omega)-\Dh^\dagger(g,\omega)\Ph(g)].
\end{align}
In order to form a valid input--output theory, the postulated dark modes must maintain the dynamics of Eq.~\eqref{eq:main_detuning_hamiltonian}, they must be orthogonal to every class mode with \(\left[\Dh(g, \omega), \Ph^\dagger(g')\right]=0\), and they must be normalized as \(\left[\Dh(g, \omega), \Dh^\dagger(g', \omega')\right]=\delta(g-g')\delta(\omega-\omega')\). These conditions give 
\begin{align}
    \Dh(g,\omega)={}&\sqrt{\frac{\Gamma}{2\pi}}\sqrt{N\rhog}\,
      \Bigg[\frac{\sigmah^-(g,\omega)}{\Gamma/2+i\omega}\nonumber\\
    &-i\int\dd{\omega'}\,
       \frac{n(\omega')\,\sigmah^-(g,\omega')}{\omega'-\omega+i0^+}\Bigg].
    \label{eq:main_dark_modes}
\end{align}

Equation~\eqref{eq:main_waveguide_hamiltonian} is precisely the equation of bosonic modes coupled to a multimode waveguide. Similar to a physical waveguide, this dark waveguide has a transverse dimension, here set by the coupling \(g\).
The bright mode occupies a single transverse mode profile \(u(g)\), and \(\Hspin\) preserves the transverse profile of any stored excitation. Therefore, the dynamics of the bright mode is determined by the projection of \(\Hspin\) onto the mode profile \(u(g)\), resulting in
\begin{align}
    \Hspin={}&\int\dd{g}\dd{\omega}\,\omega\,\Dh^\dagger(g,\omega)\Dh(g,\omega)\nonumber\\
    &+i\sqrt{\frac{\Gamma}{2\pi}}\int\dd{\omega}\,
      \qty[\Ph^\dagger\Dh(\omega)-\Dh^\dagger(\omega)\Ph]
      +\Hh_\perp,
    \label{eq:main_waveguide_hamiltonian_split}
\end{align}
where \(\Dh(\omega)=\int\dd{g}\,u(g)\,\Dh(g,\omega)\) and \(\Hh_\perp\) acts on transverse modes that are orthogonal to \(u(g)\). Dark modes with a profile orthogonal to \(u(g)\) never exchange excitations with the bright mode. This transverse structure is what allows inversion pulses to silence echoes (Sec.~\ref{sec:silencing}).

We transform the continuum of dark modes into an explicit waveguide picture by introducing the coordinate \(\tau\), the Fourier-conjugate of \(\omega\). In the Heisenberg picture, \(\Dh(g, \omega; t)=e^{i\Hh t}\Dh(g, \omega)e^{-i\Hh t}\) and the waveguide lowering operator is
\begin{align}
    \Dh(g,\tau;t)=\frac{1}{\sqrt{2\pi}}\int\dd{\omega}\,e^{-i\omega\tau}\,\Dh(g,\omega;t).
    \label{eq:main_waveguide_field}
\end{align}
These dark modes couple to the bright mode at \(\tau=0\). Away from this point, the excitation stored in the waveguide translates toward smaller \(\tau\),
\begin{align}
    \Dh(g,\tau;t+\Delta t)&=\Dh(g,\tau+\Delta t;t)
    \label{eq:main_translation}\\
    &\quad(\tau,\ \tau+\Delta t\ \text{of the same sign}).\nonumber
\end{align}
The region \(\tau>0\) thus holds excitations that will reach the bright mode after a delay \(\tau\), and the region \(\tau<0\) holds excitations that left the bright mode at a time \(\abs{\tau}\) ago [Fig.~\ref{fig:waveguide}(b)].
At \(\tau=0\), only the projection \(\Dh(\tau;t)=\int\dd{g}\,u(g)\Dh(g,\tau;t)\) interacts with the bright mode.
Components with a transverse mode profile orthogonal to \(u(g)\) propagate through \(\tau=0\) without interacting with the bright mode.

Near the fully excited state, the same construction applies to the bright mode operator \(\Pht\) (Appendix~\ref{sec:waveguide_model_appendix}).
Because the inverted TLSs precess in the opposite direction, the detuning term changes sign,
\begin{align}
    \Hspin\mathrel{=}{}&-\int\dd{g}\dd{\omega}\,\omega\,\Dht^\dagger(g,\omega)\Dht(g,\omega)\nonumber\\
    &+i\sqrt{\frac{\Gamma}{2\pi}}\int\dd{\omega}\,
      \qty[\Pht^\dagger\Dht(\omega)-\Dht^\dagger(\omega)\Pht]
      +\tilde{\Hh}_\perp,
    \label{eq:main_waveguide_hamiltonian_excited}
\end{align}
where \(\Dht(\omega)=\int\dd{g}\,u(g)\Dht(g,\omega)\) and \(\tilde{\Hh}_\perp\) acts on the mode profiles in \(g\) that are orthogonal to \(u(g)\), and the field \(\Dht(g,\tau;t)\) now translates toward larger \(\tau\): excitations reach the bright mode from \(\tau<0\) and leave it toward \(\tau>0\).

An ideal \pipulse{} inverts all TLSs the same way and maps each dark mode onto its inverted counterpart, \(\Dh(g,\omega)\to\Dht(g,\omega)\), up to a uniform phase set by the pulse and the dark-mode convention.
It therefore leaves each excitation at its position \(\tau\) and reverses its direction of propagation.
An excitation stored at \(\tau=-\tau_0\) when the pulse is applied thus travels back toward the bright mode and reaches it a time \(\tau_0\) later.
The ideal \pipulse{} acts as a mirror in time, which is our waveguide representation of the Hahn echo.
In practice, real \pipulse{}s also imprint a phase on each TLS, which changes this behavior, as we discuss in Sec.~\ref{sec:arp}.

The dark modes thus behave as a multimode waveguide that meets the bright mode at a single point, \(\tau=0\).
We summarize its properties below.

\begin{proposition}[Dark waveguide]
    Within the Holstein--Primakoff approximation, the dark modes of the ensemble have the following properties:
    \begin{itemize}
        \item[(i)] The dark modes form a waveguide whose propagation coordinate \(\tau\) is the Fourier conjugate of the TLS detuning \(\omega\), and whose transverse coordinate is the coupling \(g\).
        The field of the waveguide is given by the bosonic operators \(\Dh(g,\tau)\) (\(\Dht(g,\tau)\)) near the ground-state (excited-state).

        \item[(ii)] For a Lorentzian inhomogeneous broadening, the bright mode couples to the waveguide at \(\tau=0\), with a rate \(\Gamma\), and only to the transverse mode with profile \(u(g)\).
        Transverse modes with a profile orthogonal to \(u(g)\) cross \(\tau=0\) without interacting with the bright mode.

        \item[(iii)] The propagation direction depends on the state of the ensemble.
        Away from \(\tau=0\), the field translates toward smaller \(\tau\) near the ground state, and toward larger \(\tau\) near the fully excited state:
        \begin{align*}
          \Dh(g,\tau;t+\Delta t)&=\Dh(g,\tau+\Delta t;t) && \text{(ground)},\\
          \Dht(g,\tau;t+\Delta t)&=\Dht(g,\tau-\Delta t;t) && \text{(excited)}.
        \end{align*}

        \item[(iv)] An ideal \pipulse{} maps \(\Dh(g,\tau)\) onto \(\Dht(g,\tau)\), up to a uniform phase, and thereby reverses the direction of propagation while leaving each excitation at its position \(\tau\).
    \end{itemize}
\end{proposition}

The waveguide meets the bright mode at \(\tau=0\).
The next subsection describes the dynamics at this junction and combines it with the resonator.

\subsection{Waveguide model}\label{sec:waveguide_model}

Combining the bright mode and dark waveguide gives the \textit{waveguide model} for the ensemble--resonator system, as shown in Fig.~\ref{fig:waveguide}(b).
For a Lorentzian distribution of detunings, the bright mode interacts with the dark waveguide only at \(\tau=0\), through the field about to reach it and the field that has just left it,
\begin{subequations}\label{eq:main_boundary_fields}
\begin{align}
    \Pin(t)&=\int\dd{g}\,u(g)\,\Dh(g,0^+;t),\\
    \Pout(t)&=\int\dd{g}\,u(g)\,\Dh(g,0^-;t).
\end{align}
\end{subequations}
These are the input and output fields of standard input--output theory~\cite{gardinerInputOutputb}.
By Eq.~\eqref{eq:main_translation}, the waveguide field keeps a record of them: \(\Dh(\tau;t)\) equals \(\Pin(t+\tau)\) for \(\tau>0\) and \(\Pout(t+\tau)\) for \(\tau<0\).
Eliminating the dark modes from the Heisenberg equations of Eqs.~\eqref{eq:main_hamiltonian} and \eqref{eq:main_waveguide_hamiltonian} (Appendix~\ref{sec:waveguide_model_appendix}) gives a beam-splitter interaction between the resonator and bright mode near the ground state:
\begin{subequations}\label{eq:main_ground_waveguide}
\begin{align}
    \dot{\ah}&=-\qty(i\delta+\frac{\kappa}{2})\ah-iG\Ph+\sqrt{\kappa}\ain, \\
    \dot{\Ph}&=-\frac{\Gamma}{2}\Ph-iG\ah+\sqrt{\Gamma}\Pin, \\
    \aout&=\ain-\sqrt{\kappa}\ah, \\
    \Pout&=\Pin-\sqrt{\Gamma}\Ph.
\end{align}
\end{subequations}
Here \(\ain\) and \(\aout\) are the physical resonator's input and output fields.
The last equation is the jump of the waveguide field across \(\tau=0\): the field leaving the bright mode carries its emission at rate \(\Gamma\).
Causality is built into these definitions: the input field at time \(t\) commutes with the bright mode at all earlier times.

Near the fully excited state, the waveguide field propagates toward larger \(\tau\), so the input now arrives from \(\tau<0\) and the output leaves toward \(\tau>0\),
\begin{subequations}\label{eq:main_boundary_fields_excited}
\begin{align}
    \Ptin(t)&=\int\dd{g}\,u(g)\,\Dht(g,0^-;t),\\
    \Ptout(t)&=\int\dd{g}\,u(g)\,\Dht(g,0^+;t),
\end{align}
\end{subequations}
and the resonator--bright-mode interaction takes the two-mode-squeezing form:
\begin{subequations}\label{eq:main_excited_waveguide}
\begin{align}
    \dot{\ah}&=-\qty(i\delta+\frac{\kappa}{2})\ah-iG\Pht^\dagger+\sqrt{\kappa}\ain, \\
    \dot{\Pht}&=-\frac{\Gamma}{2}\Pht-iG\ah^\dagger+\sqrt{\Gamma}\Ptin, \\
    \aout&=\ain-\sqrt{\kappa}\ah, \\
    \Ptout&=\Ptin-\sqrt{\Gamma}\Pht.
\end{align}
\end{subequations}
In this picture, the dynamics reduce to a symmetric linear system, providing an intuitive description of collective emission, storage, and rephasing.
The key element is the construction of the dark modes as a waveguide-like storage channel with explicit input--output relations to the bright mode, rather than merely as dark collective excitations.
As a result, echo-based quantum-memory and entanglement-generation protocols, including revival of silenced echo (ROSE)~\cite{damonRevivalSilenced} and rephased amplified spontaneous emission (RASE)~\cite{awilliamsonCavityEnhanced}, can be intuitively understood within this model.

In the waveguide model, the collective cooperativity, the commonly used figure of merit for characterizing the strength of collective light--matter interactions, emerges naturally as the ratio of the relevant coupling rates, \(C=4G^2/(\kappa\Gamma)\).
For an ensemble near the ground state, \(C=1\) gives the impedance-matching condition for perfect absorption of a narrowband input~\cite{afzeliusImpedancematchedCavity}.
For an inverted ensemble, the same parameter controls the two-mode-squeezing gain, with \(C\geq1\) corresponding to laser oscillation~\cite{julsgaardDynamicalEvolution}.

The usefulness of the waveguide model becomes especially clear when realistic \pipulse{}s are considered.
An ideal \pipulse{} reverses the propagation direction of the dark waveguide and leaves every excitation at its position (Proposition~2).
A realistic \pipulse{} also imprints a phase on each TLS, which reshapes the excitation stored in the waveguide.
This makes the waveguide model a natural tool for tracking how practical inversion pulses affect rephasing dynamics.

\begin{figure}[t]
    \centering
    \includegraphics[scale=1]{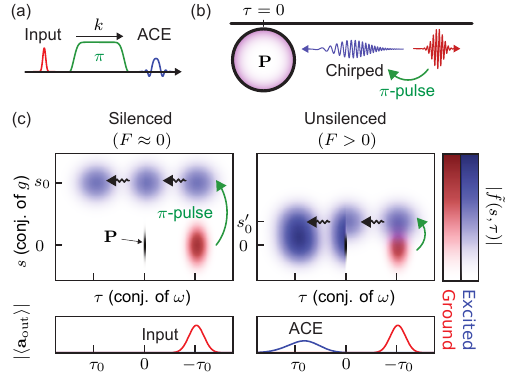}
\caption{Effects of adiabatic rapid passage (ARP).
        (a) For homogeneous coupling, an ARP \pipulse{} (chirp rate \(k\)) applied after an input pulse generates an amplified chirped echo (ACE).
        (b) Waveguide model picture: ARP inverts each frequency component \(\omega\) at its own time \((\omega-\omega_0)/k\), so the components rephase at different times and the echo is chirped at rate \(k/2\).
        (c) Illustration of echo silencing by coupling inhomogeneity in \((s,\tau)\) space [Eq.~\eqref{eq:main_st_mode_function}], obtained by numerically evolving the waveguide-model field amplitudes.
        Red (blue) denotes an ensemble near the ground (excited) state.
        We use \(C=0.25\), \(\kappa/(2\pi)=\Gamma/(2\pi)=\SI{1}{MHz}\), and a Gaussian coupling distribution with \(g_0/(2\pi)=\SI{100}{Hz}\) and \(\sigmag/(2\pi)=\SI{0.8}{Hz}\).
        A coherent-state input, a Gaussian pulse with amplitude standard deviation \(\sigma_t=\SI{3}{\micro s}\), is sent at \(t=-\tau_0\), with \(\tau_0=\SI{20}{\micro s}\).
        At \(t=0\), it is stored in the waveguide as a wavepacket centered at \(\tau=-\tau_0\).
        The ARP pulse is then applied as the instantaneous phase of Eq.~\eqref{eq:main_arp_phase_integral}, calculated for a WURST pulse of duration \(T=\SI{250}{\micro s}\), order \(n=20\), and chirp rate \(k/(2\pi)=\SI{0.02}{MHz^2}\).
        The two peak intracavity amplitudes \(A=2800\) and \(1500\) give displacements of magnitude \(s_0=\SI{0.611}{s}\) and \(\SI{0.230}{s}\), corresponding to \(F=0.0089\) and \(0.51\), respectively.
        This displacement reduces the overlap of the stored excitation with the bright profile \(u(g)\), and therefore suppresses the ACE.}
    \label{fig:chirp}
\end{figure}
\section{Adiabatic rapid passage}\label{sec:arp}
ARP pulses are commonly chosen as \pipulse{}s because they invert TLSs even in the presence of inhomogeneous broadening and coupling. Such pulses are important for the ensemble to remain in the Holstein--Primakoff regime. However, unlike the ideal \pipulse{} of Sec.~\ref{sec:dark}, ARP pulses imprint a different phase on each class of TLSs, which depends on their coupling and detuning. We use the waveguide model to analyze how this phase affects echo-based quantum memories. We show that, for homogeneous coupling, ARP chirps the echo without changing its total energy [Fig.~\ref{fig:chirp}(a)], whereas coupling inhomogeneity suppresses it, explaining the experimental observations of Refs.~\cite{osullivanRandomAccessQuantum,kamelMultimodeRandomAccess}. In Sec.~\ref{sec:rose}, we demonstrate quantitative agreement with the microwave experiment of Ref.~\cite{osullivanRandomAccessQuantum} as a representative example.

We consider an ARP pulse
\begin{align}
    \ev*{\ah(t)} = A(t)\exp[-i\qty(\omega_0 t + \frac{k}{2} t^2)],
    \label{eq:e_of_t}
\end{align}
with instantaneous frequency \(\omega_0+kt\) and Rabi frequency \(\Omega(t)=2A(t)g\).
We characterize the pulse by its adiabaticity factor \(Q=\Omega_0^2/k\), where \(\Omega_0\) is the average Rabi frequency.
In the adiabatic limit, \(Q\gg1\), each TLS adiabatically inverts while accumulating a frequency- and coupling-dependent dynamical phase
\begin{align}
    \phi(g,\omega)=-\int_{t_i}^{t_f}\dd{t}\,
    \sqrt{\qty[2gA(t)]^2+\qty[\omega-\omega_0-kt]^2}.
    \label{eq:main_arp_phase_integral}
\end{align}
This expression depends explicitly on \(g\) whenever the light--matter coupling is inhomogeneous.
To see how this phase acts in the waveguide, we describe a stored excitation by its dark-waveguide mode function \(f_D(g,\omega)\), through the wavepacket mode
\begin{align}
    \Dh_{f_D}=\int\dd{g}\dd{\omega}\,f_D^*(g,\omega)\,\Dh(g,\omega),
    \label{eq:main_wavepacket_mode}
\end{align}
normalized as \(\int\dd{g}\dd{\omega}\,\abs{f_D}^2=1\).
Its spin-wave mode function \(f(g,\omega)\), which gives its amplitude on the TLSs with coupling \(g\) and detuning \(\omega\), is related to \(f_D\) by a unitary change of transverse basis derived in Appendix~\ref{sec:waveguide_model_appendix}. This relation depends on which side of \(\tau=0\) contains the wavepacket.
An input absorbed through the resonator enters through the bright mode and is therefore stored with the bright profile, \(f(g,\omega)=u(g)f_0(\omega)\), where \(f_0(\omega)\) contains the absorption response and the basis-dependent spectral phase (Appendix~\ref{sec:silencing_factor_appendix}).
The pulse is applied while the stored excitation is away from \(\tau=0\), so this stored excitation does not interact with the resonator during the pulse, and the pulse acts on each TLS separately (Appendix~\ref{sec:ARP_appendix}).
In addition to the mirror action of an ideal \pipulse{} (Proposition~2), it therefore multiplies the spin-wave mode function by the phase of each TLS,
\begin{align}
    f(g,\omega)\to f(g,\omega)\,e^{\pm i\phi(g,\omega)},
    \label{eq:main_pulse_mode_function}
\end{align}
with the plus sign for a pulse starting from the ground state and the minus sign for a pulse starting from the fully excited state.
Consequently, a second identical ARP pulse cancels the phase accumulated under the first.

The phase acts on the stored excitation in two ways.
Its dependence on \(\omega\) displaces each frequency component along \(\tau\), which chirps the echo (Sec.~\ref{sec:ace}).
Its dependence on \(g\) changes the transverse profile of the excitation from \(u(g)\) to \(u(g)e^{i\phi(g,\omega)}\), which reduces its overlap with the bright mode and silences the echo (Sec.~\ref{sec:silencing}).
The detailed derivation of the ARP phase and its action on the dark modes is provided in Appendix~\ref{sec:ARP_appendix}.

\section{Amplified chirped echo}\label{sec:ace}
We first consider the case of homogeneous coupling, where only the TLS frequencies vary across the ensemble, as in the conventional frequency-dependent-phase explanation adopted in Refs.~\cite{osullivanRandomAccessQuantum,kamelMultimodeRandomAccess}. We show that in this case, the accumulated phase in Eq.~\eqref{eq:main_arp_phase_integral} only chirps the first echo in time, leaving its total energy and its contribution to memory noise unchanged.

For a sufficiently flat drive amplitude \(A(t)\), we have, up to a constant,
\begin{align}
    \phi(g,\omega)=-\frac{(\omega-\omega_0)^2}{k}.
    \label{eq:chirp_phase}
\end{align}
This quadratic spectral phase chirps the pulse in the dark waveguide, producing an echo with chirp rate \(k/2\), half the ARP chirp rate [Fig.~\ref{fig:chirp}(a)].
This half-rate chirping has a simple time-domain interpretation.
During ARP, each frequency component \(\omega\) is inverted when the drive crosses resonance, at \(t=(\omega-\omega_0)/k\). For an input absorbed at \(t=-\tau_0\), this component then rephases after the same additional time \(\tau_0+(\omega-\omega_0)/k\), at \(t=\tau_0+2(\omega-\omega_0)/k\).
The echo is therefore chirped at the rate \(k/2\) [Fig.~\ref{fig:chirp}(b)].

Since this chirp only spreads the echo in time, ARP preserves the total echo energy in the homogeneous-coupling case.
We call this process an \textit{amplified chirped echo} (ACE).
Thus, a frequency-dependent ARP phase alone cannot explain the echo suppression observed in Refs.~\cite{osullivanRandomAccessQuantum,kamelMultimodeRandomAccess}. Importantly, as we will see in Sec.~\ref{sec:rose}, ACE is not compatible with a noiseless quantum memory. 

\section{Suppression of ACE by inhomogeneous coupling}\label{sec:silencing}
We now show that ACE is suppressed in the presence of coupling inhomogeneity.
When the coupling is inhomogeneous, the ARP pulse imprints an additional \(g\)-dependent phase, and the stored excitation leaves the pulse with the transverse profile \(u(g)e^{i\phi(g,\omega)}\) instead of \(u(g)\).
When it returns to \(\tau=0\), only its component along the bright profile interacts with the bright mode.
The amplitude of ACE is therefore multiplied by the modulus of the overlap of the two profiles,
\begin{align}
    F=\abs{\int\dd{g}\,u(g)\qty[u(g)e^{i\phi(g,\omega)}]},
    \label{eq:silence_factor}
\end{align}
which we refer to as the \textit{silencing factor}.
Large broadening of the coupling distribution leads to \(F\to0\), and the echo is completely suppressed.

For a Gaussian coupling distribution with mean \(g_0\) and standard deviation \(\sigmag\), \(\rhog=(\sqrt{2\pi}\sigmag)^{-1}\exp[-(g-g_0)^2/(2\sigmag^2)]\), we obtain
\begin{align}
    F\sim \exp\qty[-\frac{1}{2}\qty(\frac{\sigmag}{g_0})^2Q^2],
    \label{eq:silencing_adiabaticity}
\end{align}
where \(Q\) is the adiabaticity factor of Sec.~\ref{sec:arp} (Appendix~\ref{sec:silencing_appendix}).
Thus, increasing adiabaticity improves inversion but simultaneously magnifies the phase dispersion caused by coupling inhomogeneity, so even weak inhomogeneity can strongly suppress ACE.
This mechanism explains the echo silencing observed in Refs.~\cite{osullivanRandomAccessQuantum,kamelMultimodeRandomAccess}.
To illustrate this, Fig.~\ref{fig:rose}(e) compares analytical and numerical predictions directly with the microwave data of Ref.~\cite{osullivanRandomAccessQuantum}.
Details of the comparison and parameter estimates for both experiments are provided in Appendix~\ref{sec:experiment_appendix}.

To visualize how the coupling-dependent ARP phase suppresses the echo by reducing its overlap with the bright mode, we Fourier transform the spin-wave mode function in both coordinates,
\begin{align}
    \tilde{f}(s,\tau)=\frac{1}{2\pi}\int\dd{g}\dd{\omega}\,f(g,\omega)\,e^{i(gs+\omega\tau)},
    \label{eq:main_st_mode_function}
\end{align}
where \(s\) is conjugate to \(g\) as \(\tau\) is to \(\omega\).
In this representation, the ARP-induced phase displaces the mode along the \(s\)-direction by \(s_0=\pdv{\phi}{g}\big|_{g=g_0}\) when \(\sigmag\ll g_0\). This displacement reduces its overlap with the bright mode and thereby suppresses the emission [Fig.~\ref{fig:chirp}(c)].

This silencing allows a noiseless ROSE quantum memory with ARP pulses alone, without an additional silencing mechanism (Sec.~\ref{sec:rose}).
However, it also means that ARP pulses are not readily usable for RASE, whose rephased echo carries the entanglement (Sec.~\ref{sec:rase}).

\section{ROSE protocol}\label{sec:rose}
\begin{figure}[t]
    \centering
    \includegraphics[scale=1]{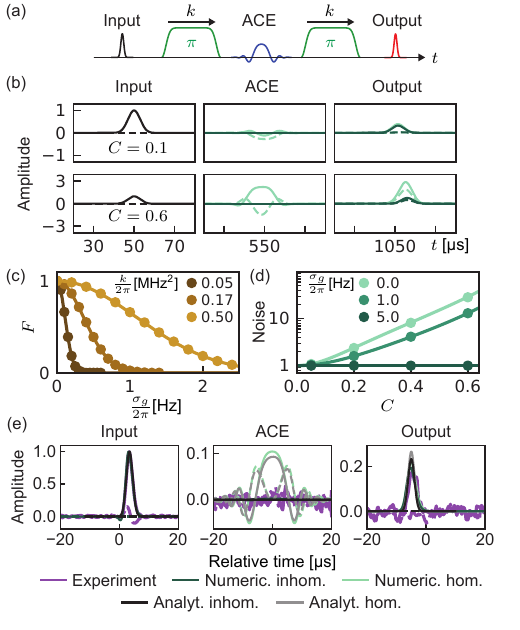}
    \caption{ROSE protocol with ARP \pipulse{}s.
        (a) Schematic: the first \pipulse{} generates an unwanted ACE, and the second \pipulse{} retrieves the stored quantum state.
        (b) Numerical simulation: amplitudes of \(\ev*{\ain}\) (``Input'') and \(\ev*{\aout}\) (``ACE,'' ``Output''). Solid (dashed) lines show real (imaginary) parts, with input peak normalized to 1. Light green: \(\sigmag=\SI{0}{Hz}\); dark green: \(\sigmag=\SI{5}{Hz}\).
        (c) Silencing factor \(F\) vs coupling inhomogeneity \(\sigmag\) for different chirp rates \(k\). Markers: numerics; lines: theory [Eq.~\eqref{eq:silence_factor}]. Numerical values are obtained by dividing the square root of the integrated first-echo intensity by the corresponding theoretical value for homogeneous coupling, including finite-bandwidth effects.
        (d) Output noise \(\ev*{\Delta \xh^2}+\ev*{\Delta \ph^2}\) vs cooperativity \(C\) for different \(\sigmag\) at fixed \(k/(2\pi)=\SI{0.50}{MHz^{2}}\). Markers: numerics; lines: theory [Eq.~\eqref{eq:noise}].
        (e) Comparison of analytical and numerical predictions with the experimental data of Ref.~\cite{osullivanRandomAccessQuantum}. ``Experiment'' refers to the measured data, while ``Numeric.'' and ``Analyt.'' denote numerical simulations and analytical predictions from the waveguide model, respectively. ``Inhom.'' accounts for the experimental coupling distribution, whereas ``hom.'' assumes homogeneous coupling. Solid and dashed curves represent the real and imaginary components, respectively. All datasets are normalized to the peak amplitude of their respective input pulses, with the same normalization applied to both echoes.}
    \label{fig:rose}
\end{figure}
This echo suppression is precisely what is needed for the revival of silenced echo (ROSE) protocol~\cite{damonRevivalSilenced}.
ROSE is an echo-based quantum memory protocol using two \pipulse{}s: the noisy echo after the first pulse must be suppressed (silenced), while the second pulse retrieves the stored state [Fig.~\ref{fig:rose}(a)].
Its dynamics are naturally captured by the waveguide model, as detailed in Appendix~\ref{sec:rose_appendix}.

Because the first \pipulse{} switches the ensemble to the near-excited-state regime, insufficient suppression of the first echo leads to two-mode-squeezing dynamics.
This creates noise both in the emitted echo and in the state stored in the dark-mode waveguide.
The quality of first-echo suppression therefore directly determines the added noise in the final output.
For a coherent-state input, the output noise is
\begin{align}
    \ev*{\Delta \xh^2}+\ev*{\Delta \ph^2}= 1 + 2F^2\qty(\frac{4C}{1-C^2})^2,\label{eq:noise}
\end{align}
where \(F\) is the silencing factor and the unity corresponds to shot noise.

In the original proposal, the first echo was silenced using phase-mismatch \cite{damonRevivalSilenced}, while later works introduced alternative mechanisms such as cavity detuning \cite{afzeliusProposalCoherent,julsgaardQuantumMemory,julsgaardFundamentalLimitations,fuCavityassistedRevival}.
More recently, experiments have reported that the first echo is suppressed when ARP pulses are used as \pipulse{}s \cite{osullivanRandomAccessQuantum,kamelMultimodeRandomAccess}.
In these works, the suppression is attributed to the prevention of rephasing due to a quadratic frequency-dependent phase.
However, as discussed above, such a frequency-dependent phase alone produces an ACE rather than suppressing it.
Our key finding is that echo silencing instead originates from coupling inhomogeneity.

In this ARP-silenced version of ROSE, two ARP \pipulse{}s with the same chirp rate are used.
The first ARP suppresses the first echo in the presence of coupling inhomogeneity, with the silencing factor \(F\), given by Eq.~\eqref{eq:silence_factor}, corresponding to the overlap with the bright mode in coupling space.
When the second ARP pulse is applied, the same phase is imprinted again but with the opposite sign, because the pulse starts from the excited state rather than the ground state.
This cancels the ARP-induced phase imprinted by the first pulse.
This not only removes the chirp of the temporal mode, but also restores mode matching with the bright mode, thereby enabling retrieval of the stored state.

To verify this mechanism, we numerically simulate the ROSE protocol with a coherent-state input using a cumulant expansion; see Appendix~\ref{sec:numerical_appendix} for details.
Unless otherwise stated, the coupling distribution is Gaussian, and we set \(\delta=0\), \(\kappa/(2\pi)=\Gamma/(2\pi)=\SI{1}{MHz}\), and \(g_0/(2\pi)=\SI{100}{Hz}\).
The ARP-pulse amplitude is chosen such that the near-resonant average Rabi frequency is \(\Omega_0/(2\pi)=\SI{4.0}{MHz}\).
These parameter values are chosen to be comparable to those of microwave spin-ensemble memory experiments~\cite{osullivanRandomAccessQuantum}.
We then vary the cooperativity \(C\) by tuning the total TLS number \(N\), and sweep the coupling inhomogeneity \(\sigmag\) and the ARP chirp rate \(k\).
We also consider cooperativities beyond those reached in current experiments, which are required for high-efficiency quantum memory, to clearly show how ACE and its suppression by coupling inhomogeneity affect memory performance.

The simulation results agree with our theoretical prediction.
For homogeneous coupling, ACE is clearly observed, especially at larger cooperativity, leading to strong unwanted amplification of the output signal [Fig.~\ref{fig:rose}(b), light-green lines].
Introducing coupling inhomogeneity significantly suppresses this effect [Fig.~\ref{fig:rose}(b), dark-green lines].
We extract \(F\) by dividing the square root of the integrated first-echo intensity by the corresponding analytical prediction for homogeneous coupling, including finite-bandwidth effects as described in Appendix~\ref{sec:rose_appendix}. The extracted values agree well with theory across different chirp rates \(k\) and coupling inhomogeneities \(\sigmag\) [Fig.~\ref{fig:rose}(c)].
In particular, the homogeneous-coupling results (\(\sigmag=0\)) give \(F\simeq1\) for all chirp rates shown in Fig.~\ref{fig:rose}(c), confirming that changing the chirp rate preserves the total first-echo energy.
As \(F \to 0\), the output noise is suppressed toward the shot-noise limit, consistent with Eq.~\eqref{eq:noise} (Fig.~\ref{fig:rose}(d)).

Figure~\ref{fig:rose}(e) compares our analytical and numerical predictions directly with the microwave experimental data of Ref.~\cite{osullivanRandomAccessQuantum}, normalized to the input-pulse peak. Both calculations reproduce the observed first-echo suppression when coupling inhomogeneity is included. Assuming homogeneous coupling instead predicts a finite, chirped first echo. With coupling inhomogeneity, the predicted mode amplitudes are consistent with the measured first echo, differing by only \(1.37\sigma\) (numerical) and \(1.58\sigma\) (analytical), where \(\sigma\) is the estimated quadrature noise standard deviation of each mode. In contrast, homogeneous coupling gives discrepancies of \(35.9\sigma\) and \(31.6\sigma\), respectively. This supports coupling inhomogeneity as the mechanism responsible for echo suppression. The analysis procedure and experimental parameter estimates are given in Appendix~\ref{sec:experiment_appendix}, which also shows that relative coupling variations of only about \(0.4\%\) and \(14\%\) explain strong silencing, \(F<10^{-3}\), in Refs.~\cite{osullivanRandomAccessQuantum,kamelMultimodeRandomAccess}, respectively.

\begin{figure}[t]
    \centering
    \includegraphics[scale=1]{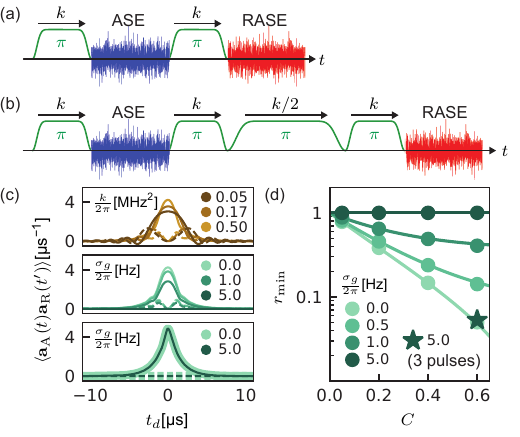}
    \caption{RASE protocol with ARP \pipulse{}s.
        (a) Schematic: the first \pipulse{} generates ASE, entangled with the RASE retrieved after the second \pipulse{}.
        (b) Composite three-\pipulse{} protocol restoring the entanglement: the second \pipulse{} is replaced by three ARP pulses to cancel the coupling-dependent phase.
        (c) Time correlation \(\ev*{\aase(t)\arase(t')}\) plotted as a function of the time offset \(\tdelay\) from the point of maximum correlation, with \(t'=2t_{0}-t+\tdelay\); see Appendix~\ref{sec:rase_appendix} for the definition.
        Solid (dashed) lines show the real (imaginary) parts.
        (upper) Homogeneous coupling (\(\frac{\sigmag}{2\pi}=\SI{0}{Hz}\)), protocol (a), varying chirp rate \(k\) (colormap as in Fig.~\ref{fig:rose}(c)).
        (middle) Protocol (a), varying \(\sigmag\) at fixed \(\frac{k}{2\pi}=\SI{0.50}{MHz^2}\) (colormap as in (d)).
        (lower) Protocol (b) at fixed \(\frac{k}{2\pi}=\SI{0.50}{MHz^2}\), with \(\frac{\sigmag}{2\pi}=\SI{0}{Hz}\) (light green) and \SI{5}{Hz} (dark green).
        (d) Duan--Simon criterion [Eq.~\eqref{eq:duan_simon}] between ASE and RASE vs cooperativity \(C\) for different \(\sigmag\) at fixed \(\frac{k}{2\pi}=\SI{0.50}{MHz^2}\). Markers: protocol (a) (circles) and protocol (b) (star); lines: theory [protocol (a)].}
    \label{fig:rase}
\end{figure}
\section{RASE protocol}\label{sec:rase}
The same echo-suppression mechanism reveals a previously unrecognized limitation in the RASE protocol for time-delayed entanglement generation~\cite{ledinghamNonclassicalPhoton,awilliamsonCavityEnhanced}.
Its waveguide-model description is given in Appendix~\ref{sec:rase_appendix}.
In the RASE protocol, the amplified spontaneous emission (ASE) generated after the first \pipulse{} forms an entangled pair with the rephased ASE (RASE) emitted after the second \pipulse{}.
Therefore, unlike in ROSE, the rephased echo is not unwanted noise to be silenced, but the desired output field.
Although robust ARP pulses are needed to provide the high-fidelity inversion required for faithful entanglement generation, our model predicts that, in the presence of coupling inhomogeneity, the same mechanism that suppresses the first echo in ROSE also limits the RASE emission and degrades the ASE--RASE entanglement.

We numerically simulate the RASE protocol with and without coupling inhomogeneity.
In the homogeneous-coupling case, the ASE--RASE time correlation acquires a chirp [Fig.~\ref{fig:rase}(c), upper].
However, this chirp does not degrade the entanglement, as it can be compensated by choosing appropriate temporal modes for ASE and RASE (see Appendix \ref{sec:rase_appendix}).

In the presence of coupling inhomogeneity, however, the entanglement is degraded, as can be seen in the reduction of correlation [Fig.~\ref{fig:rase}(c), middle].
We quantify the ASE--RASE entanglement using the Duan--Simon criterion~\cite{duanInseparabilityCriterion,simonPeresHorodeckiSeparability},
\begin{align}
    \begin{split}
        \rmin & = \min_{s}[\ev*{(\sqrt{1-s}\xh_\mrase-\sqrt{s}\xh_\mrrase)^2} \\
              & \quad+\ev*{(\sqrt{1-s}\ph_\mrase+\sqrt{s}\ph_\mrrase)^2}],
    \end{split}\label{eq:duan_simon}
\end{align}
where \((\xh_\mrase,\ph_\mrase)\) and \((\xh_\mrrase,\ph_\mrrase)\) are the quadratures of the ASE mode \(\aase\) and RASE mode \(\arase\), respectively, and \(\rmin<1\) indicates inseparability.
Increasing coupling inhomogeneity raises \(\rmin\), showing reduced entanglement in agreement with the theoretical model [Fig.~\ref{fig:rase}(d)].
This identifies a limitation of using ARP pulses for RASE.

To solve this problem, we propose a simple way to mitigate this limitation.
Specifically, we consider a composite \pipulse{} consisting of three ARP pulses [Fig.~\ref{fig:rase}(b)].
By choosing chirp rates \(+k\), \(+k/2\), and \(+k\), the coupling-dependent phase cancels out.
As a result, the ASE--RASE correlations are preserved and become insensitive to coupling inhomogeneity [Fig.~\ref{fig:rase}(c), lower], recovering the entanglement [Fig.~\ref{fig:rase}(d), star].
We note that a related three-adiabatic-pulse strategy for dynamical-phase cancellation can also be found in the earlier NMR literature on selective spin echoes~\cite{conollySelectiveAdiabatic}.

\section{Conclusion and Discussion}
We introduced an intuitive \textit{waveguide model} for inhomogeneous ensembles, providing a unified framework for echo-based quantum memory and entanglement-generation protocols.
Using this model, we show that ARP does not intrinsically suppress echoes.
In fact, in the homogeneous-coupling limit, it produces ACE, while true suppression arises from coupling inhomogeneity.
This mechanism explains the experimental observations of Refs.~\cite{osullivanRandomAccessQuantum,kamelMultimodeRandomAccess}. Our analytical and numerical calculations reproduce the echo suppression observed in the microwave experiment of Ref.~\cite{osullivanRandomAccessQuantum}, providing quantitative support for this explanation.

We demonstrated two consequences of this mechanism.
In ROSE, coupling-induced echo suppression removes the intermediate noisy echo and drives the added-noise contribution in Eq.~\eqref{eq:noise} toward the shot-noise limit, enabling high-fidelity quantum memory without additional experimental overhead.
In RASE, the same mechanism degrades entanglement, revealing a trade-off between robustness and coherence.
We overcome this limitation with a composite three-pulse ARP sequence that cancels the coupling-dependent phase and restores entanglement while retaining robustness.

More broadly, the waveguide model provides a predictive framework for controlling inhomogeneous ensembles in quantum memories, entanglement generation, and collective light--matter dynamics.
Although the main text focuses on Lorentzian broadening, Appendix~\ref{sec:non_lorentzian} presents the general construction for arbitrary inhomogeneous broadening, where a frequency-dependent coupling to the dark waveguide is considered.
This formulation provides an intuitive interpretation of cavity protection~\cite{dinizStronglyCoupling}.

A further direction is to go beyond the Holstein--Primakoff approximation.
Including higher-order corrections would extend the waveguide model to strongly driven and nonlinear collective dynamics~\cite{kitagawaSqueezedSpin,schmidtGiantKerr,guptaCavityNonlinear,yangSteadyentangledstateGeneration}.
Such regimes are relevant to quantum metrology~\cite{pezzeQuantumMetrology}, quantum information processing~\cite{firstenbergNonlinearQuantum,motesEncodingQubits}, and quantum many-body physics~\cite{morschDynamicsBoseEinstein,browaeysManybodyPhysics,kirtonIntroductionDicke}.

\bibliography{main.bib}

@article{afzeliusImpedancematchedCavity,
  title     = {Impedance-Matched Cavity Quantum Memory},
  author    = {Afzelius, Mikael and Simon, Christoph},
  year      = {2010},
  month     = aug,
  journal   = {Physical Review A},
  volume    = {82},
  number    = {2},
  pages     = {022310},
  issn      = {1050-2947, 1094-1622},
  doi       = {10.1103/PhysRevA.82.022310},
  urldate   = {2025-12-15},
  copyright = {http://link.aps.org/licenses/aps-default-license},
  langid    = {english}
}

@article{afzeliusProposalCoherent,
  title     = {Proposal for a Coherent Quantum Memory for Propagating Microwave Photons},
  author    = {Afzelius, M and Sangouard, N and Johansson, G and Staudt, M U and Wilson, C M},
  year      = {2013},
  month     = jun,
  journal   = {New Journal of Physics},
  volume    = {15},
  number    = {6},
  pages     = {065008},
  publisher = {IOP Publishing},
  issn      = {1367-2630},
  doi       = {10.1088/1367-2630/15/6/065008},
  urldate   = {2025-12-15},
  langid    = {english}
}

@article{awilliamsonCavityEnhanced,
  title     = {Cavity Enhanced Rephased Amplified Spontaneous Emission},
  author    = {A Williamson, Lewis and J Longdell, Jevon},
  year      = {2014},
  month     = jul,
  journal   = {New Journal of Physics},
  volume    = {16},
  number    = {7},
  pages     = {073046},
  publisher = {IOP Publishing},
  issn      = {1367-2630},
  doi       = {10.1088/1367-2630/16/7/073046},
  urldate   = {2025-12-15},
  langid    = {english}
}

@article{damonRevivalSilenced,
  title    = {Revival of Silenced Echo and Quantum Memory for Light},
  author   = {Damon, V and Bonarota, M and {Louchet-Chauvet}, A and Chaneli{\`e}re, T and Le Gou{\"e}t, J-L},
  year     = {2011},
  month    = sep,
  journal  = {New Journal of Physics},
  volume   = {13},
  number   = {9},
  pages    = {093031},
  issn     = {1367-2630},
  doi      = {10.1088/1367-2630/13/9/093031},
  urldate  = {2025-04-04},
  langid   = {english}
}

@article{fuCavityassistedRevival,
  title    = {Cavity-Assisted Revival of Silenced Echo Quantum Memory},
  author   = {Fu, Yan and Wang, Ming-Feng and Zheng, Yi-Zhuang},
  year     = {2014},
  month    = jun,
  journal  = {Optics Communications},
  volume   = {321},
  pages    = {162--166},
  issn     = {00304018},
  doi      = {10.1016/j.optcom.2014.01.036},
  urldate  = {2025-04-04},
  langid   = {english}
}

@article{greggioOptimalAbsorption,
  title     = {Optimal Absorption and Emission of Itinerant Fields into a Spin-Ensemble Memory},
  author    = {Greggio, Linda and Lorriaux, Tristan and Petrescu, Alexandru and Mirrahimi, Mazyar and Bienfait, Audrey},
  year      = 2026,
  month     = aug,
  journal   = {Physical Review A},
  volume    = {114},
  number    = {2},
  pages     = {022615},
  publisher = {American Physical Society},
  doi       = {10.1103/6xjr-tflh},
  urldate   = {2026-10-17}
}

@article{grezesMultimodeStorage,
  title     = {Multimode {{Storage}} and {{Retrieval}} of {{Microwave Fields}} in a {{Spin Ensemble}}},
  author    = {Grezes, C. and Julsgaard, B. and Kubo, Y. and Stern, M. and Umeda, T. and Isoya, J. and Sumiya, H. and Abe, H. and Onoda, S. and Ohshima, T. and Jacques, V. and Esteve, J. and Vion, D. and Esteve, D. and M{\o}lmer, K. and Bertet, P.},
  year      = {2014},
  month     = jun,
  journal   = {Physical Review X},
  volume    = {4},
  number    = {2},
  pages     = {021049},
  issn      = {2160-3308},
  doi       = {10.1103/PhysRevX.4.021049},
  urldate   = {2025-12-15},
  copyright = {http://creativecommons.org/licenses/by/3.0/},
  langid    = {english}
}

@article{julsgaardDynamicalEvolution,
  title      = {Dynamical Evolution of an Inverted Spin Ensemble in a Cavity: {{Inhomogeneous}} Broadening as a Stabilizing Mechanism},
  shorttitle = {Dynamical Evolution of an Inverted Spin Ensemble in a Cavity},
  author     = {Julsgaard, Brian and M{\o}lmer, Klaus},
  year       = {2012},
  month      = dec,
  journal    = {Physical Review A},
  volume     = {86},
  number     = {6},
  pages      = {063810},
  publisher  = {American Physical Society},
  doi        = {10.1103/PhysRevA.86.063810},
  urldate    = {2025-09-08}
}

@article{julsgaardFundamentalLimitations,
  title     = {Fundamental Limitations in Spin-Ensemble Quantum Memories for Cavity Fields},
  author    = {Julsgaard, Brian and M{\o}lmer, Klaus},
  year      = {2013},
  month     = dec,
  journal   = {Physical Review A},
  volume    = {88},
  number    = {6},
  pages     = {062324},
  issn      = {1050-2947, 1094-1622},
  doi       = {10.1103/PhysRevA.88.062324},
  urldate   = {2025-12-15},
  copyright = {http://link.aps.org/licenses/aps-default-license},
  langid    = {english}
}

@article{julsgaardQuantumMemory,
  title     = {Quantum {{Memory}} for {{Microwave Photons}} in an {{Inhomogeneously Broadened Spin Ensemble}}},
  author    = {Julsgaard, Brian and Grezes, C{\'e}cile and Bertet, Patrice and M{\o}lmer, Klaus},
  year      = {2013},
  month     = jun,
  journal   = {Physical Review Letters},
  volume    = {110},
  number    = {25},
  pages     = {250503},
  issn      = {0031-9007, 1079-7114},
  doi       = {10.1103/PhysRevLett.110.250503},
  urldate   = {2025-12-15},
  copyright = {http://link.aps.org/licenses/aps-default-license},
  langid    = {english}
}

@article{kamelMultimodeRandomAccess,
  title     = {Multimode and {{Random-Access Optical Quantum Memory}} via {{Adiabatic Phase Imprinting}}},
  author    = {Gohari Kamel, Nasser and Kumar, Sourabh and Gautam, Ujjwal and Saglamyurek, Erhan and Salari, Vahid and Oblak, Daniel},
  year      = 2026,
  month     = sep,
  journal   = {Communications Physics},
  publisher = {Nature Publishing Group},
  issn      = {2399-3650},
  doi       = {10.1038/s42005-026-02876-1},
  urldate   = {2026-10-15},
  copyright = {2026 The Author(s)},
  langid    = {english}
}

@article{ledinghamNonclassicalPhoton,
  title     = {Nonclassical Photon Streams Using Rephased Amplified Spontaneous Emission},
  author    = {Ledingham, Patrick M. and Naylor, William R. and Longdell, Jevon J. and Beavan, Sarah E. and Sellars, Matthew J.},
  year      = {2010},
  month     = jan,
  journal   = {Physical Review A},
  volume    = {81},
  number    = {1},
  pages     = {012301},
  issn      = {1050-2947, 1094-1622},
  doi       = {10.1103/PhysRevA.81.012301},
  urldate   = {2025-12-16},
  copyright = {http://link.aps.org/licenses/aps-default-license},
  langid    = {english}
}

@article{mcauslanPhotonechoQuantum,
  title     = {Photon-Echo Quantum Memories in Inhomogeneously Broadened Two-Level Atoms},
  author    = {McAuslan, D. L. and Ledingham, P. M. and Naylor, W. R. and Beavan, S. E. and Hedges, M. P. and Sellars, M. J. and Longdell, J. J.},
  year      = {2011},
  month     = aug,
  journal   = {Physical Review A},
  volume    = {84},
  number    = {2},
  pages     = {022309},
  issn      = {1050-2947, 1094-1622},
  doi       = {10.1103/PhysRevA.84.022309},
  urldate   = {2025-12-15},
  copyright = {http://link.aps.org/licenses/aps-default-license},
  langid    = {english}
}

@article{osullivanRandomAccessQuantum,
  title    = {Random-{{Access Quantum Memory Using Chirped Pulse Phase Encoding}}},
  author   = {O'Sullivan, James and Kennedy, Oscar W. and Debnath, Kamanasish and Alexander, Joseph and Zollitsch, Christoph W. and {\v S}im{\.e}nas, Mantas and Hashim, Akel and Thomas, Christopher N. and Withington, Stafford and Siddiqi, Irfan and M{\o}lmer, Klaus and Morton, John J. L.},
  year     = {2022},
  month    = nov,
  journal  = {Physical Review X},
  volume   = {12},
  number   = {4},
  pages    = {041014},
  issn     = {2160-3308},
  doi      = {10.1103/PhysRevX.12.041014},
  urldate  = {2025-04-04},
  langid   = {english}
}

@article{wesenbergDynamicsCollective,
  title     = {Dynamics of the Collective Modes of an Inhomogeneous Spin Ensemble in a Cavity},
  author    = {Wesenberg, Janus H. and Kurucz, Zoltan and M{\o}lmer, Klaus},
  year      = {2011},
  month     = feb,
  journal   = {Physical Review A},
  volume    = {83},
  number    = {2},
  pages     = {023826},
  issn      = {1050-2947, 1094-1622},
  doi       = {10.1103/PhysRevA.83.023826},
  urldate   = {2025-12-20},
  copyright = {http://link.aps.org/licenses/aps-default-license},
  langid    = {english}
}

@article{malinovskyGeneralTheory,
  title    = {General Theory of Population Transfer by Adiabatic Rapid Passage with Intense, Chirped Laser Pulses},
  author   = {Malinovsky, V.S. and Krause, J.L.},
  year     = 2001,
  month    = may,
  journal  = {The European Physical Journal D - Atomic, Molecular, Optical and Plasma Physics},
  volume   = {14},
  number   = {2},
  pages    = {147--155},
  issn     = {1434-6079},
  doi      = {10.1007/s100530170212},
  urldate  = {2026-04-14},
  langid   = {english}
}

@article{LandauEnergy,
  author  = {Landau, L. D.},
  title   = {Zur Theorie der Energieübertragung. II},
  journal = {Physikalische Zeitschrift der Sowjetunion},
  volume  = {2},
  pages   = {46--51},
  year    = {1932}
}

@article{rosenDoubleSternGerlach,
  title     = {Double {{Stern-Gerlach Experiment}} and {{Related Collision Phenomena}}},
  author    = {Rosen, N. and Zener, C.},
  year      = 1932,
  month     = may,
  journal   = {Physical Review},
  volume    = {40},
  number    = {4},
  pages     = {502--507},
  issn      = {0031-899X},
  doi       = {10.1103/PhysRev.40.502},
  urldate   = {2026-04-14},
  copyright = {http://link.aps.org/licenses/aps-default-license},
  langid    = {english}
}

@article{degraafAdiabaticRf,
  title      = {Adiabatic Rf Pulses: {{Applications}} to in Vivo {{NMR}}},
  shorttitle = {Adiabatic Rf Pulses},
  author     = {De Graaf, Robin A. and Nicolay, Klaas},
  year       = 1997,
  journal    = {Concepts in Magnetic Resonance},
  volume     = {9},
  number     = {4},
  pages      = {247--268},
  issn       = {1099-0534},
  doi        = {10.1002/(SICI)1099-0534(1997)9:4<247::AID-CMR4>3.0.CO;2-Z},
  urldate    = {2026-04-15},
  copyright  = {Copyright \copyright{} 1997 John Wiley \& Sons, Inc.},
  langid     = {english}
}

@book{abragam1961principles,
  title     = {The {{Principles}} of {{Nuclear Magnetism}}},
  author    = {Abragam, A. and Abragam, A.},
  year      = 1983,
  month     = oct,
  series    = {International {{Series}} of {{Monographs}} on {{Physics}}},
  publisher = {Oxford University Press},
  address   = {Oxford, New York},
  isbn      = {978-0-19-852014-6}
}

@article{kimbleQuantumInterneta,
  title     = {The Quantum Internet},
  author    = {Kimble, H. J.},
  year      = 2008,
  month     = jun,
  journal   = {Nature},
  volume    = {453},
  number    = {7198},
  pages     = {1023--1030},
  publisher = {Nature Publishing Group},
  issn      = {1476-4687},
  doi       = {10.1038/nature07127},
  urldate   = {2026-04-15},
  copyright = {2008 Springer Nature Limited},
  langid    = {english}
}

@article{kuruczSpectroscopicPropertiesa,
  title     = {Spectroscopic Properties of Inhomogeneously Broadened Spin Ensembles in a Cavity},
  author    = {Kurucz, Z. and Wesenberg, J. H. and M{\o}lmer, K.},
  year      = 2011,
  month     = may,
  journal   = {Physical Review A},
  volume    = {83},
  number    = {5},
  pages     = {053852},
  issn      = {1050-2947, 1094-1622},
  doi       = {10.1103/PhysRevA.83.053852},
  urldate   = {2025-04-06},
  copyright = {http://link.aps.org/licenses/aps-default-license},
  langid    = {english}
}

@article{dinizStronglyCoupling,
  title      = {Strongly Coupling a Cavity to Inhomogeneous Ensembles of Emitters: {{Potential}} for Long-Lived Solid-State Quantum Memories},
  shorttitle = {Strongly Coupling a Cavity to Inhomogeneous Ensembles of Emitters},
  author     = {Diniz, I. and Portolan, S. and Ferreira, R. and G{\'e}rard, J. M. and Bertet, P. and Auff{\`e}ves, A.},
  year       = 2011,
  month      = dec,
  journal    = {Physical Review A},
  volume     = {84},
  number     = {6},
  pages      = {063810},
  issn       = {1050-2947, 1094-1622},
  doi        = {10.1103/PhysRevA.84.063810},
  urldate    = {2026-01-04},
  copyright  = {http://link.aps.org/licenses/aps-default-license},
  langid     = {english}
}

@article{duttaAtomicFrequency,
  title     = {An {{Atomic Frequency Comb Memory}} in {{Rare-Earth-Doped Thin-Film Lithium Niobate}}},
  author    = {Dutta, Subhojit and Zhao, Yuqi and Saha, Uday and Farfurnik, Demitry and Goldschmidt, Elizabeth A. and Waks, Edo},
  year      = 2023,
  month     = apr,
  journal   = {ACS Photonics},
  volume    = {10},
  number    = {4},
  pages     = {1104--1109},
  publisher = {American Chemical Society},
  doi       = {10.1021/acsphotonics.2c01835},
  urldate   = {2026-06-12}
}

@article{dyteStoringQuantum,
  title     = {Storing Quantum Coherence in a Quantum Dot Nuclear Spin Ensemble for over 100 Milliseconds},
  author    = {Dyte, Harry E. and Manna, Santanu and {Covre da Silva}, Saimon F. and Rastelli, Armando and Chekhovich, Evgeny A.},
  year      = 2025,
  month     = dec,
  journal   = {Nature Communications},
  volume    = {17},
  number    = {1},
  pages     = {239},
  publisher = {Nature Publishing Group},
  issn      = {2041-1723},
  doi       = {10.1038/s41467-025-66948-6},
  urldate   = {2026-04-16},
  copyright = {2025 The Author(s)},
  langid    = {english}
}

@article{hedgesEfficientQuantum,
  title     = {Efficient Quantum Memory for Light},
  author    = {Hedges, Morgan P. and Longdell, Jevon J. and Li, Yongmin and Sellars, Matthew J.},
  year      = 2010,
  month     = jun,
  journal   = {Nature},
  volume    = {465},
  number    = {7301},
  pages     = {1052--1056},
  publisher = {Nature Publishing Group},
  issn      = {1476-4687},
  doi       = {10.1038/nature09081},
  urldate   = {2026-04-16},
  copyright = {2010 Springer Nature Limited},
  langid    = {english}
}

@article{hsiaoHighlyEfficient,
  title   = {Highly {{Efficient Coherent Optical Memory Based}} on {{Electromagnetically Induced Transparency}}},
  author  = {Hsiao, Ya-Fen and Tsai, Pin-Ju and Chen, Hung-Shiue and Lin, Sheng-Xiang and Hung, Chih-Chiao and Lee, Chih-Hsi and Chen, Yi-Hsin and Chen, Yong-Fan and Yu, Ite A. and Chen, Ying-Cheng},
  year    = 2018,
  month   = may,
  journal = {Physical Review Letters},
  volume  = {120},
  number  = {18},
  pages   = {183602},
  issn    = {0031-9007, 1079-7114},
  doi     = {10.1103/PhysRevLett.120.183602},
  urldate = {2026-04-16},
  langid  = {english}
}

@article{maOnehourCoherent,
  title     = {One-Hour Coherent Optical Storage in an Atomic Frequency Comb Memory},
  author    = {Ma, Yu and Ma, You-Zhi and Zhou, Zong-Quan and Li, Chuan-Feng and Guo, Guang-Can},
  year      = 2021,
  month     = apr,
  journal   = {Nature Communications},
  volume    = {12},
  number    = {1},
  pages     = {2381},
  publisher = {Nature Publishing Group},
  issn      = {2041-1723},
  doi       = {10.1038/s41467-021-22706-y},
  urldate   = {2026-04-16},
  copyright = {2021 The Author(s)},
  langid    = {english}
}

@article{plankensteiner2022quantumcumulants,
  title     = {QuantumCumulants. jl: A Julia framework for generalized mean-field equations in open quantum systems},
  author    = {Plankensteiner, David and Hotter, Christoph and Ritsch, Helmut},
  journal   = {Quantum},
  volume    = {6},
  pages     = {617},
  year      = {2022},
  publisher = {Verein zur F{\"o}rderung des Open Access Publizierens in den Quantenwissenschaften},
  doi       = {10.22331/q-2022-01-04-617}
}

@article{kubo1962generalized,
  title     = {Generalized cumulant expansion method},
  author    = {Kubo, Ryogo},
  journal   = {Journal of the Physical Society of Japan},
  volume    = {17},
  number    = {7},
  pages     = {1100--1120},
  year      = {1962},
  publisher = {The Physical Society of Japan},
  doi       = {10.1143/JPSJ.17.1100}
}

@book{gardiner2004quantum,
  title     = {Quantum noise: a handbook of Markovian and non-Markovian quantum stochastic methods with applications to quantum optics},
  author    = {Gardiner, Crispin and Zoller, Peter},
  year      = {2004},
  publisher = {Springer Science \& Business Media}
}

@article{khan2022quantum,
  title     = {Quantum regression theorem for multi-time correlators: A detailed analysis in the Heisenberg picture},
  author    = {Khan, Sakil and Agarwalla, Bijay Kumar and Jain, Sachin},
  journal   = {Physical Review A},
  volume    = {106},
  number    = {2},
  pages     = {022214},
  year      = {2022},
  publisher = {APS},
  doi       = {10.1103/PhysRevA.106.022214}
}

@article{kaluznyObservationSelfInduced,
  title      = {Observation of {{Self-Induced Rabi Oscillations}} in {{Two-Level Atoms Excited Inside}} a {{Resonant Cavity}}: {{The Ringing Regime}} of {{Superradiance}}},
  shorttitle = {Observation of {{Self-Induced Rabi Oscillations}} in {{Two-Level Atoms Excited Inside}} a {{Resonant Cavity}}},
  author     = {Kaluzny, Y. and Goy, P. and Gross, M. and Raimond, J. M. and Haroche, S.},
  year       = 1983,
  month      = sep,
  journal    = {Physical Review Letters},
  volume     = {51},
  number     = {13},
  pages      = {1175--1178},
  publisher  = {American Physical Society},
  doi        = {10.1103/PhysRevLett.51.1175},
  urldate    = {2026-05-03}
}

@article{kurmanPoweringQuantum,
  title     = {Powering {{Quantum Computation}} with {{Quantum Batteries}}},
  author    = {Kurman, Yaniv and Hymas, Kieran and Fedorov, Arkady and Munro, William J. and Quach, James},
  year      = 2026,
  month     = jan,
  journal   = {Physical Review X},
  volume    = {16},
  number    = {1},
  pages     = {011016},
  publisher = {American Physical Society},
  doi       = {10.1103/l39v-jwwz},
  urldate   = {2026-06-02}
}

@article{pezzeQuantumMetrology,
  title   = {Quantum Metrology with Nonclassical States of Atomic Ensembles},
  author  = {Pezz{\`e}, Luca and Smerzi, Augusto and Oberthaler, Markus K. and Schmied, Roman and Treutlein, Philipp},
  year    = 2018,
  month   = sep,
  journal = {Reviews of Modern Physics},
  volume  = {90},
  number  = {3},
  pages   = {035005},
  issn    = {0034-6861, 1539-0756},
  doi     = {10.1103/RevModPhys.90.035005},
  urldate = {2026-05-10},
  langid  = {english}
}

@article{motesEncodingQubits,
  title     = {Encoding Qubits into Oscillators with Atomic Ensembles and Squeezed Light},
  author    = {Motes, Keith R. and Baragiola, Ben Q. and Gilchrist, Alexei and Menicucci, Nicolas C.},
  year      = 2017,
  month     = may,
  journal   = {Physical Review A},
  volume    = {95},
  number    = {5},
  pages     = {053819},
  publisher = {American Physical Society},
  doi       = {10.1103/PhysRevA.95.053819},
  urldate   = {2026-05-10}
}

@article{schmidtGiantKerr,
  title     = {Giant {{Kerr}} Nonlinearities Obtained by Electromagnetically Induced Transparency},
  author    = {Schmidt, H. and Imamoglu, A.},
  year      = 1996,
  month     = dec,
  journal   = {Optics Letters},
  volume    = {21},
  number    = {23},
  pages     = {1936},
  issn      = {0146-9592, 1539-4794},
  doi       = {10.1364/OL.21.001936},
  urldate   = {2026-05-10},
  copyright = {https://doi.org/10.1364/OA\_License\_v1\#VOR},
  langid    = {english}
}

@article{yangSteadyentangledstateGeneration,
  title     = {Steady-Entangled-State Generation via the Cross-{{Kerr}} Effect in a Ferrimagnetic Crystal},
  author    = {Yang, Zhi-Bo and Wu, Wei-Jiang and Li, Jie and Wang, Yi-Pu and You, J. Q.},
  year      = 2022,
  month     = jul,
  journal   = {Physical Review A},
  volume    = {106},
  number    = {1},
  pages     = {012419},
  publisher = {American Physical Society},
  doi       = {10.1103/PhysRevA.106.012419},
  urldate   = {2026-05-10}
}

@article{guptaCavityNonlinear,
  title     = {Cavity {{Nonlinear Optics}} at {{Low Photon Numbers}} from {{Collective Atomic Motion}}},
  author    = {Gupta, Subhadeep and Moore, Kevin L. and Murch, Kater W. and {Stamper-Kurn}, Dan M.},
  year      = 2007,
  month     = nov,
  journal   = {Physical Review Letters},
  volume    = {99},
  number    = {21},
  pages     = {213601},
  publisher = {American Physical Society},
  doi       = {10.1103/PhysRevLett.99.213601},
  urldate   = {2026-05-10}
}

@article{kitagawaSqueezedSpin,
  title     = {Squeezed Spin States},
  author    = {Kitagawa, Masahiro and Ueda, Masahito},
  year      = 1993,
  month     = jun,
  journal   = {Physical Review A},
  volume    = {47},
  number    = {6},
  pages     = {5138--5143},
  publisher = {American Physical Society},
  doi       = {10.1103/PhysRevA.47.5138},
  urldate   = {2026-05-10}
}

@article{firstenbergNonlinearQuantum,
  title     = {Nonlinear Quantum Optics Mediated by {{Rydberg}} Interactions},
  author    = {Firstenberg, O and Adams, C S and Hofferberth, S},
  year      = 2016,
  month     = jun,
  journal   = {Journal of Physics B: Atomic, Molecular and Optical Physics},
  volume    = {49},
  number    = {15},
  pages     = {152003},
  publisher = {IOP Publishing},
  issn      = {0953-4075},
  doi       = {10.1088/0953-4075/49/15/152003},
  urldate   = {2026-05-10},
  langid    = {english}
}

@article{morschDynamicsBoseEinstein,
  title     = {Dynamics of {{Bose-Einstein}} Condensates in Optical Lattices},
  author    = {Morsch, Oliver and Oberthaler, Markus},
  year      = 2006,
  month     = feb,
  journal   = {Reviews of Modern Physics},
  volume    = {78},
  number    = {1},
  pages     = {179--215},
  publisher = {American Physical Society},
  doi       = {10.1103/RevModPhys.78.179},
  urldate   = {2026-05-10}
}

@article{browaeysManybodyPhysics,
  title     = {Many-Body Physics with Individually Controlled {{Rydberg}} Atoms},
  author    = {Browaeys, Antoine and Lahaye, Thierry},
  year      = 2020,
  month     = feb,
  journal   = {Nature Physics},
  volume    = {16},
  number    = {2},
  pages     = {132--142},
  publisher = {Nature Publishing Group},
  issn      = {1745-2481},
  doi       = {10.1038/s41567-019-0733-z},
  urldate   = {2026-05-10},
  copyright = {2020 Springer Nature Limited},
  langid    = {english}
}

@article{kirtonIntroductionDicke,
  title      = {Introduction to the {{Dicke Model}}: {{From Equilibrium}} to {{Nonequilibrium}}, and {{Vice Versa}}},
  shorttitle = {Introduction to the {{Dicke Model}}},
  author     = {Kirton, Peter and Roses, Mor M. and Keeling, Jonathan and Dalla Torre, Emanuele G.},
  year       = 2019,
  journal    = {Advanced Quantum Technologies},
  volume     = {2},
  number     = {1-2},
  pages      = {1800043},
  issn       = {2511-9044},
  doi        = {10.1002/qute.201800043},
  urldate    = {2026-05-10},
  copyright  = {\copyright{} 2018 WILEY-VCH Verlag GmbH \& Co. KGaA, Weinheim},
  langid     = {english}
}

@article{duanInseparabilityCriterion,
  title     = {Inseparability {{Criterion}} for {{Continuous Variable Systems}}},
  author    = {Duan, Lu-Ming and Giedke, G. and Cirac, J. I. and Zoller, P.},
  year      = 2000,
  month     = mar,
  journal   = {Physical Review Letters},
  volume    = {84},
  number    = {12},
  pages     = {2722--2725},
  publisher = {American Physical Society},
  doi       = {10.1103/PhysRevLett.84.2722},
  urldate   = {2026-05-10}
}

@article{simonPeresHorodeckiSeparability,
  title     = {Peres-{{Horodecki Separability Criterion}} for {{Continuous Variable Systems}}},
  author    = {Simon, R.},
  year      = 2000,
  month     = mar,
  journal   = {Physical Review Letters},
  volume    = {84},
  number    = {12},
  pages     = {2726--2729},
  publisher = {American Physical Society},
  doi       = {10.1103/PhysRevLett.84.2726},
  urldate   = {2026-05-10}
}

@article{holsteinFieldDependence,
  title     = {Field {{Dependence}} of the {{Intrinsic Domain Magnetization}} of a {{Ferromagnet}}},
  author    = {Holstein, T. and Primakoff, H.},
  year      = 1940,
  month     = dec,
  journal   = {Physical Review},
  volume    = {58},
  number    = {12},
  pages     = {1098--1113},
  publisher = {American Physical Society},
  doi       = {10.1103/PhysRev.58.1098},
  urldate   = {2026-05-10}
}

@article{woodsMappingsOpen,
  title   = {Mappings of Open Quantum Systems onto Chain Representations and {{Markovian}} Embeddings},
  author  = {Woods, M. P. and Groux, R. and Chin, A. W. and Huelga, S. F. and Plenio, M. B.},
  year    = 2014,
  month   = mar,
  journal = {Journal of Mathematical Physics},
  volume  = {55},
  number  = {3},
  pages   = {032101},
  issn    = {0022-2488},
  doi     = {10.1063/1.4866769},
  urldate = {2026-05-16}
}

@article{odellWURSTKind,
  title    = {The {{WURST}} Kind of Pulses in Solid-State {{NMR}}},
  author   = {O'Dell, Luke A.},
  year     = 2013,
  month    = oct,
  journal  = {Solid State Nuclear Magnetic Resonance},
  volume   = {55--56},
  pages    = {28--41},
  issn     = {0926-2040},
  doi      = {10.1016/j.ssnmr.2013.10.003},
  urldate  = {2026-05-18}
}

@article{zenerNonadiabaticCrossing,
  title   = {Non-Adiabatic Crossing of Energy Levels},
  author  = {Zener, Clarence},
  year    = 1932,
  month   = sep,
  journal = {Proceedings of the Royal Society of London. Series A, Containing Papers of a Mathematical and Physical Character},
  volume  = {137},
  number  = {833},
  pages   = {696--702},
  issn    = {0950-1207},
  doi     = {10.1098/rspa.1932.0165},
  urldate = {2026-05-19}
}

@article{ranjanPulsedElectron,
  title   = {Pulsed Electron Spin Resonance Spectroscopy in the {{Purcell}} Regime},
  author  = {Ranjan, V. and Probst, S. and Albanese, B. and Doll, A. and Jacquot, O. and Flurin, E. and Heeres, R. and Vion, D. and Esteve, D. and Morton, J.J.L. and Bertet, P.},
  year    = 2020,
  month   = jan,
  journal = {Journal of Magnetic Resonance},
  volume  = {310},
  pages   = {106662},
  issn    = {10907807},
  doi     = {10.1016/j.jmr.2019.106662},
  urldate = {2025-12-15},
  langid  = {english}
}

@article{yangPhotonicIntegration,
  title      = {Photonic Integration of {{Er}} {\textsuperscript{3+}} :{{Y}} {\textsubscript{2}} {{SiO}} {\textsubscript{5}} with Thin-Film Lithium Niobate by Flip Chip Bonding},
  shorttitle = {Photonic Integration of {{Er}} {\textsuperscript{3+}}},
  author     = {Yang, Likai and Wang, Sihao and Shen, Mohan and Xu, Yuntao and Xie, Jiacheng and Tang, Hong X.},
  year       = 2021,
  month      = may,
  journal    = {Optics Express},
  volume     = {29},
  number     = {10},
  pages      = {15497},
  issn       = {1094-4087},
  doi        = {10.1364/OE.423659},
  urldate    = {2023-02-28},
  langid     = {english}
}

@article{hetetMultimodalProperties,
  title     = {Multimodal {{Properties}} and {{Dynamics}} of {{Gradient Echo Quantum Memory}}},
  author    = {H{\'e}tet, G. and Longdell, J. J. and Sellars, M. J. and Lam, P. K. and Buchler, B. C.},
  year      = 2008,
  month     = nov,
  journal   = {Physical Review Letters},
  volume    = {101},
  number    = {20},
  pages     = {203601},
  publisher = {American Physical Society},
  doi       = {10.1103/PhysRevLett.101.203601},
  urldate   = {2026-05-21}
}

@article{krausQuantumMemory,
  title     = {Quantum Memory for Nonstationary Light Fields Based on Controlled Reversible Inhomogeneous Broadening},
  author    = {Kraus, B. and Tittel, W. and Gisin, N. and Nilsson, M. and Kr{\"o}ll, S. and Cirac, J. I.},
  year      = 2006,
  month     = feb,
  journal   = {Physical Review A},
  volume    = {73},
  number    = {2},
  pages     = {020302},
  publisher = {American Physical Society},
  doi       = {10.1103/PhysRevA.73.020302},
  urldate   = {2026-05-21}
}

@article{maEliminationNoise,
  title     = {Elimination of Noise in Optically Rephased Photon Echoes},
  author    = {Ma, You-Zhi and Jin, Ming and Chen, Duo-Lun and Zhou, Zong-Quan and Li, Chuan-Feng and Guo, Guang-Can},
  year      = 2021,
  month     = jul,
  journal   = {Nature Communications},
  volume    = {12},
  number    = {1},
  pages     = {4378},
  publisher = {Nature Publishing Group},
  issn      = {2041-1723},
  doi       = {10.1038/s41467-021-24679-4},
  urldate   = {2026-05-21},
  copyright = {2021 The Author(s)},
  langid    = {english}
}

@article{hahnSpinEchoes,
  title     = {Spin {{Echoes}}},
  author    = {Hahn, E. L.},
  year      = 1950,
  month     = nov,
  journal   = {Physical Review},
  volume    = {80},
  number    = {4},
  pages     = {580--594},
  publisher = {American Physical Society},
  doi       = {10.1103/PhysRev.80.580},
  urldate   = {2026-05-22}
}

@article{gardinerInputOutputb,
  title      = {Input and Output in Damped Quantum Systems: {{Quantum}} Stochastic Differential Equations and the Master Equation},
  shorttitle = {Input and Output in Damped Quantum Systems},
  author     = {Gardiner, C. W. and Collett, M. J.},
  year       = 1985,
  month      = jun,
  journal    = {Physical Review A},
  volume     = {31},
  number     = {6},
  pages      = {3761--3774},
  publisher  = {American Physical Society},
  doi        = {10.1103/PhysRevA.31.3761},
  urldate    = {2026-05-25}
}

@article{osullivanSpinResonanceLinewidthsb,
  title   = {Spin-{{Resonance Linewidths}} of {{Bismuth Donors}} in {{Silicon Coupled}} to {{Planar Microresonators}}},
  author  = {O'Sullivan, James and Kennedy, Oscar W. and Zollitsch, Christoph W. and {\v S}im{\.e}nas, Mantas and Thomas, Christopher N. and Abdurakhimov, Leonid V. and Withington, Stafford and Morton, John J.L.},
  year    = 2020,
  month   = dec,
  journal = {Physical Review Applied},
  volume  = {14},
  number  = {6},
  pages   = {064050},
  issn    = {2331-7019},
  doi     = {10.1103/PhysRevApplied.14.064050},
  urldate = {2026-10-14},
  langid  = {english}
}

@article{rhoderickCurrentDistribution,
  title     = {Current {{Distribution}} in {{Thin Superconducting Films}}},
  author    = {Rhoderick, E. H. and Wilson, E. M.},
  year      = 1962,
  month     = jun,
  journal   = {Nature},
  volume    = {194},
  number    = {4834},
  pages     = {1167--1168},
  publisher = {Nature Publishing Group},
  issn      = {1476-4687},
  doi       = {10.1038/1941167b0},
  urldate   = {2026-10-14},
  copyright = {1962 Springer Nature Limited},
  langid    = {english}
}

@article{conollySelectiveAdiabatic,
  title   = {A Selective Adiabatic Spin-Echo Pulse},
  author  = {Conolly, Steven and Nishimura, Dwight and Macovski, Albert},
  year    = 1989,
  month   = jun,
  journal = {Journal of Magnetic Resonance (1969)},
  volume  = {83},
  number  = {2},
  pages   = {324--334},
  issn    = {0022-2364},
  doi     = {10.1016/0022-2364(89)90194-7},
  urldate = {2026-10-16}
}

@software{codeInhomogeneousDynamics,
  author = {Hanamura, Fumiya and Bao, Sicheng and Yoo, Jie Lerk and Auff{\`e}ves, Alexia and Touzard, Steven},
  title  = {\url{https://github.com/Qovelab/InhomogeneousSpinCavityDynamics.jl}},
  year   = {2026}
}
\begin{acknowledgments}
This research is supported by the Ministry of Education, Singapore, the National Research Foundation, Singapore, under grant ID NRFF14-2022-0002, and through the National Quantum Office, hosted in A*STAR, under its Centre for Quantum Technologies Funding Initiative (S24Q2d0009). We also acknowledge the funding support from The University of Sydney - National University of Singapore 2026 Ignition Grants, and from NUS Central HPC facility. S.B. acknowledges the support of the Singapore National Quantum Scholarship Scheme (NQSS). We thank John Morton and James O'Sullivan for permission to reproduce their experimental data.
\end{acknowledgments}

\section*{Data Availability}
The code used to generate the numerical results and figures presented in this work is publicly available in Ref.~\cite{codeInhomogeneousDynamics}.

\newpage
\appendix
\section{Waveguide model}\label{sec:waveguide_model_appendix}
This appendix derives the near-ground- and near-excited-state input--output equations summarized in Eqs.~\eqref{eq:main_ground_waveguide} and \eqref{eq:main_excited_waveguide}.

The coarse-grained operators \(\sigmah^\pm(g,\omega)\) and \(\sigmah^z(g,\omega)\), introduced in Sec.~\ref{sec:system}, are the average spin operators of TLSs within narrow coupling and detuning intervals around \((g,\omega)\). A cell of widths \(\dd{g}\) and \(\dd{\omega}\) contains \(N\rhog\nomega\,\dd{g}\dd{\omega}\) TLSs. Averaging rather than summing gives the ground- and excited-state values \(\sigmah^z(g,\omega)=\mp1\) used in Sec.~\ref{sec:bright}. In the continuum limit, the commutation relations are
\begin{subequations}\label{eq:coarse_grained_commutators}
\begin{align}
    &[\sigmah^+(g,\omega),\sigmah^-(g',\omega')]\nonumber\\
    &\quad=\frac{\delta(g-g')\delta(\omega-\omega')}{N\rhog\nomega}
             \sigmah^z(g,\omega),\\
    &[\sigmah^z(g,\omega),\sigmah^\pm(g',\omega')]\nonumber\\
    &\quad=\pm\frac{2\delta(g-g')\delta(\omega-\omega')}{N\rhog\nomega}
             \sigmah^\pm(g,\omega).
\end{align}
\end{subequations}

\subsection{Construction of the dark waveguide}
These commutators give the bright-mode commutator in Eq.~\eqref{eq:main_bright_commutator}.
Starting from the continuous Hamiltonian in Eq.~\eqref{eq:main_hamiltonian}, we use the Holstein--Primakoff regime described in Sec.~\ref{sec:bright} to construct the dark field and derive its input--output equations.

\subsubsection{Near-ground-state ensemble}

Near the ground state, we introduce the normalized spin-wave field
\begin{align}
    \bop{b}(g,\omega)=\sqrt{N\rhog\nomega}\sigmah^-(g,\omega).
\end{align}
Equation~\eqref{eq:coarse_grained_commutators} and \(\sigmah^z(g,\omega)\simeq-1\) give the bosonic relation
\begin{align}
    [\bop b(g,\omega),\bop b^\dagger(g',\omega')]
    =\delta(g-g')\delta(\omega-\omega').
\end{align}
Up to a constant energy offset, the detuning Hamiltonian is
\begin{align}
    \Hspin=\int\dd{g}\dd{\omega}\,\omega\,\bop{b}^\dagger(g,\omega)\bop{b}(g,\omega).
    \label{eq:spin_wave_detuning}
\end{align}
In this basis, the collective mode for coupling \(g\), defined in Eq.~\eqref{eq:main_class_mode}, is
\begin{align}
    \Ph(g)=\int\dd{\omega}\,\sqrt{n(\omega)}\,\bop b(g,\omega).
    \label{eq:coupling_class_mode}
\end{align}
The spin waves are not orthogonal to this mode:
\begin{align}
    [\bop b(g,\omega),\Ph^\dagger(g')]
    =\delta(g-g')\sqrt{n(\omega)}.
\end{align}
We therefore seek a field \(\Dh(g,\omega)\) satisfying the waveguide Hamiltonian in Eq.~\eqref{eq:main_waveguide_hamiltonian}, orthogonality to every collective mode \(\Ph(g)\), and canonical normalization, as specified in Sec.~\ref{sec:dark}. Here we assume Lorentzian broadening for \(\nomega\) and show that the coupling rate is frequency independent and equals its full width at half maximum \(\Gamma\).

We now determine \(\Dh(g,\omega)\) as a collective combination of the original spin waves:
\begin{align}
    \Dh(g,\omega)=\int\dd{g'}\dd{\omega'}\,K(g,\omega;g',\omega')\bop{b}(g',\omega').
    \label{eq:dark_kernel_ansatz}
\end{align}
Here \(K\) is a distribution, allowing singular kernels such as delta functions and principal values.
Equation~\eqref{eq:main_waveguide_hamiltonian} requires
\begin{align}
    [\Dh(g,\omega),\Hspin]
    =\omega\Dh(g,\omega)-i\sqrt{\frac{\Gamma}{2\pi}}\bop P(g).
\end{align}
The same commutator can also be evaluated using the original spin-wave Hamiltonian, Eq.~\eqref{eq:spin_wave_detuning}. Equating the coefficients of each spin-wave operator gives an equation for the kernel:
\begin{align}
    (\omega'-\omega)K(g,\omega;g',\omega')
    =-i\sqrt{\frac{\Gamma}{2\pi}}\delta(g-g')\sqrt{n(\omega')}.
\end{align}
To solve this equation, we use the fact that a distributional equation of the form \(xf(x)=g(x)\) has the general solution
\(f(x)=g(x)/(x+i0^+)+c\delta(x)\), where \(c\) is an arbitrary constant. The first term gives the required result when multiplied by \(x\), while the second vanishes under this multiplication and is therefore not fixed by the equation. Applying this identity with \(x=\omega'-\omega\), we obtain
\begin{align}
    K(g,\omega;g',\omega')={}&-i\sqrt{\frac{\Gamma}{2\pi}}\,
    \frac{\delta(g-g')\sqrt{n(\omega')}}{\omega'-\omega+i0^+}\nonumber\\
    &+\delta(\omega'-\omega)U_\omega(g,g').
    \label{eq:dark_kernel_general}
\end{align}
where \(U_\omega(g,g')\) is an undetermined kernel that mixes coupling coordinates at the same frequency \(\omega\).
To determine \(U_\omega\), we impose \([\Dh(g,\omega),\Ph^\dagger(g')]=0\). Substituting the kernel gives
\begin{align}
    &\sqrt{n(\omega)}U_\omega(g,g')\nonumber\\
    &\quad=i\sqrt{\frac{\Gamma}{2\pi}}\delta(g-g')
       \int\dd{\omega'}\,\frac{n(\omega')}{\omega'-\omega+i0^+}.
\end{align}
Thus the coupling coordinate is preserved:
\begin{align}
    U_\omega(g,g')=z(\omega)\delta(g-g'),
    \label{eq:dark_transverse_basis}
\end{align}
where
\begin{align}
    z(\omega)=i\sqrt{\frac{\Gamma}{2\pi n(\omega)}}
       \int\dd{\omega'}\,\frac{n(\omega')}{\omega'-\omega+i0^+}.
    \label{eq:dark_profile_condition}
\end{align}
The kernel therefore becomes
\begin{align}
    K(g,\omega;g',\omega')={}&\delta(g-g')\nonumber\\
       &\times\Biggl[z(\omega)\delta(\omega'-\omega)\nonumber\\
       &\qquad-i\sqrt{\frac{\Gamma}{2\pi}}\,
       \frac{\sqrt{n(\omega')}}{\omega'-\omega+i0^+}\Biggr].
\end{align}

The commutator follows directly by substituting the kernel, using Eq.~\eqref{eq:dark_profile_condition}, and combining the pole terms by partial fractions:
\begin{align}
    &[\Dh(g,\omega),\Dh^\dagger(g',\omega')]\nonumber\\
    &=\int\dd{g''}\dd{\omega''}\,K(g,\omega;g'',\omega'')
          K^*(g',\omega';g'',\omega'')\nonumber\\
    &=\delta(g-g')\Biggl\{
       z(\omega)z^*(\omega')\delta(\omega-\omega')\nonumber\\
    &\quad+\frac{\Gamma}{2\pi}\int\dd{\omega''}\,
       \frac{n(\omega'')}
       {(\omega''-\omega+i0^+)(\omega''-\omega'-i0^+)}
       \nonumber\\
    &\quad+\frac{i\sqrt{\Gamma/(2\pi)}
       \bigl[z(\omega)\sqrt{n(\omega)}
       +z^*(\omega')\sqrt{n(\omega')}\bigr]}
       {\omega-\omega'-i0^+}
       \Biggr\}\nonumber\\
    &=\delta(g-g')\Biggl\{
       |z(\omega)|^2\delta(\omega-\omega')
       +\frac{\Gamma}{2\pi}\lim_{\epsilon\to0^+}
       \int\dd{\omega''}\,n(\omega'')\nonumber\\
    &\qquad\times\Biggl[
       \frac{1}{(\omega''-\omega+i\epsilon)(\omega''-\omega'-i\epsilon)}
       \nonumber\\
    &\qquad\quad-\frac{1}{\omega-\omega'-2i\epsilon}\nonumber\\
    &\qquad\qquad\times\qty(
       \frac{1}{\omega''-\omega+i\epsilon}
       -\frac{1}{\omega''-\omega'-i\epsilon})
       \Biggr]\Biggr\}\nonumber\\
    &=|z(\omega)|^2\delta(g-g')\delta(\omega-\omega').
    \label{eq:dark_kernel_commutator}
\end{align}
The \(i0^+\) expressions are understood as limits taken after integration, retaining the equal-frequency contribution.
Canonical normalization therefore requires \(|z(\omega)|=1\).

When the broadening is Lorentzian, this condition is naturally satisfied:
\begin{align}
    n(\omega)=\frac{\Gamma/(2\pi)}{\omega^2+(\Gamma/2)^2}.
\end{align}
Contour integration gives, for \(\epsilon>0\),
\begin{align}
    i\sqrt{\frac{\Gamma}{2\pi}}
    \int\dd{\omega'}\,\frac{n(\omega')}{\omega'-\omega+i\epsilon}
    =\frac{\sqrt{\Gamma/(2\pi)}}{\Gamma/2+\epsilon+i\omega}.
\end{align}
Taking \(\epsilon\to0^+\) in Eq.~\eqref{eq:dark_profile_condition} gives
\begin{align}
    z(\omega)=\frac{\Gamma/2-i\omega}{\sqrt{\omega^2+(\Gamma/2)^2}},
    \qquad |z(\omega)|=1.
    \label{eq:dark_kernel_phase}
\end{align}
This verifies canonical normalization and completes the construction.
For a general broadening profile, these conditions require a frequency-dependent coupling rate \(\Gamma(\omega)\), derived in Appendix~\ref{sec:non_lorentzian}.

The resulting dark field is
\begin{align}
    \Dh(g,\omega)={}&z(\omega)\bop{b}(g,\omega)\nonumber\\
    &-i\sqrt{\frac{\Gamma}{2\pi}}
      \int\dd{\omega'}\,
      \frac{\sqrt{n(\omega')}\bop{b}(g,\omega')}{\omega'-\omega+i0^+}.
    \label{eq:appendix_dark_modes}
\end{align}
Substituting the spin-wave definition recovers Eq.~\eqref{eq:main_dark_modes} in the main text.

Using the waveguide coordinate of Eq.~\eqref{eq:main_waveguide_field}, we Fourier transform the detuning:
\begin{align}
    \Dh(g,\tau;t)=\frac{1}{\sqrt{2\pi}}\int\dd{\omega}\,
    e^{-i\omega\tau}\Dh(g,\omega;t).
\end{align}
The Hamiltonian in Eq.~\eqref{eq:main_waveguide_hamiltonian} and the commutation relations give
\begin{align}
    (\partial_t-\partial_\tau)\Dh(g,\tau;t)
    =-\sqrt{\Gamma}\delta(\tau)\bop P(g,t).
    \label{eq:dark_propagation}
\end{align}
Thus \(\tau\) plays the role of position along the waveguide: away from \(\tau=0\), wavepackets propagate toward decreasing \(\tau\), and each channel interacts with \(\bop P(g)\) at \(\tau=0\). The region \(\tau>0\) is the incoming side. The resonator couples to these channels only through the collective profile \(u(g)\).

We next relate wavepacket mode functions in the spin-wave basis \(\bop b(g,\omega)\) and the dark-waveguide basis \(\Dh(g,\omega)\). For an envelope \(\tilde f_D(g,\tau)\), define
\begin{align}
    f_D(g,\omega)=\frac{1}{\sqrt{2\pi}}\int\dd{\tau}\,
       e^{i\omega\tau}\tilde f_D(g,\tau).
\end{align}
The same wavepacket operator can be written as
\begin{align}
    &\int\dd{g}\dd{\tau}\,\tilde f_D^*(g,\tau)\Dh(g,\tau)\nonumber\\
    &\quad=\int\dd{g}\dd{\omega}\,f_D^*(g,\omega)\Dh(g,\omega)\nonumber\\
    &\quad=\int\dd{g}\dd{\omega}\,f_b^*(g,\omega)\bop b(g,\omega),
\end{align}
so the two spectral mode functions satisfy
\begin{align}
    f_b(g',\omega')=\int\dd{g}\dd{\omega}\,
    K^*(g,\omega;g',\omega')f_D(g,\omega).
    \label{eq:dark_mode_kernel_relation}
\end{align}
To evaluate this relation on either side of the junction, we use
\begin{align}
    \frac{1}{\sqrt{2\pi}}\int\dd{\omega}\,
       \frac{e^{-i\omega\tau}}{\omega'-\omega+i0^+}
    =-i\sqrt{2\pi}\Theta(-\tau)e^{-i\omega'\tau}.
    \label{eq:dark_pole_support}
\end{align}
Consequently,
\begin{align}
    \Dh(g,\tau)={}&\frac{1}{\sqrt{2\pi}}\int\dd{\omega}\,
       z(\omega)\bop b(g,\omega)e^{-i\omega\tau}\nonumber\\
    &-\sqrt{\Gamma}\Theta(-\tau)
       \int\dd{\omega'}\,\sqrt{n(\omega')}\bop b(g,\omega')e^{-i\omega'\tau}.
    \label{eq:dark_field_pulse_support}
\end{align}
For a wavepacket supported in \(\tau>0\), the second term does not contribute. For support in \(\tau<0\), it contributes in full. Using
\(z(\omega)-\Gamma/\sqrt{\omega^2+(\Gamma/2)^2}=-z^*(\omega)\),
we obtain
\begin{subequations}\label{eq:dark_wavepacket_basis_change}
\begin{align}
    f_b(g,\omega)&=z^*(\omega)f_D(g,\omega),&&\tau>0,\\
    f_b(g,\omega)&=-z(\omega)f_D(g,\omega),&&\tau<0.
\end{align}
\end{subequations}
Here the inequalities specify the support of the wavepacket envelope. On either side, every transverse profile acquires the same scalar spectral phase.

Integrating Eq.~\eqref{eq:dark_propagation} across the junction gives
\begin{align}
    \Dh(g,0^-;t)=\Dh(g,0^+;t)-\sqrt{\Gamma}\bop P(g,t).
\end{align}
Projecting this jump onto \(u(g)\) and using the boundary fields in Eq.~\eqref{eq:main_boundary_fields} gives the bright-mode input--output relation. To obtain the collective-mode dynamics, we use the symmetric boundary value \(\Dh(g,0;t)=[\Dh(g,0^+;t)+\Dh(g,0^-;t)]/2\), which gives
\begin{align}
    \dot{\bop P}(g,t)=-\frac{\Gamma}{2}\bop P(g,t)-iG u(g)\ah(t)
    +\sqrt{\Gamma}\Dh(g,0^+;t).
\end{align}
Projecting onto \(u(g)\) and including the resonator yields Eq.~\eqref{eq:main_ground_waveguide}.

\subsubsection{Near-excited-state ensemble}
The same construction can be applied near the fully excited state, where the bosonic excitations describe small departures from population inversion. We start by defining
\begin{align}
    \tilde{\bop b}(g,\omega)&=\sqrt{N\rhog\nomega}\sigmah^+(g,\omega),\\
    \tilde{\bop P}(g)&=\int\dd{\omega}\,\sqrt{n(\omega)}\tilde{\bop b}(g,\omega),\\
    \Pht&=\int\dd{g}\,u(g)\tilde{\bop P}(g).
\end{align}
Here \(\tilde{\bop b}\), \(\tilde{\bop P}\), and \(\Pht\) annihilate missing excitations. Removing an excitation reverses the sign of its detuning energy, so the detuning Hamiltonian becomes
\begin{align}
    \Hspin=-\int\dd{g}\dd{\omega}\,\omega\,
        \tilde{\bop b}^\dagger(g,\omega)\tilde{\bop b}(g,\omega),
\end{align}
up to a constant. To retain the positive coupling convention for the collective modes and their continua, choose
\begin{align}
    \Dht(g,\omega)=-\int\dd{g'}\dd{\omega'}\,
             K(g,\omega;g',\omega')\tilde{\bop b}(g',\omega').
    \label{eq:excited_dark_kernel}
\end{align}
The overall minus sign is a phase convention for the dark continuum. Then
\begin{align}
    \Hspin={}&-\int\dd{g}\dd{\omega}\,\omega\,\Dht^\dagger\Dht\nonumber\\
    &+i\sqrt{\frac{\Gamma}{2\pi}}\int\dd{g}\dd{\omega}\,\nonumber\\
    &\quad\times\qty[\tilde{\bop P}^\dagger(g)\Dht(g,\omega)
       -\Dht^\dagger(g,\omega)\tilde{\bop P}(g)],
\end{align}
with canonical commutators and \([\Dht(g,\omega),\tilde{\bop P}^\dagger(g')]=0\). Projecting the collective modes onto \(u(g)\) recovers the resonator-coupled mode \(\Pht\).
The change in sign of the free Hamiltonian reverses propagation. In the \(\tau\) representation, the equation of motion is
\begin{align}
    (\partial_t+\partial_\tau)\Dht(g,\tau;t)
      =-\sqrt{\Gamma}\delta(\tau)\tilde{\bop P}(g,t).
\end{align}
The opposite propagation direction exchanges the input and output sides, as specified in Eq.~\eqref{eq:main_boundary_fields_excited}. Using the symmetric boundary value gives
\begin{align}
    \dot{\tilde{\bop P}}(g,t)={}&-\frac{\Gamma}{2}\tilde{\bop P}(g,t)
       -iG u(g)\ah^\dagger(t)\nonumber\\
    &+\sqrt{\Gamma}\Dht(g,0^-;t).
\end{align}
Projecting onto \(u(g)\) and including the resonator yields Eq.~\eqref{eq:main_excited_waveguide}.

\subsubsection{\pipulse{}}
An ideal \pipulse{} exchanges \(\sigmah^-\) and \(\sigmah^+\), up to a constant phase. It maps the ground-state dark modes onto their excited-state counterparts and reverses propagation. With Eq.~\eqref{eq:excited_dark_kernel}, this map includes a uniform minus sign, which can be absorbed into the global pulse phase.
Explicitly, since the kernel \(K\) is unchanged and \(\bop b(g,\omega)\to\tilde{\bop b}(g,\omega)\),
\begin{align}
    \Dh(g,\omega)&\to-\Dht(g,\omega),\\
    \Dht(g,\omega)&\to-\Dh(g,\omega).
\end{align}
Fourier transforming gives the same exchange at each position \(\tau\). Thus an ideal pulse leaves the excitation at its position in the waveguide while reversing its propagation direction, as described in Proposition~2 of the main text.
For a realistic pulse,
\begin{align}
    \sigmah^-(g,\omega)\leftrightarrow
    \sigmah^+(g,\omega)e^{i\phi(g,\omega)},
    \label{eq:sigma_phase}
\end{align}
so the spin-wave mode function acquires the phase of each TLS:
\begin{align}
    f_b(g,\omega)\to e^{\pm i\phi(g,\omega)}f_b(g,\omega).
    \label{eq:f_trans_compact}
\end{align}
The plus and minus signs denote ground-to-excited and excited-to-ground inversion, respectively; the uniform minus sign from the definition of \(\Dht\) is absorbed into the pulse phase. For a wavepacket supported entirely on either side of the junction, Eq.~\eqref{eq:dark_wavepacket_basis_change} relates the two bases by a scalar spectral phase. This phase cancels when converting to the spin-wave basis and back, giving
\begin{align}
    f_D(g,\omega)\to e^{\pm i\phi(g,\omega)}f_D(g,\omega).
    \label{eq:dark_pulse_mode_function}
\end{align}
The same pointwise rule therefore holds for every transverse profile on either side. It assumes that pulse-induced reshaping keeps the packet on that side and away from the junction, so that no collective-mode component must be retained; small tails give the corresponding approximation. For an initial wavepacket proportional to \(u(g)\), its subsequent bright-profile overlap is multiplied by \(\int\dd{g}\,u(g)^2e^{i\phi(g,\omega)}\), whose modulus is the silencing factor used in the main text.

\subsection{Frequency response of the system and cooperativity}\label{sec:freq_response_appendix}

The frequency response follows from Eqs.~\eqref{eq:main_ground_waveguide} and \eqref{eq:main_excited_waveguide} for the near-ground- and near-excited-state ensembles, respectively.
We use the Fourier-transform convention
\begin{align}
    \Oh(t)=\frac{1}{\sqrt{2\pi}}\int \dd{\omega} e^{-i\omega t}\Oh(\omega).
\end{align}

For the near-ground-state case, the frequency-domain input--output relation takes the beam-splitter form
\begin{align}
    \aout(\omega) & = r(\omega)\ain(\omega)+t(\omega)\Pin(\omega), \label{eq:bs_a}\\
    \Pout(\omega) & = t'(\omega)\ain(\omega)+r'(\omega)\Pin(\omega),\label{eq:bs_P}
\end{align}
where
\begin{align}
    r(\omega)  & = \frac{-\qty(\frac{\kappa}{2}+i(\omega-\delta))\qty(\frac{\Gamma}{2}-i\omega)+G^2}
    {\qty(\frac{\kappa}{2}-i(\omega-\delta))\qty(\frac{\Gamma}{2}-i\omega)+G^2},\label{eq:ground_r}  \\
    t(\omega)  & = i\frac{\sqrt{\kappa\Gamma}G}
    {\qty(\frac{\kappa}{2}-i(\omega-\delta))\qty(\frac{\Gamma}{2}-i\omega)+G^2}, \label{eq:ground_t} \\
    t'(\omega) & = i\frac{\sqrt{\kappa\Gamma}G}
    {\qty(\frac{\kappa}{2}-i(\omega-\delta))\qty(\frac{\Gamma}{2}-i\omega)+G^2},\label{eq:ground_tp} \\
    r'(\omega) & = \frac{-\qty(\frac{\kappa}{2}-i(\omega-\delta))\qty(\frac{\Gamma}{2}+i\omega)+G^2}
    {\qty(\frac{\kappa}{2}-i(\omega-\delta))\qty(\frac{\Gamma}{2}-i\omega)+G^2}.\label{eq:ground_rp}
\end{align}

For the near-excited-state case, we obtain the two-mode-squeezing relation
\begin{align}
    \aout(\omega)          & = \tilde{r}(\omega)\ain(\omega)+\tilde{t}(\omega)\Ptin^\dagger(\omega), \label{eq:ts_a}  \\
    \Ptout^\dagger(\omega) & = \tilde{t}'(\omega)\ain(\omega)+\tilde{r}'(\omega)\Ptin^\dagger(\omega),\label{eq:ts_P}
\end{align}
where
\begin{align}
    \tilde{r}(\omega)  & = \frac{-\qty(\frac{\kappa}{2}+i(\omega-\delta))\qty(\frac{\Gamma}{2}-i\omega)-G^2}
    {\qty(\frac{\kappa}{2}-i(\omega-\delta))\qty(\frac{\Gamma}{2}-i\omega)-G^2}, \label{eq:excited_r}       \\
    \tilde{t}(\omega)  & = i\frac{\sqrt{\kappa\Gamma}G}
    {\qty(\frac{\kappa}{2}-i(\omega-\delta))\qty(\frac{\Gamma}{2}-i\omega)-G^2}, \label{eq:excited_t}       \\
    \tilde{t}'(\omega) & = -i\frac{\sqrt{\kappa\Gamma}G}
    {\qty(\frac{\kappa}{2}-i(\omega-\delta))\qty(\frac{\Gamma}{2}-i\omega)-G^2}, \label{eq:excited_tp}      \\
    \tilde{r}'(\omega) & = \frac{-\qty(\frac{\kappa}{2}-i(\omega-\delta))\qty(\frac{\Gamma}{2}+i\omega)-G^2}
    {\qty(\frac{\kappa}{2}-i(\omega-\delta))\qty(\frac{\Gamma}{2}-i\omega)-G^2}.\label{eq:excited_rp}
\end{align}

At resonance, \(\delta=0\) and \(\omega=0\), the response of the system is characterized by the collective cooperativity
\begin{align}
    C = \frac{4G^2}{\kappa\Gamma}.
\end{align}
For the near-ground-state case, the response reduces to
\begin{align}
    \aout & =-\frac{1-C}{1+C}\ain+i\frac{2\sqrt{C}}{1+C}\Pin, \label{eq:resonant_inout_ground_a} \\
    \Pout & =i\frac{2\sqrt{C}}{1+C}\ain-\frac{1-C}{1+C}\Pin.\label{eq:resonant_inout_ground_P}
\end{align}
Especially when \(C=1\), the resonator input is completely absorbed into the dark waveguide, corresponding to the impedance-matched case \cite{afzeliusImpedancematchedCavity}.

For the near-excited-state case, the resonant response is given by
\begin{align}
    \aout          & =-\frac{1+C}{1-C}\ain+i\frac{2\sqrt{C}}{1-C}\Ptin^\dagger, \label{eq:resonant_inout_excited_a} \\
    \Ptout^\dagger & =-i\frac{2\sqrt{C}}{1-C}\ain-\frac{1+C}{1-C}\Ptin^\dagger.\label{eq:resonant_inout_excited_P}
\end{align}
The power gain of the two-mode squeezing, \(\qty[(1+C)/(1-C)]^2\), diverges when \(C\to1\), corresponding to the laser oscillation \cite{julsgaardDynamicalEvolution}.

\subsection{Mode function and silencing factor}\label{sec:silencing_factor_appendix}
We use the normalized wavepacket operator of Eq.~\eqref{eq:main_wavepacket_mode}, writing its dark-waveguide mode function as \(f(g,\omega)\) in this subsection. Its bosonic commutator is \([\Dh_f,\Dh_f^\dagger]=1\).
A state stored near the ground state can be expanded as
\begin{align}
    \sum_{n=0}^\infty\frac{c_n}{\sqrt{n!}}
        (\Dh_f^\dagger)^n\ket{0}.
\end{align}
Near the fully excited state, the corresponding vacuum and creation operator describe missing excitations.
Unlike the bright mode, these wavepacket operators are built from modes orthogonal to \(\Ph\). Their interaction with the bright mode is determined by their transverse overlap with \(u(g)\) when they reach the junction \(\tau=0\).

As an example, consider an input wavepacket with mode function \(f_\mrin(\omega)\), normalized as
\begin{align}
    \int \dd{\omega} |f_\mrin(\omega)|^2 = 1.
\end{align}
The corresponding input field operator is
\begin{align}
    \ah_{\mrin,f}= \int \dd{\omega} f_\mrin^*(\omega)\ain(\omega).
\end{align}
When this wavepacket is absorbed into the ensemble through the resonator, the mode function of the stored quantum state takes the form
\begin{align}
    f(g,\omega) = \frac{1}{\sqrt{\eta}} u(g)t'(\omega)f_\mrin(\omega),\label{eq:absorbed_mode}
\end{align}
where \(t'(\omega)\) is the frequency-dependent transmittance defined in Eq.~\eqref{eq:ground_tp}. The total absorption efficiency of the state is given by
\begin{align}
    \eta=\int \dd{\omega} |t'(\omega)f_\mrin(\omega)|^2.
\end{align}
The factor \(1/\sqrt{\eta}\) normalizes the stored mode function, so that
\begin{align}
    \int\dd{g}\int\dd{\omega} |f(g,\omega)|^2=1.
\end{align}

For a \pipulse{}, Eq.~\eqref{eq:dark_pulse_mode_function} gives \(f_\pi(g,\omega)=e^{i\phi(g,\omega)}f(g,\omega)\), provided the packet remains away from the junction on the same side. For each frequency, define the complex bright-profile overlap and its modulus:
\begin{align}
    \mathcal F(\omega)&=\int\dd{g}\,|u(g)|^2e^{i\phi(g,\omega)},\\
    F(\omega)&=|\mathcal F(\omega)|,\qquad 0\leq F(\omega)\leq1.
    \label{eq:silence_factor_appendix}
\end{align}
Here \(F(\omega)\) is the silencing factor defined in Eq.~\eqref{eq:silence_factor}. We omit the frequency argument when its variation is negligible. For a wavepacket initially stored in the bright-mode profile,
\begin{align}
    \int\dd{g}\,u(g)f_\pi(g,\omega)
    =\mathcal F(\omega)\int\dd{g}\,u(g)f(g,\omega).
\end{align}
The echo amplitude is then
\begin{align}
    \ev*{\ah_{\mrout,\mathrm{e}}(\omega)}
    =\tilde{t}'(\omega)t'(\omega)f_\mrin(\omega)\mathcal F(\omega)
       \ev*{\ah_{\mrin,f}}.
    \label{eq:silencing_effect}
\end{align}
Here \(\tilde{t}'(\omega)\) is defined in Eq.~\eqref{eq:excited_tp}. A fully quantum description of the echo mode operator is given in Appendix~\ref{sec:rose_appendix}.

\section{Effect of ARP}\label{sec:ARP_appendix}
\subsection{Adiabatic rapid passage}
Here we derive the dynamical phase stated in Eq.~\eqref{eq:main_arp_phase_integral} and its frequency- and coupling-dependent contributions.
Adiabatic rapid passage (ARP) is a widely used technique for robust inversion of an ensemble of two-level systems~\cite{malinovskyGeneralTheory}.
It provides a natural implementation of a \pipulse{} in the presence of inhomogeneous broadening, since it can achieve near-perfect inversion across the full inhomogeneous profile and keep the system in the regime where the waveguide model applies.
An ARP pulse is a chirped pulse of the form
\begin{align}
    \ev*{\ah(t)} & =A(t) \exp[i\qty(\omega_0 t + \frac{k}{2} t^2)],\label{eq:arp_pulse}
\end{align}
where \(A(t)\) is the pulse envelope, \(\omega_0\) is the initial frequency, and \(k\) is the chirp rate.
The instantaneous frequency of the pulse is given by \(\omega_0 + k t\).
We consider the case where the frequency sweep is sufficiently large to cover the entire inhomogeneous broadening.

We assume that there is no initial excitation in the bright mode before the ARP pulse is applied, as is usually the case in quantum memory protocols where the ARP pulse is applied neither during absorption nor during echo emission.
For a sufficiently strong ARP pulse, the contribution from the bright mode then remains negligible.
The dynamics can therefore be treated as independent single-TLS dynamics, with each TLS driven by the classical ARP field.
The coupling \(g\) and the amplitude of the pulse \(A(t)\) set the instantaneous Rabi frequency \(\Omega(t) =2 g A (t)\) for each TLS.
Assuming the pulse satisfies the adiabatic condition
\begin{align}
    \frac{\Omega^2}{k} & \gg 1
\end{align}
we consider the dynamics of a single TLS at frequency \(\omega\) and coupling \(g\). In the rotating frame of the drive, the effective Hamiltonian is given by
\begin{align}
    \Hh(t) & =\frac12 \qty(\omega - \omega_0 - k t) \sigmah^z + \frac{1}{2}\Omega (\sigmah^-+\sigmah^+)
\end{align}
The eigenstates of the Hamiltonian are given by
\begin{align}
    \ket{+} & =\cos\frac{\theta(t)}{2} \ket{e} + \sin\frac{\theta(t)}{2} \ket{g}  \\
    \ket{-} & =-\sin\frac{\theta(t)}{2} \ket{e} + \cos\frac{\theta(t)}{2} \ket{g}
\end{align}
where \(\ket{e},\ket{g}\) are the excited and ground states, respectively, and \(\theta(t)\) is defined as
\begin{align}
    \theta(t) & =\arctan\qty(\frac{\Omega(t)}{\Delta\omega(t)})
\end{align}
where \(\Delta\omega(t)=\omega - \omega_0 - k t\).
The corresponding instantaneous energy splitting is given by
\begin{align}
    \Omeff(t) & =\sqrt{\Omega(t)^2 +\Delta\omega(t)^2}
\end{align}
\(\Omeff(t)\) can be seen as the instantaneous angular velocity of the rotation of the Bloch vector.
If the system is initially in the ground state \(\ket{g}\), it will be adiabatically transferred to the excited state \(\ket{e}\), and vice versa, hence achieving a \pipulse{}. However, during this transfer the state also accumulates dynamical phase between the ground and excited states, which is given by the integral of \(\Omeff(t)\) over time as
\begin{align}
    \phi(t) & =-\int_{t_0}^{t} \Omeff(t') \dd{t'}=-\frac{1}{k}\int_{\omega(t_0)}^{\omega(t)} \Omeff(\omega') \dd{\omega'}.\label{eq:dynamical_phase_int}
\end{align}
This phase depends on both the coupling \(g\) and the TLS detuning \(\omega\), and gives the phase appearing in Eq.~\eqref{eq:sigma_phase}.

In particular, when the amplitude is constant \(A(t)=A_0\) over the range of interest, we have
\begin{align}
    \phi(t) & = \frac{1}{2k}\Bigg[\Delta\omega(t)\sqrt{\Omega^2 + \Delta\omega(t)^2}\nonumber\\
            & \qquad+ \Omega^2\sinh^{-1}\qty(\frac{\Delta\omega(t)}{\Omega})\Bigg]_{t_0}^{t}.
\end{align}
In the limit of \(t\to \infty\) and \(t_0\to -\infty\), this phase is approximately given by
\begin{align}
    \begin{split}
    \phi & \approx \frac{\Omega^2}{2k}+\frac{\Delta\omega(t)^2+\Delta\omega(t_0)^2}{2k}\\
    &\quad +\frac{\Omega^2}{2k}\qty[\ln\qty|\frac{2\Delta\omega(t)}{\Omega}|+\ln\qty|\frac{2\Delta\omega(t_0)}{\Omega}|]
    \end{split} \\
         & = \int_{t_0}^0\Delta\omega(t') dt'-\int_0^t\Delta\omega(t') dt'\label{eq:dynamical_phase_linear_freq}                                                                                     \\
         & \quad+\frac{(\omega-\omega_0)^2}{k}\label{eq:dynamical_phase_quad_freq}                                                                                                                   \\
         & \quad+\frac{\Omega^2}{2k}\qty(1+\ln\qty|\frac{2\Delta\omega(t)}{\Omega}|+\ln\qty|\frac{2\Delta\omega(t_0)}{\Omega}|)\label{eq:dynamical_phase_coupling}
\end{align}
The first term, Eq.~\eqref{eq:dynamical_phase_linear_freq}, is the phase accumulated before and after the \pipulse{} in the idealized case where all TLSs are instantaneously flipped at \(t=0\).
The second term, Eq.~\eqref{eq:dynamical_phase_quad_freq}, is the quadratic frequency-dependent phase arising from the chirp.
The third term, Eq.~\eqref{eq:dynamical_phase_coupling}, is the nontrivial dynamical phase contribution that depends on both the coupling \(g\) and the detuning \(\omega\).
\subsection{Amplified chirped echo}\label{sec:ace_appendix}
We first consider the case of homogeneous coupling, \(g_j=g\) for all \(j\).
In this case, the only relevant phase term is the quadratic frequency-dependent term Eq.~\eqref{eq:dynamical_phase_quad_freq}:
\begin{align}
    \phi(\omega)=\frac{(\omega-\omega_0)^2}{k}.
\end{align}
Thus, according to Eq.~\eqref{eq:main_pulse_mode_function}, for a narrow-band input wavepacket with frequency-mode function \(f(\omega)\), the mode function of the echo becomes
\begin{align}
    f_\mathrm{e}(\omega)=f^*(\omega)\exp\qty[\pm i\phi(\omega)].
\end{align}
Thus, the echo is chirped with chirp rate \(k/2\), which is half the chirp rate of the ARP pulse.
This can also be understood in the time-domain picture: a TLS with detuning \(\omega-\omega_0\) is inverted at time \(t=(\omega-\omega_0)/k\), and its echo is emitted at time \(t=2(\omega-\omega_0)/k\).
The echo frequency therefore changes in time at rate \(k/2\).

For example, for a Gaussian input envelope
\begin{align}
    \ev*{\ah_{\mrin}(t)}
    =
    \alpha_0\exp\qty[-\frac{t^2}{2\sigma^2}],
\end{align}
the echo amplitude is given by
\begin{align}
\begin{split}
    \ev*{\ah_{\mrout,\mathrm{e}}(t)}
    &=
    t_\mathrm{e}
    \frac{\alpha_0\sigma}{\sqrt{\sigma^2\mp 2i/k}}
    \exp\qty[-\frac{t^2}{2\qty(\sigma^2+\frac{4}{\sigma^2k^2})}]\\
    &\quad \cdot\exp\qty[\mp i\frac{k t^2}{\sigma^4k^2+4}],
\end{split}
\end{align}
where \(t_\mathrm{e}\) is the echo-amplitude coefficient in the absence of the chirp-induced spectral phase, determined by the frequency response discussed in Appendix~\ref{sec:freq_response_appendix}.
The final quadratic phase factor describes the chirp of the echo.

Importantly, although the chirp spreads the echo in time and suppresses its amplitude at any fixed time, it does not reduce the total echo energy or its frequency bandwidth.
Therefore, in the homogeneous-coupling case, an ARP \pipulse{} does not silence the first echo.
Instead, it produces a noisy first echo with an additional chirp, which we refer to as the \textit{amplified chirped echo (ACE)}.
\subsection{Coupling inhomogeneity silences echoes}\label{sec:silencing_appendix}
If the coupling is inhomogeneous, i.e., \(g_j\) is not constant, the coupling-dependent term Eq.~\eqref{eq:dynamical_phase_coupling} also contributes to the dynamical phase, in addition to the chirping effect discussed above.
As a result, the silencing factor \(F(\omega)\) in Eq.~\eqref{eq:silence_factor_appendix} is reduced, and the echo is suppressed.

We now evaluate how the silencing factor scales with the width of the coupling distribution.
We assume a Gaussian distribution
\begin{align}
    \rhog=\frac{1}{\sqrt{2\pi}\sigmag}\exp\qty[-\frac{(g-g_0)^2}{2\sigmag^2}],
\end{align}
and consider the limit of small variation, \(\sigmag\ll \bar{g}\).
We approximate the coupling-dependent phase in Eq.~\eqref{eq:dynamical_phase_coupling} as
\begin{align}
    \phi(g)\sim\phi(g_0)+s_0(g-g_0),
\end{align}
where
\begin{align}
    s_0=\eval{\pdv{\phi}{g}}_{g=g_0}.
\end{align}
Using \(\Omega(g)=2Ag\), the leading-order scaling is
\begin{align}
    s_0\sim\frac{4A^2 g_0}{k},\label{eq:approx_s0}
\end{align}
up to logarithmic factors that vary slowly with \(g\).

Then, up to an overall phase factor, the complex overlap becomes
\begin{align}
    \mathcal F
     & \sim \frac{1}{\bar{g}^2}\int \dd{g} g^2 \rhog e^{is_0 g} \\
     & \sim \exp\qty[-\frac{\sigmag^2s_0^2}{2}+is_0 g_0].
\end{align}
Therefore, the silencing factor is approximately
\begin{align}
    F\sim \exp\qty[-\frac{\sigmag^2s_0^2}{2}],
\end{align}
or, using Eq.~\eqref{eq:approx_s0}, in terms of the average Rabi frequency \(\Omega_0=2A g_0\) and the relative coupling inhomogeneity \(\sigma_g/g_0\),
\begin{align}
    F\sim \exp\qty[-\frac{1}{2}\qty(\frac{\sigma_g}{g_0}\frac{\Omega_0^2}{k})^2].
\end{align}
Introducing the ARP adiabaticity factor
\begin{align}
    Q=\frac{\Omega_0^2}{k},
\end{align}
this expression becomes
\begin{align}
    F\sim \exp\qty[-\frac{1}{2}\qty(\frac{\sigma_g}{g_0})^2Q^2].\label{eq:silencing_adaibaticity}
\end{align}
Thus, for a fixed relative coupling inhomogeneity \(\sigma_g/g_0\), a more adiabatic pulse produces a larger coupling-dependent phase dispersion and therefore stronger echo silencing.

In the \(s,\tau\) representation introduced in Eq.~\eqref{eq:main_st_mode_function}, the same transformation corresponds to a displacement of the mode function in the \(s\) direction:
\begin{align}
    \tilde{f}(s,\tau)\rightarrow\tilde{f}(s\pm s_0,\tau),
\end{align}
up to an overall phase factor.
In this picture, echo silencing can be understood as a reduction of the mode overlap caused by this displacement, as shown in Fig.~2 of the main text.

In a realistic implementation, the ARP pulse has a finite duration and is filtered by the resonator, so the intracavity amplitude \(A(t)\) is generally time dependent.
For the theoretical curves in Figs.~3(c) and 3(d) of the main text, we numerically calculate the intracavity amplitude \(A(t)\) and obtain the coefficient \(s_0\) by evaluating the integral in Eq.~\eqref{eq:dynamical_phase_int} at zero detuning, \(\omega=\delta=0\).

\section{ROSE protocol}\label{sec:rose_appendix}
\begin{figure*}[htb]
    \centering
    \includegraphics[scale=1]{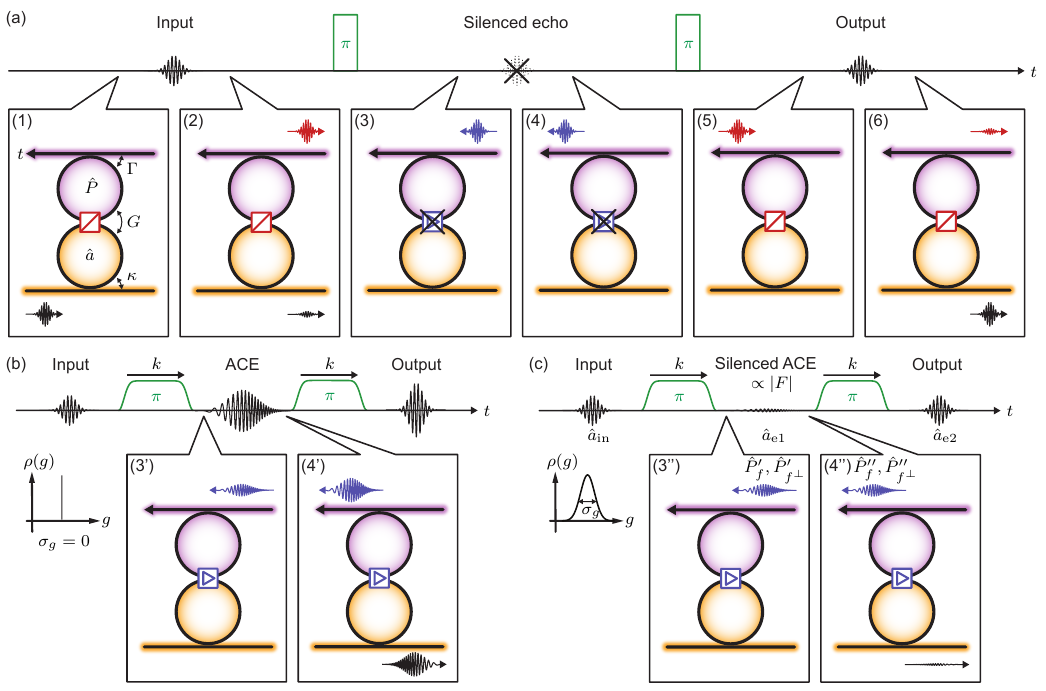}
    \caption{Description of the ROSE protocol in the waveguide model.
        (a) Ideal \pipulse{}s with an additional echo-silencing mechanism.
        (1),(2) Part of the input pulse is absorbed into the dark waveguide, with an efficiency determined by the cooperativity [Eq.~\eqref{eq:rose_efficiency}].
        (3) A \pipulse{} reverses the propagation direction of the excitation in the dark waveguide.
        (4) When the excitation returns to the bright mode \(\Ph\), the excited ensemble produces a noisy echo through the two-mode-squeezing interaction.
        This echo must be silenced by an external mechanism, such as phase mismatching or cavity detuning.
        (5) A second \pipulse{} reverses the propagation direction again.
        (6) The stored state is re-emitted with the same efficiency as the writing process.
        (b) ARP pulses used as \pipulse{}s without an additional silencing mechanism, for homogeneous coupling.
        (3') An ARP \pipulse{} chirps the excitation in the dark waveguide by imprinting a frequency-dependent phase \(\phi(\omega)\) [Appendix~\ref{sec:ace_appendix}].
        (4') The excited-state two-mode-squeezing interaction emits an amplified chirped echo (ACE), which amplifies the stored state and adds ASE noise.
        Although this chirp is cancelled by the second \pipulse{}, this process also adds noise to the final output.
        (c) ARP pulses used as \pipulse{}s without an additional silencing mechanism, for inhomogeneous coupling.
        (3'') The first ARP pulse still chirps the excitation, but also imprints a coupling-dependent phase \(\phi(g,\omega)\) [Appendix~\ref{sec:silencing_appendix}].
        (4'') The ACE is suppressed by the silencing factor \(F\), which corresponds to the overlap between the phase-imprinted mode and the original bright mode [Appendix~\ref{sec:ARP_appendix}].
        The coupling-dependent phase is cancelled by the second \pipulse{}, allowing the final echo to be retrieved with no additional noise in the limit \(F\to0\) and \(C\to1\).}
    \label{fig:rose_appendix}
\end{figure*}

The revival of silenced echo (ROSE) protocol~\cite{damonRevivalSilenced} is an echo-based quantum memory protocol that uses two \pipulse{}s to retrieve the stored quantum state in a second echo, while the noisy first echo is suppressed by a silencing mechanism.
In the waveguide model, the protocol can be understood as the back-and-forth propagation of an excitation pulse in the dark waveguide, as illustrated in Fig.~\ref{fig:rose_appendix}(a).

After the first \pipulse{}, the excitation rephases while the ensemble is inverted.
The resulting first echo is therefore generated by the two-mode-squeezing interaction, which entangles the dark modes with the resonator output mode and destroys the stored quantum state.
The key idea of the ROSE protocol is to silence this first echo and retrieve the quantum state only through the second echo.
In the original proposal, this silencing was achieved by phase mismatching~\cite{damonRevivalSilenced}, while cavity detuning has also been used in later implementations~\cite{afzeliusProposalCoherent,julsgaardQuantumMemory,julsgaardFundamentalLimitations,fuCavityassistedRevival}.
However, recent experiments~\cite{osullivanRandomAccessQuantum,kamelMultimodeRandomAccess} have observed that ROSE can be performed with ARP pulses even without an explicit echo-silencing mechanism.
In the present work, we show that this behavior arises from coupling-inhomogeneity-induced echo silencing as discussed in Appendix~\ref{sec:ARP_appendix}, which provides an intrinsic silencing mechanism for ARP-based ROSE protocols.

We consider a ROSE protocol using this mechanism, implemented with two identical ARP pulses [Fig.~\ref{fig:rose_appendix}(b) for homogeneous coupling and Fig.~\ref{fig:rose_appendix}(c) for inhomogeneous coupling].
The dynamics are described by the frequency-dependent beam-splitter and two-mode-squeezing coefficients in Eqs.~\eqref{eq:bs_a}, \eqref{eq:bs_P}, \eqref{eq:ts_a}, and \eqref{eq:ts_P}.
For simplicity, we mostly focus on the resonant limit, where the resonator is on resonance with the ensemble, \(\delta=0\), and the wavepacket bandwidth is much smaller than the relevant dynamical rates, \(\kappa\), \(\Gamma\), and \(G\).
Finite-bandwidth effects can be incorporated by retaining the full frequency dependence of these coefficients.
Similarly, for the non-Lorentzian case discussed in Appendix~\ref{sec:non_lorentzian}, the frequency dependence of the effective coupling rate \(\Gamma\) can be included in the same way.

The input quantum state is described by \(\ah_\mrin\). 
When this pulse is sent into the system, part of the state is absorbed into the dark waveguide with efficiency \(\eta\), determined by the beam-splitter relations in Eqs.~\eqref{eq:bs_a} and \eqref{eq:bs_P}.
In particular, in the resonant narrowband limit, Eq.~\eqref{eq:resonant_inout_ground_P} gives the writing efficiency
\begin{align}
    \eta=\frac{4C}{(1+C)^2}.
    \label{eq:rose_efficiency}
\end{align}
This efficiency is maximized at the impedance-matching condition \(C=1\).
For finite-bandwidth wavepackets, additional loss arises from spectral filtering by the system.

After the writing process, the state is stored in the dark waveguide with mode function \(f(g,\omega)\), given by Eq.~\eqref{eq:absorbed_mode}.
The corresponding mode operator is
\begin{align}
    \Dh_f=\sqrt{\eta}\ah_\mrin+\sqrt{1-\eta}\Dh_0,
\end{align}
where \(\Dh_0\) is a vacuum mode operator associated with the loss channel, mainly corresponding to the input from the dark waveguide.

We then apply an ARP pulse, which imprints a coupling-dependent phase on the excitation and maps the mode \(f\) to a new mode \(f_\pi\), with a complex overlap \(\mathcal F\) with the original bright mode and silencing factor \(F=|\mathcal F|\) (see Appendix~\ref{sec:ARP_appendix}).
It is useful to decompose \(f_\pi\) into a component parallel to \(f\) and an orthogonal component,
\begin{align}
    f^\perp=\frac{1}{\sqrt{1-F^2}}\qty(f_\pi-\mathcal F f).
\end{align}
The action of the ARP pulse on the mode operators can then be written as the beamsplitter transformation
\begin{align}
    \Dht_{f}'       & =\mathcal F\Dh_f+\sqrt{1-F^2}\Dht_0',  \\
    \Dht_{f^\perp}' & =-\sqrt{1-F^2}\Dh_f+\mathcal F^*\Dht_0',
\end{align}
where \(\Dht_0'\) is a vacuum mode operator associated with the mode mismatch.

The first echo (ACE) is then described by
\begin{align}
    \ah_{\mathrm{e1}} & =\frac{1+C}{1-C}\ah_0+i\frac{2\sqrt{C}}{1-C}\Dht_{f}'^\dagger \\
                      & =\frac{4C}{1-C^2}\mathcal F^*\ah_\mrin^\dagger+\Nh,
\end{align}
where \(\ah_0\) is the vacuum input mode entering the resonator, and \(\Nh\) is a noise operator satisfying
\begin{align}
    \ev*{\Nh^\dagger\Nh}=\frac{4C}{(1-C)^2}\qty[1-\frac{4C}{(1+C)^2}F^2], \ev*{\Nh}=0.
\end{align}
We define the noise of a bosonic operator \(\ah\) as
\begin{align}
    \ev*{\Delta\ah^\dagger\Delta\ah}
    =
    \ev*{\ah^\dagger\ah}-\ev*{\ah^\dagger}\ev*{\ah}.
\end{align}
For a coherent-state input, the first-echo noise becomes
\begin{align}
    \ev*{\Delta\ah^\dagger_{\mathrm{e1}}\Delta\ah_{\mathrm{e1}}}
     & =\qty|\frac{4C}{1-C^2}\mathcal F^*|^2+\ev*{\Nh^\dagger\Nh}\\
    &=\frac{4C}{(1-C)^2}.
\end{align}
The noise is therefore independent of \(F\), reflecting the ASE produced by the inverted ensemble.

During the emission of the first echo, the dark-waveguide operator in the mode \(f\) transforms as
\begin{align}
    \Dht_{f}'' & =-i\frac{2\sqrt{C}}{1-C}\ah_0^\dagger+\frac{1+C}{1-C}\Dht_{f}' \\
             & =i\frac{2\sqrt{C}}{1-C}\mathcal F\ah_\mrin+\Nh',
\end{align}
where
\begin{align}
    \ev*{\Nh'^\dagger\Nh'}=\frac{4C}{(1-C)^2},\qquad \ev*{\Nh'}=0.
\end{align}
On the other hand, the orthogonal mode \(f^\perp\) is unaffected in the resonant limit.
For finite-bandwidth wavepackets, however, it acquires a phase factor associated with reflection from the bright-mode resonator, or equivalently with the transformation between the near-excited-state input and output modes \(\Ptin\) and \(\Ptout\).
Because the near-excited-state dark modes are time-reversed relative to the near-ground-state modes, this frequency dependence is conjugated compared with the near-ground-state case.
In the frequency representation, the phase factor is
\begin{align}
    \Dht_{f^\perp}''(\omega)=-\frac{i\omega-\Gamma/2}{i\omega+\Gamma/2}\Dht_{f^\perp}'(\omega),\label{eq:rose_reflection}
\end{align}
which agrees with the phase factor introduced in the quantum cascade model of Ref.~\cite{greggioOptimalAbsorption}.

If the second \pipulse{} is implemented using the same ARP waveform, the coupling-dependent phase accumulated during the first pulse is cancelled.
The mode \(f_\pi\) is therefore mapped back to \(f\), giving
\begin{align}
    \Dh_f'''         & =\mathcal F^*\Dht_{f}''-\sqrt{1-F^2}\Dht_{f^\perp}'', \\
    \Dh_{f^\perp}''' & =\sqrt{1-F^2}\Dht_{f}''+\mathcal F\Dht_{f^\perp}'' .
\end{align}
This mode is then retrieved as the echo according to Eq.~\eqref{eq:resonant_inout_ground_a}.
The retrieval efficiency is therefore the same as the writing efficiency in Eq.~\eqref{eq:rose_efficiency}.
As a result, the second echo is described by
\begin{align}
    \ah_{\mathrm{e2}}  &=\frac{1-C}{1+C}\ah_0'' +i\frac{2\sqrt{C}}{1+C}\Dh_f'''\\
                      & =-\frac{4C}{(1+C)^2}\qty(\frac{2C}{1-C}F^2+1)\ah_\mrin+\Nh'',
\end{align}
where \(\ah_0''\) is the vacuum input mode entering the resonator, and \(\Nh''\) is a noise operator satisfying
\begin{align}
    \ev*{\Nh''^\dagger\Nh''}=\qty(\frac{4C}{1-C^2})^2F^2,\quad \ev*{\Nh''}=0.
\end{align}

Thus, the output state of the quantum memory is subject to Gaussian noise.
The total memory efficiency, defined as the squared amplitude ratio between the retrieved and input fields, is therefore
\begin{align}
    \eta_{\mathrm{tot}}=\qty[\frac{4C}{(1+C)^2}\qty(\frac{2C}{1-C}F^2+1)]^2.
\end{align}
For a coherent-state input, the noise of the memory output is
\begin{align}
    \ev*{\Delta\ah^\dagger_{\mathrm{e2}}\Delta\ah_{\mathrm{e2}}}
    =\ev*{\Nh''^\dagger\Nh''}
    =\qty(\frac{4C}{1-C^2})^2F^2.
\end{align}
Note that the efficiency defined here can exceed 1 because the retrieval process includes amplification.
However, the added noise also increases as \(C\to1\).
Therefore, high-fidelity quantum memory requires the silencing factor to vanish, \(F\to0\), achieving perfect silencing while approaching the impedance-matching condition \(C\to1\).

In experimental works on ROSE protocols using ARP pulses, the observed echo silencing was attributed to the frequency-dependent phase imprinted by the ARP pulse~\cite{osullivanRandomAccessQuantum,kamelMultimodeRandomAccess}.
This explanation corresponds to the homogeneous-coupling case [Fig.~\ref{fig:rose_appendix}(b)].
However, the analysis above shows that frequency-dependent chirping alone does not silence the intermediate echo: it produces an amplified chirped echo (ACE), whose emission and associated ASE noise are equivalent to those of the unsilenced protocol.
True silencing instead requires the ARP pulse to reduce the overlap between the phase-imprinted mode and the original bright mode.
This occurs when coupling inhomogeneity is included, because the ARP pulse imprints an additional coupling-dependent phase and suppresses the ACE amplitude by the silencing factor \(F\) [Fig.~\ref{fig:rose_appendix}(c)].
\section{RASE protocol}\label{sec:rase_appendix}
\begin{figure*}[htb]
    \centering
    \includegraphics[scale=1]{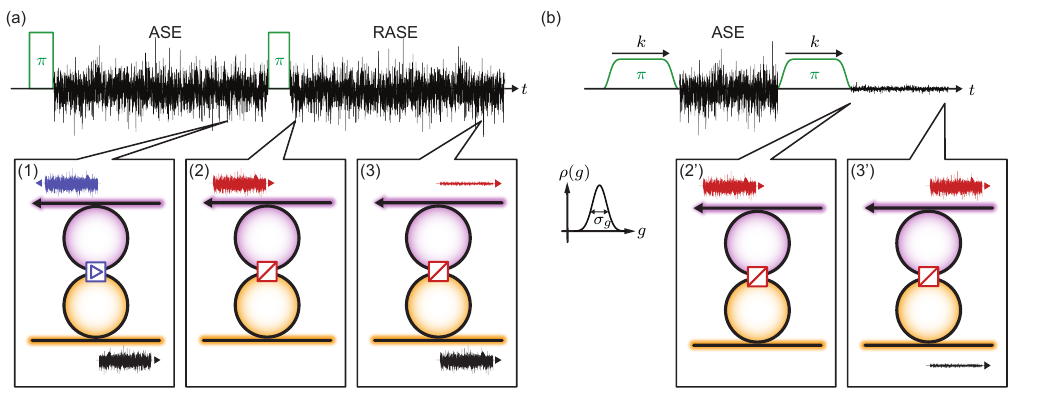}
    \caption{Description of the RASE protocol in the waveguide model.
        (a) Ideal \pipulse{}s.
        (1) After the ensemble is inverted by the first \pipulse{}, the two-mode-squeezing interaction between the resonator and the bright mode creates a two-mode squeezed vacuum state shared between the dark waveguide and the resonator output waveguide (ASE).
        (2) The second \pipulse{} reverses the propagation direction in the dark waveguide.
        (3) The RASE field, entangled with the previously emitted ASE field, is emitted through the beam-splitter interaction.
        (b) ARP pulses used as \pipulse{}s for inhomogeneous coupling.
        (2'),(3') The second \pipulse{} imprints a coupling-dependent phase, which suppresses the RASE emission by the silencing factor \(F\) [Appendix~\ref{sec:ARP_appendix}].
        This phase can be cancelled using our three-ARP-pulse protocol.}
    \label{fig:rase_appendix}
\end{figure*}
\subsection{Overview}
The rephased amplified spontaneous emission (RASE) protocol \cite{ledinghamNonclassicalPhoton} is an entanglement-generation protocol that produces a time-delayed two-mode squeezed state using two \pipulse{}s.
In the waveguide model, the protocol can be understood in a similar way to the ROSE protocol, as illustrated in Fig.~\ref{fig:rase_appendix}(a).
Indeed, RASE may be viewed as the ROSE protocol with vacuum input: the noise emitted in the first echo corresponds to amplified spontaneous emission (ASE), while the noise emitted in the second echo corresponds to RASE.

After the first \pipulse{}, two-mode squeezed states are generated for each frequency component according to Eqs.~\eqref{eq:ts_a} and \eqref{eq:ts_P}.
The photon numbers in the resonator output mode \(\aase(\omega)\) and the dark-waveguide output mode \(\Ptout(\omega)\) are
\begin{align}
    \ev*{\aase^\dagger(\omega)\aase(\omega')}&=\ev*{\Ptout^\dagger(\omega)\Ptout(\omega')}\nonumber\\
    &=|\tilde{t}(\omega)|^2\delta(\omega-\omega'),
\end{align}
where \(\tilde{t}(\omega)\) is defined in Eq.~\eqref{eq:excited_t}.
The cross-correlation is
\begin{align}
    \ev*{\aase(\omega)\Ptout(\omega')}=\tilde{r}(\omega)\tilde{t}'^*(\omega)\delta(\omega-\omega'),
\end{align}
with \(\ev*{\aase^\dagger(\omega)\Ptout(\omega')}=0\).
In the time domain, this correlation is given by the Fourier transform
\begin{align}
    C_{aP}(\tdelay)&=\ev*{\aase(t)\Ptout(t+\tdelay)}\\
    &=\frac{1}{\sqrt{2\pi}}\int \dd{\omega}\tilde{r}(\omega)\tilde{t}'^*(\omega)e^{i\omega\tdelay}.
    \label{eq:raw_time_correlation}
\end{align}

The field \(\Ptout(\omega)\) is then retrieved as the RASE field \(\arase(\omega)\) by applying a second \pipulse{}.
If this second pulse is an ideal \pipulse{} with no additional frequency- or coupling-dependent phase, the ASE--RASE correlation is
\begin{align}
    \ev*{\arase^\dagger(\omega)\arase(\omega')} & =|t(\omega)|^2|\tilde{t}(\omega)|^2\delta(\omega-\omega'),           \\
    \ev*{\aase(\omega)\arase(\omega')}          & =t(\omega)\tilde{r}(\omega)\tilde{t}'^*(\omega)\delta(\omega-\omega'),
\end{align}
where \(t(\omega)\) is the retrieval coefficient defined in Eq.~\eqref{eq:ground_t}.

\subsection{Chirped correlation}
We now consider the case where an ARP pulse is used as the second \pipulse{}, first assuming homogeneous coupling.
As discussed in Appendix~\ref{sec:ace_appendix}, the ARP pulse imprints a quadratic frequency-dependent phase on the retrieved field.
The ASE--RASE correlation therefore becomes
\begin{align}
    \begin{split}        
    &\ev*{\aase(\omega)\arase(\omega')}\\
    &=t(\omega)\tilde{r}(\omega)\tilde{t}'^*(\omega)\exp\qty[i\frac{\omega^2}{k}]\delta(\omega-\omega').
    \end{split}
\end{align}
Here, we omit constant and linear phases, which correspond only to a global phase and a time shift, respectively.
The corresponding time-domain correlation is
\begin{align}
    C_{aa}(\tdelay) & =\ev*{\aase(t)\arase(2t_0-t+\tdelay)}                                                                                              \\
                    & =\frac{1}{\sqrt{2\pi}}\int \dd{\omega}t(\omega)\tilde{r}(\omega)\tilde{t}'^*(\omega)\exp\qty[i\frac{\omega^2}{k}]e^{i\omega\tdelay}.
\end{align}
Here, \(t_0\) is the reference time of the second \pipulse{}, and \(\tdelay\) is the delay measured from the expected echo time.
Thus, the ASE--RASE correlation is chirped relative to the original correlation in Eq.~\eqref{eq:raw_time_correlation}.

Importantly, this chirping does not by itself degrade the entanglement.
The entanglement between the ASE and RASE modes can be quantified using the Duan--Simon criterion~\cite{duanInseparabilityCriterion,simonPeresHorodeckiSeparability}, which is directly accessible experimentally through homodyne measurements.
We define the quadratures of the ASE and RASE modes by
\begin{align}
    \xh_j=\frac{\ah_j+\ah_j^\dagger}{\sqrt{2}},\qquad
    \ph_j=\frac{\ah_j-\ah_j^\dagger}{\sqrt{2}i},
\end{align}
where \(j\in\{\mrase,\mrrase\}\), so that \([\xh_j,\ph_k]=i\delta_{jk}\).
For two temporal modes, we define
\begin{align}
    \bop{u}&=\sqrt{1-s}\xh_\mrase-\sqrt{s}\xh_\mrrase,\\
    \bop{v}&=\sqrt{1-s}\ph_\mrase+\sqrt{s}\ph_\mrrase,
\end{align}
where \(s\) is a free parameter.
The Duan--Simon value is
\begin{align}
    r=\ev*{\bop{u}^2}+\ev*{\bop{v}^2}.
\end{align}
The state is inseparable when \(r<1\).
This can also be written as
\begin{align}
    \begin{split}
    r&=1+2(1-s)\ev*{\aase^\dagger\aase}+2s\ev*{\arase^\dagger\arase}\\
    &\quad-4\sqrt{s(1-s)}\Re\ev*{\aase\arase},        
    \end{split}
    \label{eq:duan_simon_moments}
\end{align}
and minimizing it over \(s\) gives
\begin{align}
    r_\mathrm{min}
    &= 1+\ev*{\aase^\dagger\aase}+\ev*{\arase^\dagger\arase} \nonumber \\
    &\quad
    -\sqrt{
    \qty[\ev*{\aase^\dagger\aase}-\ev*{\arase^\dagger\arase}]^2
    +4\qty(\Re\ev*{\aase\arase})^2
    }.
    \label{eq:duan_simon_min}
\end{align}

To evaluate this criterion, we introduce temporal mode functions \(f_\mrase(t)\) and \(f_\mrrase(t)\) normalized by \(\int \dd{t} |f_\mrase(t)|^2=\int \dd{t} |f_\mrrase(t)|^2=1\) for the ASE and RASE fields, respectively, and define
\begin{align}
    \aase=\int \dd{t}\aase(t)f_\mrase(t),\quad \arase=\int \dd{t}\arase(t)f_\mrrase(t).
\end{align}
The correlation between these temporal modes is
\begin{align}
    \ev*{\aase\arase}=\int \dd{t}\int \dd{t'}f_\mrase(t)f_\mrrase(t')C_{aa}(t-t').\label{eq:correlation_wavepackets}
\end{align}
Similarly, the photon numbers are
\begin{align}
    \ev*{\aase^\dagger\aase}   & =\int \dd{t}\int \dd{t'}f_\mrase^*(t)f_\mrase(t')\ev*{\aase^\dagger(t)\aase(t')},     \\
    \ev*{\arase^\dagger\arase} & =\int \dd{t}\int \dd{t'}f_\mrrase^*(t)f_\mrrase(t')\ev*{\arase^\dagger(t)\arase(t')}.
\end{align}

For a fixed ASE mode \(f_\mrase(t)\), the correlation in Eq.~\eqref{eq:correlation_wavepackets} is maximized by choosing the RASE mode as the matched mode
\begin{align}
    f_\mrrase(t)\propto\qty[\int \dd{t'}f_\mrase(t')C_{aa}(t'-t)]^*.
\end{align}
With this choice, the maximum correlation becomes
\begin{align}
    \ev*{\aase\arase} & =\sqrt{\int \dd{t}\qty|\int \dd{t'}f_\mrase(t')C_{aa}(t'-t)|^2}                                       \\
                      & =\sqrt{\int \dd{\omega}\qty|\tilde{f}_\mrase(\omega)t(\omega)\tilde{r}(\omega)\tilde{t}'^*(\omega)|^2}.
\end{align}
The final expression is independent of the ARP-induced frequency-dependent phase.
Thus, in the homogeneous-coupling case, the ARP only changes the optimal temporal mode of the RASE field, and does not reduce the ASE--RASE correlation.

Similarly, the photon numbers become
\begin{align}
    \ev*{\aase^\dagger\aase}   & =\int \dd{\omega}\qty|\tilde{f}_\mrase(\omega)\tilde{t}(\omega)|^2,           \\
    \ev*{\arase^\dagger\arase} & =\int \dd{\omega}\qty|\tilde{f}_\mrrase(\omega)t(\omega)\tilde{t}(\omega)|^2.
\end{align}
In the narrow-band limit, where the ASE mode is centered on resonance and has bandwidth much smaller than \(\kappa\), \(G\), and \(\Gamma\), we may approximate \(|\tilde{f}_\mrase(\omega)|^2\sim\delta(\omega)\).
This gives
\begin{align}
    \ev*{\aase\arase}          & =\frac{4C}{(1-C)^2},      \\
    \ev*{\aase^\dagger\aase}   & =\frac{4C}{(1-C)^2},      \\
    \ev*{\arase^\dagger\arase} & =\frac{16C^2}{(1-C^2)^2}.
\end{align}
Substituting these expressions into Eq.~\eqref{eq:duan_simon_min} gives the cooperativity dependence of the Duan--Simon value in the homogeneous-coupling limit.

\subsection{Degradation of entanglement by coupling inhomogeneity}
The situation changes qualitatively when the coupling is inhomogeneous.
As discussed in Appendix~\ref{sec:ARP_appendix}, an ARP pulse imprints a coupling-dependent phase on the ensemble.
In the ROSE protocol, this phase suppresses the first echo by reducing the overlap between the mode of the stored state and the original bright mode, as we see in Appendix~\ref{sec:rose_appendix}.
The same mechanism suppresses the ASE--RASE entanglement in the RASE protocol [Fig.~\ref{fig:rase_appendix}(b)].

Writing the complex overlap as \(\mathcal F\) and its modulus as \(F\), the correlation and photon numbers are given by
\begin{align}
    \ev*{\aase\arase}          & =\frac{4C}{(1-C)^2}\mathcal F,          \\
    \ev*{\aase^\dagger\aase}   & =\frac{4C}{(1-C)^2},           \\
    \ev*{\arase^\dagger\arase} & =\frac{16C^2}{(1-C^2)^2}F^2.
\end{align}
A larger coupling inhomogeneity reduces the silencing factor \(F\), thereby reducing the correlation and increasing the Duan--Simon value in Eq.~\eqref{eq:duan_simon_min}.
In the completely silenced limit, \(F=0\), the correlation \(\ev*{\aase\arase}\) vanishes and the Duan--Simon value reaches the classical bound \(r_\mathrm{min}=1\).
Thus, while ARP pulses provide robustness against frequency inhomogeneity, in the presence of coupling inhomogeneity, they degrade the entanglement generated by the RASE protocol.

\subsection{Three-pulse protocol}
To avoid this degradation, we propose a modified three-pulse RASE protocol, illustrated in Fig.~\ref{fig:rase}(b) of the main text.
The goal is to preserve the robust inversion provided by ARP pulses while cancelling the coupling-dependent dynamical phase that suppresses the ASE--RASE correlation.

This protocol replaces the second ARP \pipulse{} by a composite sequence of three ARP pulses with chirp rates \(+k\), \(+k/2\), and \(+k\), where the middle pulse has twice the duration.
Explicitly, the drive is
\begin{align}
    \begin{split}
        \ev*{\ah(t)} & =A(t)\exp[i\qty(\omega_0 t+\frac{k}{2}t^2)]                             \\
                     & \;+A\qty(\frac{t-T}{2})\exp[i\qty(\omega_0(t-T)+\frac{k}{4}(t-T)^2)] \\
                     & \;+A(t-3T)\exp[i\qty(\omega_0(t-3T)+\frac{k}{2}(t-3T)^2)],
    \end{split}
\end{align}
where \(A(t)\) is the slowly varying envelope.
The pulse durations are \(T\), \(2T\), and \(T\), respectively.

The cancellation follows from the fact that the dynamical phase in Eq.~\eqref{eq:dynamical_phase_int} scales as \(1/k\).
Because the first and third pulses transfer the TLSs from the excited state to the ground state, their dynamical phases have the opposite sign, as follows from Eq.~\eqref{eq:dynamical_phase_int}.
The total phase is therefore proportional to
\begin{align}
    -\frac{1}{k}+\frac{1}{k/2}-\frac{1}{k}=0.
\end{align}
Thus, both the coupling-dependent and frequency-dependent dynamical phases cancel out.

This three-pulse sequence therefore implements a robust \pipulse{} without imprinting the ARP-induced phase on the ensemble.
As a result, the ASE--RASE correlation and the entanglement are preserved even in the presence of coupling inhomogeneity, as shown in Figs.~\ref{fig:rase}(c) and \ref{fig:rase}(d) of the main text.

\section{Numerical simulation}\label{sec:numerical_appendix}
In the main text, numerical simulations are used to verify the theoretical predictions.
In this section, we describe the simulation method in detail.
We use a mean-field approach based on a cumulant expansion, with GPU acceleration for the numerical time evolution. The mean-value simulations in Fig.~\ref{fig:rose}(b,c), as well as the direct experimental comparison in Fig.~\ref{fig:rose}(e), use a first-order truncation. The noise and correlation calculations in Fig.~\ref{fig:rose}(d) and Fig.~\ref{fig:rase} retain second-order correlations. Below we present the second-order equations; the first-order simulations evolve only the mean values, factorizing spin--cavity moments into products of their means.
\subsection{Displaced-frame Hamiltonian}
The numerical equations follow from the continuous Hamiltonian in Eq.~\eqref{eq:main_hamiltonian}, including resonator damping and driving.
Since we only consider vacuum or
coherent-state inputs to the resonator, the input field can be decomposed into
its coherent amplitude and quantum fluctuation as
\begin{equation}
    \ain(t) = E(t) + \delta \ain(t),
\end{equation}
where
\begin{equation}
    E(t) = \ev{\ain(t)}
\end{equation}
is the coherent input envelope, and \(\delta \ain(t)\) has vacuum noise
statistics.
Equivalently, one may work in a displaced frame in which the coherent part of the input field is absorbed into a classical drive term acting
on the resonator.
In this frame, the remaining input field \(\delta \ain(t)\) is vacuum, which we denote again by \(\ain(t)\) for notational simplicity.

The coherent input then appears as the standard input--output drive term
\begin{equation}
    H_{\mathrm{drive}}(t)=i\sqrt{\kappa}\qty[E(t)\ahdag - E^*(t)\ah].
\end{equation}
Combining this drive with the original Hamiltonian gives
\begin{align}
    \begin{split}
        H & = \delta \ahdag \ah + i \sqrt{\kappa} \qty( E(t)\ahdag - E^*(t)\ah )\\
         &\quad + \sum_j g_j \qty(\sigmah^-_{j}\ahdag+\sigmah^+_j\ah) + \frac12\sum_j \omega_j \sigmah^{z}_{j}.
    \end{split}
    \label{eq:displaced_frame_hamiltonian}
\end{align}
Here, \(E(t)\) is the coherent input field, and \(\ain(t)\) now denotes the residual vacuum input noise.
With this convention, the Langevin equation has the same form as before, except that the coherent input has been moved from the input operator into the Hamiltonian drive term, leaving only vacuum noise in \(\ain(t)\).

\subsection{Bin discretization of the ensemble}
In the Hamiltonian Eq.~\eqref{eq:displaced_frame_hamiltonian}, the summations \(\sum_j g_j (\sigmah^-_{j}\ahdag+\sigmah^+_j\ah)\) and \(\frac12 \sum_j \omega_j \sigmah^{z}_{j}\) are taken over all TLSs, namely from \(1\) to \(N\), where \(N\) is the total number of TLSs. However, in simulations, \(N\) can be very large (e.g., \(10^6\) to \(10^7\) in our case), which makes direct computation infeasible.

Therefore, a common approach is to divide the TLSs into \(M\) bins (\(M \le N\)), and assign a pair of parameters \((g_j,\omega_j)\) to each bin. For each bin, we define the collective operators \(\bop{S}^{z}_j\) and \(\bop{S}^{\pm}_j\) as
\begin{align}
    \bop{S}^{z}_j   & = \frac12 \sum_k \bm{\sigma}^{z}_k, \\
    \bop{S}^{\pm}_j & = \sum_k \bm{\sigma}^{\pm}_k,
\end{align}
where \(k\) labels the individual TLSs within a given bin.

The Hamiltonian after bin discretization becomes
\begin{equation}
    \begin{split}
    H &=\delta \ahdag \ah + \sum_{j=1}^{M} \omega_j \bop{S}^{z}_{j} + \sum_{j=1}^{M} g_j \left( \bop{S}^-_{j}\ahdag + \bop{S}^+_j\ah \right) \\
    &\quad+ i \sqrt{\kappa} \left( E(t)\ahdag - E^*(t)\ah \right).        
    \end{split}
\end{equation}

For simulations corresponding to the figures in the main text, we set the cavity detuning \(\delta = 0\), the decay rate \(\kappa/(2\pi) = 1\,\mathrm{MHz}\), the full width at half maximum (FWHM) of the frequency broadening \(\Gamma/(2\pi) = 1\,\mathrm{MHz}\), and the average coupling strength \(\bar{g}/(2\pi) = 100\,\mathrm{Hz}\). The total number of TLSs \(N\) can be used to tune the cooperativity \(C\) of the ensemble while keeping the other parameters fixed. These parameters are comparable to microwave spin-ensemble memory experiments.
For example, Ref.~\cite{osullivanRandomAccessQuantum} used Bi donor spins in silicon coupled to a superconducting resonator, with \(\kappa/(2\pi) = \SI{0.4}{MHz}\), \(\Gamma/(2\pi) = \SI{2.4}{MHz}\), and \(\bar{g}/(2\pi) = \SI{80}{Hz}\), and reported a cooperativity \(C=0.07\).

In the Hamiltonian, there are two types of inhomogeneity: the broadening of the resonant frequency \(\omega\) (assumed Lorentzian) and the broadening of the TLS--cavity coupling strength \(g\) (assumed Gaussian). Accordingly, we assign \(M_\omega\) bins to the former and \(M_g\) bins to the latter. Since these two broadenings are assumed to be uncorrelated, the total number of bins is \(M = M_\omega M_g\). For the frequency distribution, we truncate the simulated bandwidth to \(\xi\Gamma\), where \(\xi\) is a dimensionless truncation factor that determines how much of the inhomogeneous broadening is included.
In our simulations, we choose \(\xi=2.5\), which includes approximately \(75.8\%\) of the TLSs from the Lorentzian distribution.

Although bin discretization makes the simulation tractable, it can introduce artifacts if \(M\) is chosen too small.
For the frequency discretization, the bin number should be large enough that the artificial rephasing time lies outside the simulated time window.
The frequency-bin spacing is approximately
\begin{equation}
    \Delta\omega \simeq \frac{\xi\Gamma}{M_\omega},
\end{equation}
which gives an artificial rephasing time
\begin{equation}
    T_0 \sim \frac{1}{\Delta\omega}
    = \frac{M_\omega}{\xi\Gamma}.
\end{equation}
We therefore require
\begin{equation}
    T_0 \gtrsim T_{\mathrm{total}},
\end{equation}
or, equivalently,
\begin{equation}
    M_\omega \gtrsim \xi\Gamma T_{\mathrm{total}},
\end{equation}
where \(T_{\mathrm{total}}\) is the total simulation time.
If this condition is not satisfied, the phases of the discretized frequency bins can artificially rephase within the simulation window, producing a collective emission artifact analogous to an AFC echo.

A similar discretization artifact can also occur in the coupling distribution, which sets a lower bound on \(M_g\).
If \(M_g\) is too small, the spacing between \(g\) bins becomes large, producing artificial periodic structure in the Fourier-conjugate \(s\)-space, similarly to the effect of frequency discretization.
These artifacts can overlap with the bright mode, leading to unphysical evolution.
In practice, we gradually increase \(M_g\) and monitor whether the dynamics has converged to determine the minimum required value of \(M_g\).

In our second-order mean-field simulation, both memory usage and computation time scale as \(O(M^2)\).
The equation-solving process is substantially accelerated using a graphics processing unit (GPU).
The simulations were carried out on an NVIDIA H200 GPU on the NUS HPC system, with 141 GB of GPU memory.
In practice, the available GPU memory limits the maximum achievable bin number in our simulations to approximately \(M=11200\).

\subsection{Equations of dynamics}
The Quantum Langevin equations (QLEs) \cite{gardiner2004quantum} governing the ensemble--cavity system are given by
\begin{equation}
    \frac{d\bop{O}}{dt} = i \left[ \bop{H}, \bop{O} \right] + \frac{\kappa}{2} \left( 2\ahdag\bop{O}\ah - \ahdag\ah\bop{O} - \bop{O}\ahdag\ah \right) + \text{noise},
\end{equation}
where \(\bop{O}\) is an arbitrary operator.

Solving the evolution of the density matrix under the QLEs is infeasible for large \(M\), so we instead focus on the second-order mean-field equations. Under the white-noise assumption, one can show that the noise does not contribute to any first- and second-order terms except \(\langle \ah\ahdag \rangle\). However, a subtlety arises in that we do not explicitly evolve \(\langle \ah\ahdag \rangle\), but instead compute it using the commutation relation \(\left[ \ah, \ahdag \right] = 1\) during post-processing.

Our second-order mean-field equations are derived with the aid of the Julia framework QuantumCumulants.jl \cite{plankensteiner2022quantumcumulants}, which is particularly well suited for deriving equations of motion for average values in open quantum systems. We use it to generate the equations for the several-bin case and then manually generalize them to the arbitrary-bin-number case. The equations are then transcribed into Julia code with assistance from Chat-GPT (OpenAI).

The equations of motion for the first-order mean values, including the cavity field \(\langle \ah \rangle\) and the spin operators \(\langle \bop{S}^{z}_{j} \rangle\) and \(\langle \bop{S}^{+}_{j} \rangle\) for \(j \in \{1,\dots,M\}\), are given by
\begin{align}
    \dv{t} \langle \bop{a} \rangle
    &=
    - i \delta \langle \bop{a} \rangle
    - \frac{\kappa}{2} \langle \bop{a} \rangle
    + \sqrt{\kappa} E(t)
    - i \sum_{j=1}^{M} g_j \langle \bop{S}_j^- \rangle ,
    \\
    \dv{t} \langle \bop{S}_j^+ \rangle
    &=
    i \omega_j \langle \bop{S}_j^+ \rangle
    - 2 i g_j \langle \bop{a}^\dagger \bop{S}_j^z \rangle ,
    \\
    \dv{t} \langle \bop{S}_j^z \rangle
    &=
    - i g_j \langle \bop{a} \bop{S}_j^+ \rangle
    + i g_j \langle \bop{a}^\dagger \bop{S}_j^- \rangle .
\end{align}
The mean value of the lowering operator \(\bop{S}_j^-\) need not be evolved independently, since it is related to \(\langle \bop{S}_j^+ \rangle\) through the relation
\begin{equation}
    \langle \bop{S}_j^- \rangle = \langle \bop{S}_j^+ \rangle^* .
\end{equation}

Meanwhile, the second-order cavity moments \(\langle \ahdag \ah \rangle\) and \(\langle \ahdag \ahdag \rangle\), together with the spin--cavity correlations \(\langle \ahdag \bop{S}^{+}_{j} \rangle\), \(\langle \ahdag \bop{S}^{-}_{j} \rangle\), and \(\langle \ahdag \bop{S}^{z}_{j} \rangle\), obey the equations
\begin{widetext}
\begin{align}
    \dv{t} \langle \bop{a}^\dagger \bop{a}^\dagger \rangle &= 2 i \sum_{j=1}^{M} g_j \langle \bop{a}^\dagger \bop{S}_j^+ \rangle + 2 i \delta \langle \bop{a}^\dagger \bop{a}^\dagger \rangle - \kappa \langle \bop{a}^\dagger \bop{a}^\dagger \rangle + 2 \sqrt{\kappa} E^*(t) \langle \bop{a}^\dagger \rangle, \\
    \dv{t} \langle \bop{a}^\dagger \bop{a} \rangle &= i \sum_{j=1}^{M} g_j \langle \bop{a} \bop{S}_j^+ \rangle - i \sum_{j=1}^{M} g_j \langle \bop{a}^\dagger \bop{S}_j^- \rangle - \kappa \langle \bop{a}^\dagger \bop{a} \rangle + \sqrt{\kappa}\left[E(t) \langle \bop{a}^\dagger \rangle + E^*(t) \langle \bop{a} \rangle\right], \\
    \dv{t} \langle \bop{a}^\dagger \bop{S}_j^+ \rangle &= i \sum_{j'=1}^{M} g_{j'} \langle \bop{S}_j^+ \bop{S}_{j'}^+ \rangle + i \delta \langle \bop{a}^\dagger \bop{S}_j^+ \rangle + i \omega_j \langle \bop{a}^\dagger \bop{S}_j^+ \rangle - \frac{\kappa}{2} \langle \bop{a}^\dagger \bop{S}_j^+ \rangle + \sqrt{\kappa} E^*(t) \langle \bop{S}_j^+ \rangle - 2 i g_j \langle \bop{a}^\dagger \bop{a}^\dagger \bop{S}_j^z \rangle, \\
    \dv{t} \langle \bop{a}^\dagger \bop{S}_j^- \rangle &= i \sum_{j'=1}^{M} g_{j'} \langle \bop{S}_j^- \bop{S}_{j'}^+ \rangle + i \delta \langle \bop{a}^\dagger \bop{S}_j^- \rangle - i \omega_j \langle \bop{a}^\dagger \bop{S}_j^- \rangle + 2 i g_j \langle \bop{S}_j^z \rangle - \frac{\kappa}{2} \langle \bop{a}^\dagger \bop{S}_j^- \rangle + \sqrt{\kappa} E^*(t) \langle \bop{S}_j^- \rangle + 2 i g_j \langle \bop{a}^\dagger \bop{a} \bop{S}_j^z \rangle, \\
    \dv{t} \langle \bop{a}^\dagger \bop{S}_j^z \rangle &= i \sum_{j'=1}^{M} g_{j'} \langle \bop{S}_j^z \bop{S}_{j'}^+ \rangle + i \delta \langle \bop{a}^\dagger \bop{S}_j^z \rangle - \frac{\kappa}{2} \langle \bop{a}^\dagger \bop{S}_j^z \rangle + \sqrt{\kappa} E^*(t) \langle \bop{S}_j^z \rangle - i g_j \left(\langle \bop{S}_j^+ \rangle + \langle \bop{a}^\dagger \bop{a} \bop{S}_j^+ \rangle\right) + i g_j \langle \bop{a}^\dagger \bop{a}^\dagger \bop{S}_j^- \rangle.
\end{align}
Additionally, for spin correlations of the same bin where \(j = j'\), the collective spin operators obey the commutation relations of angular momentum operators. The dynamical equations for same-bin correlations are expressed by
\begin{align}
    \dv{t} \langle \bop{S}_j^+ \bop{S}_j^+ \rangle &= 2 i \omega_j \langle \bop{S}_j^+ \bop{S}_j^+ \rangle + 2 i g_j \langle \bop{a}^\dagger \bop{S}_j^+ \rangle - 4 i g_j \langle \bop{a}^\dagger \bop{S}_j^z \bop{S}_j^+ \rangle, \\
    \dv{t} \langle \bop{S}_j^z \bop{S}_j^+ \rangle &= i \omega_j \langle \bop{S}_j^z \bop{S}_j^+ \rangle - i g_j \langle \bop{a} \bop{S}_j^+ \bop{S}_j^+ \rangle + i g_j \langle \bop{a}^\dagger \bop{S}_j^- \bop{S}_j^+ \rangle - 2 i g_j \langle \bop{a}^\dagger \bop{S}_j^z \bop{S}_j^z \rangle, \\
    \dv{t} \langle \bop{S}_j^- \bop{S}_j^+ \rangle &= 2 i g_j \langle \bop{a} \bop{S}_j^z \bop{S}_j^+ \rangle - 2 i g_j \langle \bop{a}^\dagger \bop{S}_j^- \bop{S}_j^z \rangle, \\
    \dv{t} \langle \bop{S}_j^z \bop{S}_j^z \rangle &= i g_j \langle \bop{a} \bop{S}_j^+ \rangle - i g_j \langle \bop{a}^\dagger \bop{S}_j^- \rangle - 2 i g_j \langle \bop{a} \bop{S}_j^z \bop{S}_j^+ \rangle + 2 i g_j \langle \bop{a}^\dagger \bop{S}_j^- \bop{S}_j^z \rangle.
\end{align}

Finally, collective spin operators from two different bins \(j\) and \(j'\) commute, and the equations for different-bin correlations are given by
\begin{align}
    \dv{t} \langle \bop{S}_j^+ \bop{S}_{j'}^+ \rangle &= i(\omega_j+\omega_{j'}) \langle \bop{S}_j^+ \bop{S}_{j'}^+ \rangle - 2 i g_j \langle \bop{a}^\dagger \bop{S}_j^z \bop{S}_{j'}^+ \rangle - 2 i g_{j'} \langle \bop{a}^\dagger \bop{S}_j^+ \bop{S}_{j'}^z \rangle, \\
    \dv{t} \langle \bop{S}_j^z \bop{S}_{j'}^+ \rangle &= i\omega_{j'} \langle \bop{S}_j^z \bop{S}_{j'}^+ \rangle - i g_j \langle \bop{a} \bop{S}_j^+ \bop{S}_{j'}^+ \rangle + i g_j \langle \bop{a}^\dagger \bop{S}_j^- \bop{S}_{j'}^+ \rangle - 2 i g_{j'} \langle \bop{a}^\dagger \bop{S}_j^z \bop{S}_{j'}^z \rangle, \\
    \dv{t} \langle \bop{S}_j^- \bop{S}_{j'}^+ \rangle &= - i\omega_j \langle \bop{S}_j^- \bop{S}_{j'}^+ \rangle + i\omega_{j'} \langle \bop{S}_j^- \bop{S}_{j'}^+ \rangle + 2 i g_j \langle \bop{a} \bop{S}_j^z \bop{S}_{j'}^+ \rangle - 2 i g_{j'} \langle \bop{a}^\dagger \bop{S}_j^- \bop{S}_{j'}^z \rangle, \\
    \dv{t} \langle \bop{S}_j^z \bop{S}_{j'}^z \rangle &= - i g_j \langle \bop{a} \bop{S}_j^+ \bop{S}_{j'}^z \rangle + i g_j \langle \bop{a}^\dagger \bop{S}_j^- \bop{S}_{j'}^z \rangle - i g_{j'} \langle \bop{a} \bop{S}_j^z \bop{S}_{j'}^+ \rangle + i g_{j'} \langle \bop{a}^\dagger \bop{S}_j^z \bop{S}_{j'}^- \rangle.
\end{align}
\end{widetext}

To make the mean-field equations closed, the cumulant expansion \cite{kubo1962generalized} is used to express higher-order terms as combinations of lower-order terms. For instance, since we truncate at second order, the following cumulant expansion can be introduced:
\begin{equation}
    \begin{split}
        \langle \bop{A}\bop{B}\bop{C} \rangle
         & \approx \langle \bop{A}\bop{B} \rangle \langle \bop{C} \rangle
        + \langle \bop{A}\bop{C} \rangle \langle \bop{B} \rangle  \\
        &\quad + \langle \bop{A} \rangle \langle \bop{B}\bop{C} \rangle
        - 2 \langle \bop{A} \rangle \langle \bop{B} \rangle \langle \bop{C} \rangle,
    \end{split}
\end{equation}
where \(\bop{A}\), \(\bop{B}\), and \(\bop{C}\) are arbitrary operators.

The initial conditions of the second-order mean-field equations are determined
under the following assumptions at \(t=0\):

\begin{enumerate}[label=(\roman*)]
    \item The spin ensemble is initially either fully polarized in the ground state or fully inverted, such that
    \begin{equation}
        \begin{split}
            \langle \bop{S}_+^j \rangle &= 0, \\
            \langle \bop{S}_z^j \rangle &=
            \begin{cases}
                -N_j/2, & \text{fully polarized ground state},\\
                +N_j/2, & \text{fully inverted state},
            \end{cases}
        \end{split}
    \end{equation}
    where \(N_j\) is the number of TLSs in the \(j\)-th bin.

    \item All cavity-related terms, including
    \(\langle \ah \rangle\),
    \(\langle \ahdag \ah \rangle\),
    \(\langle \ahdag \ahdag \rangle\),
    \(\langle \bop{a}^\dagger \bop{S}_j^+ \rangle\),
    \(\langle \bop{a}^\dagger \bop{S}_j^- \rangle\), and
    \(\langle \bop{a}^\dagger \bop{S}_j^z \rangle\),
    are zero.

    \item For two bins \(j\) and \(j'\), which may coincide, the initial spin
    correlations are given by
    \begin{equation}
        \begin{split}
            \langle \bop{S}_+^j \bop{S}_+^{j'} \rangle &= 0, \\
            \langle \bop{S}_z^j \bop{S}_+^{j'} \rangle &= 0, \\
            \langle \bop{S}_-^j \bop{S}_+^{j'} \rangle &=
            \begin{cases}
                N_j \delta_{jj'}, & \text{fully polarized ground state},\\
                0, & \text{fully inverted state},
            \end{cases} \\
            \langle \bop{S}_z^j \bop{S}_z^{j'} \rangle
            &= \frac{N_j N_{j'}}{4}.
        \end{split}
    \end{equation}
\end{enumerate}

\subsection{Coherent drive}\label{sec:coherent_drive}
Both the input and ARP pulses are sent into the cavity through the coherent drive \(E(t)\). The input pulse has a Gaussian shape, given by
\begin{equation}
    E(t) = A_{0s} e^{-\frac{(t - t_0)^2}{2\sigma_t^2}},
\end{equation}
where \(A_{0s}\) is an amplitude constant, \(t_0\) is the center time of the Gaussian pulse, and \(\sigma_t\) is the temporal width. To ensure that the input pulse is sufficiently weak, we set \(A_{0s} = 0.001 A_{0\pi}\), where \(A_{0\pi}\) is the amplitude for a \pipulse{}. Its value is given by
\begin{equation}
    A_{0\pi} = \sqrt{\frac{\kappa}{2\pi}} \frac{\pi}{4\bar{g}\sigma_t},
\end{equation}
where the temporal width is set to \(\sigma_t = 3\,\mu\text{s}\), corresponding to a bandwidth much narrower than frequency broadening bandwidth \(\Gamma\).

For the ARP pulse, we use a wideband uniform rate smooth truncation (WURST) pulse~\cite{odellWURSTKind}.
The input amplitude is given by
\begin{equation}
    A_\mrin (t) = A_{0,\mrin} \left( 1 - \left| \sin\left( \frac{\pi (t - T_w/2)}{T_w} \right) \right|^n \right),
\end{equation}
where \(A_{0,\mrin}\), \(T_w\), and \(n\) are the input-amplitude scale, the ARP-pulse duration, and a parameter controlling the smoothness of the pulse edges, respectively.
In our simulation, we set \(A_{0,\mrin} = 2.5 \times 10^7 \sqrt{\mathrm{Hz}}\) and \(n = 20\).
The intracavity amplitude \(A(t)\) appearing in Eq.~\eqref{eq:e_of_t} of the main text is obtained by filtering this input waveform through the cavity response.
Within a steady-state approximation, it is
\begin{align}
    A(t) &= \sqrt{\frac{\kappa}{\qty[\omega(t)]^2+\qty(\kappa/2)^2}}A_\mrin(t),
\end{align}
where \(\omega(t)=\omega_0+kt\) is the instantaneous detuning of the ARP pulse from the cavity resonance.
Near resonance, the intracavity amplitude is therefore approximately
\begin{equation}
    A(t) \approx A_0 = \frac{2}{\sqrt{\kappa}} A_{0,\mrin}.
\end{equation}
For the parameters used here, this gives \(A_0 = \SI{2e4}{}\), corresponding to an average Rabi frequency \(\Omega_0 = 2A_0 g_0 = 2\pi\times\SI{4}{MHz}\).

\subsection{Correlation functions}
Solutions to the second-order mean-field equations contain first-order terms and equal-time second-order terms. However, to analyze both the echo noise in the ROSE protocol and the ASE--RASE correlation in the RASE protocol, two-time correlation functions, such as \(\langle \ahdag(t)\ah(t') \rangle\) and \(\langle \ahdag(t)\ahdag(t') \rangle\) (for \(t \neq t'\)), are also required.

Fortunately, the quantum regression theorem (QRT) \cite{gardiner2004quantum, khan2022quantum} provides a practical method for calculating two-time correlation functions in Markovian systems by stating that their time evolution is directly related to the evolution of the corresponding mean values.

In the simulation, we first solve the second-order mean-field QLEs. The resulting mean values are then used as both the initial conditions and the time-dependent coefficients in the QRT equations, whose solutions yield the desired two-time correlation functions.

In order to obtain the linear equations required by the QRT, we separate the fluctuations from the mean values by defining
\begin{equation}
    \delta \bop{A} = \bop{A} - \langle \bop{A} \rangle,
\end{equation}
where \(\bop{A}\) is an arbitrary operator. Accordingly, the product of any two operators can be expressed as
\begin{equation}
    \begin{split}
        \bop{A} \bop{B}
         & = \left( \langle \bop{A} \rangle + \delta \bop{A} \right)
        \left( \langle \bop{B} \rangle + \delta \bop{B} \right)      \\
         & = \langle \bop{A} \rangle \langle \bop{B} \rangle
        + \langle \bop{A} \rangle \delta \bop{B}
        + \langle \bop{B} \rangle \delta \bop{A}
        + \delta \bop{A} \delta \bop{B}.
    \end{split}
\end{equation}
It follows that the correlation function of any two operators, \(\langle \bop{A}(t_0)\, \bop{B}(t) \rangle\), can be expressed as
\begin{equation}
    \langle \bop{A}(t_0)\, \bop{B}(t) \rangle
    =
    \langle \delta \bop{A}(t_0)\,\delta \bop{B}(t) \rangle
    +
    \langle \bop{A}(t_0) \rangle
    \langle \bop{B}(t) \rangle.
\end{equation}

To formulate the QRT equations, we first define the following correlation functions:
\begin{equation}
    \begin{split}
        C^a(t_0;t)
        &\equiv
        \langle \delta \ahdag(t_0)\,\delta \ah(t) \rangle,
        \\
        C^{a^{\dagger}}(t_0;t)
        &\equiv
        \langle \delta \ahdag(t_0)\,\delta \ahdag(t) \rangle,
        \\
        C^{S^+}_j(t_0;t)
        &\equiv
        \langle \delta \ahdag(t_0)\,\delta \bop{S}^+_j(t) \rangle,
        \\
        C^{S^-}_j(t_0;t)
        &\equiv
        \langle \delta \ahdag(t_0)\,\delta \bop{S}^-_j(t) \rangle,
        \\
        C^{S^z}_j(t_0;t)
        &\equiv
        \langle \delta \ahdag(t_0)\,\delta \bop{S}^z_j(t) \rangle.
    \end{split}
\end{equation}

Then, after omitting all third-order correlation functions, the QRT yields the dynamical equations for the correlation functions for \(t > t_0\):
\begin{widetext}
\begin{align}
    \dv{t} C^a(t_0;t) &= \left(-\frac{\kappa}{2}-i\delta\right) C^a(t_0;t) - i \sum_{j=1}^{M} g_j\, C^{S^-}_j(t_0;t), \\
    \dv{t} C^{a^{\dagger}}(t_0;t) &= \left(-\frac{\kappa}{2}+i\delta\right) C^{a^{\dagger}}(t_0;t) + i \sum_{j=1}^{M} g_j\, C^{S^+}_j(t_0;t), \\
    \dv{t} C^{S^+}_j(t_0;t) &= i\omega_j\, C^{S^+}_j(t_0;t) -2i g_j \left[\langle \ahdag(t)\rangle\, C^{S^z}_j(t_0;t) + \langle \bop{S}^z_j(t)\rangle\, C^{a^\dagger}(t_0;t)\right], \\
    \dv{t} C^{S^-}_j(t_0;t) &= -i\omega_j\, C^{S^-}_j(t_0;t) +2i g_j \left[\langle \ah(t)\rangle\, C^{S^z}_j(t_0;t) + \langle \bop{S}^z_j(t)\rangle\, C^{a}(t_0;t)\right], \\
    \dv{t} C^{S^z}_j(t_0;t) &= -i g_j \left[\langle \ah(t)\rangle\, C^{S^+}_j(t_0;t) + \langle \bop{S}^+_j(t)\rangle\, C^{a}(t_0;t)\right] + i g_j \left[\langle \ahdag(t)\rangle\, C^{S^-}_j(t_0;t) + \langle \bop{S}^-_j(t)\rangle\, C^{a^\dagger}(t_0;t)\right].
\end{align}
\end{widetext}

\subsection{From cavity to output fields}
In experiments, measured observables are typically associated with the output field. We therefore express the mean values and correlation functions of the output field in terms of the corresponding quantities of the cavity field.

The mean value of the output field \(\aout(t)\) can be written directly from the input--output relation, given by
\begin{equation}
    \langle \ah_{out}(t) \rangle = \langle \ah_{in}(t) \rangle - \sqrt{\kappa} \langle \ah(t) \rangle.
\end{equation}

When calculating output-field amplitudes, we project the output field under a temporal mode function \(f(t)\), satisfying
\(\int |f(t)|^2 \dd{t}=1\), as
\begin{equation}
    \langle \bop{a}_{\mathrm{out}} \rangle=\int f(t)\langle \bop{a}_{\mathrm{out}}(t) \rangle\dd{t}.
\end{equation}
Although this definition is valid for any choice of \(f(t)\), for analyzing memory protocols we choose the temporal mode that matches the echo waveform,
\begin{equation}
    f(t)
    =
    \frac{a_{\mathrm{out}}^*(t)}
    {\sqrt{\int |a_{\mathrm{out}}(t)|^2 \dd{t}}},
\end{equation}
which maximizes \(|\langle \bop{a}_{\mathrm{out}} \rangle|\).

For the correlation functions of the output field, we begin the analysis from the expression for the quantum fluctuation operator of the output field, given by
\begin{equation}
    \delta \aout(t) = \delta \ain(t) - \sqrt{\kappa} \delta\ah(t).
\end{equation}
By assuming vacuum quantum noise for the input field, we obtain
\begin{equation}
    \begin{split}
        \langle \delta \ain(t) \rangle
        &= 0,
        \\
        \langle \delta \ain^\dagger(t)\delta \ain(t') \rangle
        &= 0,
        \\
        \langle \delta \ain(t)\delta \ain^\dagger(t') \rangle
        &= \delta \left( t-t'\right).
    \end{split}
\end{equation}
Therefore, for the output field, we have
\begin{align}
    \langle \delta\aout^\dagger(t)\,\delta\aout(t') \rangle
    &= \kappa \,\langle \delta\ah^{\dagger}(t)\,\delta\ah(t') \rangle, \\
    \begin{split}        
    \langle \delta\aout(t)\,\delta\aout(t') \rangle
    &= -\sqrt{\kappa} \, \langle \delta \ain(t)\,\delta\ah(t') \rangle\\
    &\quad+ \kappa \,\langle \delta\ah(t)\,\delta\ah(t') \rangle.
    \end{split}
\end{align}

The term \( \langle \delta \ain(t)\,\delta\ah(t') \rangle \) is zero when \( t > t' \), since the future input field is uncorrelated with the cavity in the past. When \( t < t' \), although the term \( \langle \delta \ain(t)\,\delta\ah(t') \rangle \) can be nonzero, the value of \( \langle \delta\aout(t)\,\delta\aout(t') \rangle \) can be obtained from the bosonic commutation relation of the output field, given by 
\begin{equation}
    \left[\delta\ah_{out}(t),\ \delta\ah_{out}(t')\right]=0.
\end{equation}

As a consequence, the correlation function \(\langle \delta\bop{a}_{\mathrm{out}}(t)\,\delta\bop{a}_{\mathrm{out}}(t') \rangle\) can be summarized as
\begin{equation}
    \langle \delta\aout(t)\,\delta\aout(t') \rangle =
    \begin{cases}
        \kappa \,\langle \delta\ah(t')\,\delta\ah(t) \rangle & \text{for } t < t', \\
        \kappa \,\langle \delta\ah(t)\,\delta\ah(t') \rangle & \text{for } t > t'.
    \end{cases}
\end{equation}

The formulas for the output field correlation functions are essential for the ASE--RASE cross-correlation calculation, defined by
\begin{equation}
    C(\tau) \equiv \langle \bop{a}_{\mathrm{out}}(t)\, \bop{a}_{\mathrm{out}}(\tau - t) \rangle,
\end{equation}
where the center of the \pipulse{} is set as the origin of time, \(t\) lies in the ASE region, and \(\tau - t\) lies in the RASE region.

For the echo noise, we first compute the output photon number under the mode function \(f(t)\),
\begin{equation}
    \langle \bop{a}^\dagger_{\mathrm{out}} \, \bop{a}_{\mathrm{out}} \rangle
    = \iint f^*(t) f(t') \,\langle \bop{a}^\dagger_{\mathrm{out}}(t) \, \bop{a}_{\mathrm{out}}(t') \rangle \, dt \, dt',
\end{equation}
where \(f(t)\) is taken to be a boxcar mode with a window duration of \(65\,\mu\text{s}\) in the noise analysis.

With this, we calculate the quantum noise as
\begin{equation}
    \mathrm{noise} = 2 \left( \langle \bop{a}^\dagger_{\mathrm{out}} \, \bop{a}_{\mathrm{out}} \rangle
    - \langle \bop{a}^\dagger_{\mathrm{out}} \rangle \, \langle \bop{a}_{\mathrm{out}} \rangle \right) + 1,
\end{equation}
where the vacuum noise is normalized to \(1\).

\section{Non-Lorentzian case}\label{sec:non_lorentzian}
\begin{figure*}[htb]
    \centering
    \includegraphics[scale=1]{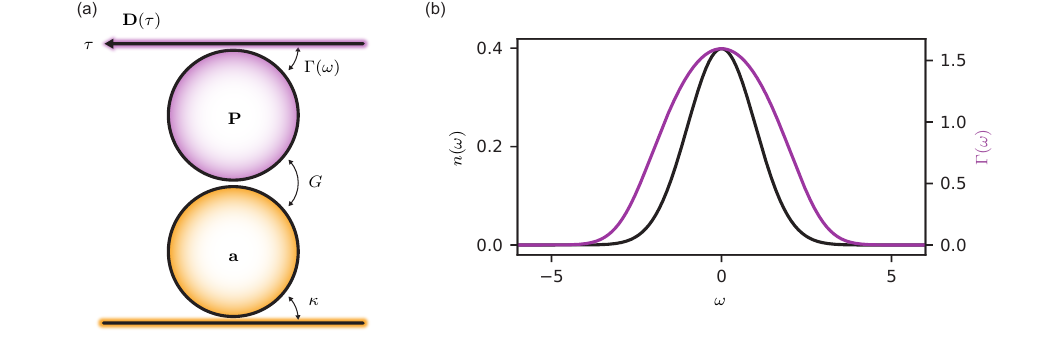}
    \caption{Non-Lorentzian waveguide model.
        (a) A non-Lorentzian inhomogeneous broadening profile \(n(\omega)\) is mapped to a frequency-dependent coupling rate \(\Gamma(\omega)\) to the dark waveguide, obtained from Eq.~\eqref{eq:n_to_Gamma}.
        (b) Example of a non-flat \(\Gamma(\omega)\) [Eq.~\eqref{eq:gaussian_coupling}] corresponding to a Gaussian broadening profile \(n(\omega)\) [Eq.~\eqref{eq:gaussian_broadening}] with \(\sigma=1\).
        }
    \label{fig:nonlorentzian_appendix}
\end{figure*}

For Lorentzian broadening, each collective mode \(\Ph(g)\) couples to its dark continuum at a constant rate. Non-Lorentzian broadening can be accommodated by a non-Markovian dark waveguide with a frequency-dependent rate \(\Gamma(\omega)\). We derive this rate using the construction of Appendix~\ref{sec:waveguide_model_appendix}, then examine its consequences for the dynamics and cavity protection.

\subsection{Frequency-dependent coupling to the dark waveguide}

We replace \(\Gamma\) by \(\Gamma(\omega)\) in Eq.~\eqref{eq:main_waveguide_hamiltonian} and impose the same dynamical, orthogonality, and normalization conditions as in Appendix~\ref{sec:waveguide_model_appendix}.

We retain the collective modes defined in Eq.~\eqref{eq:coupling_class_mode} and choose the frequency origin so that \(\int\dd{\omega}\,\omega n(\omega)=0\), using a symmetric principal value when needed.
The detuning Hamiltonian then takes the form
\begin{align}
    \Hspin={}&\int\dd{g}\dd{\omega}\,\omega\Dh^\dagger(g,\omega)\Dh(g,\omega)\nonumber\\
    &+i\int\dd{g}\dd{\omega}\,\sqrt{\frac{\Gamma(\omega)}{2\pi}}\nonumber\\
    &\quad\times\qty[\Ph^\dagger(g)\Dh(g,\omega)
                   -\Dh^\dagger(g,\omega)\Ph(g)],
    \label{eq:nonlorentzian_hamiltonian}
\end{align}
where \(\Gamma(\omega)\geq0\) is the frequency-dependent coupling rate.
The dark field must obey the same canonical commutators and orthogonality to every \(\Ph(g)\) as in Appendix~\ref{sec:waveguide_model_appendix}.

The kernel and phase factor follow from Eqs.~\eqref{eq:dark_kernel_general}--\eqref{eq:dark_profile_condition} with \(\Gamma\) replaced by \(\Gamma(\omega)\). All pole terms again cancel as in Eq.~\eqref{eq:dark_kernel_commutator}. The remaining normalization condition \(|z(\omega)|=1\) fixes
\begin{align}
    \Gamma(\omega)=\frac{2\pi n(\omega)}
    {\left|\displaystyle\int\dd{\omega'}\,
       \frac{n(\omega')}{\omega'-\omega+i0^+}\right|^2}.
\end{align}
Using the Sokhotski--Plemelj relation \(1/(x+i0^+)=\mathcal P(1/x)-i\pi\delta(x)\), where \(\mathcal P\) denotes the Cauchy principal value, we obtain
\begin{align}
    \Gamma(\omega)
    =\frac{2\pi n(\omega)}
    {\qty[\mathcal P\int\dd{\omega'}\frac{n(\omega')}{\omega-\omega'}]^2
     +\pi^2 n(\omega)^2}.
    \label{eq:n_to_Gamma}
\end{align}
Equation~\eqref{eq:n_to_Gamma} determines the coupling rate directly from the broadening profile, consistently with the collective-mode mapping of Ref.~\cite{woodsMappingsOpen}. The dark field follows from Eq.~\eqref{eq:appendix_dark_modes} by the same replacement \(\Gamma\to\Gamma(\omega)\).

\subsection{Non-Markovian dynamics}

The frequency dependence of the coupling changes both the bright-mode dynamics and the spatial extent of its interaction with the continuum. Eliminating the continuum from Eq.~\eqref{eq:nonlorentzian_hamiltonian} and projecting onto \(u(g)\) gives
\begin{align}
    \dot{\Ph}(t)={}&-iG\ah(t)\nonumber\\
    &+\int\dd{\omega}\,\sqrt{\frac{\Gamma(\omega)}{2\pi}}\Dh(\omega;t_0)
                         e^{-i\omega(t-t_0)}\nonumber\\
    &-\int_{t_0}^{t}\dd{t'}\,\mathcal K(t-t')\Ph(t'),
\end{align}
where \(\Dh(\omega)=\int\dd{g}\,u(g)\Dh(g,\omega)\) and
\begin{align}
    \mathcal K(t)=\int\frac{\dd{\omega}}{2\pi}\,
                       \Gamma(\omega)e^{-i\omega t}.
\end{align}
The memory kernel \(\mathcal K\) makes the current evolution depend on the collective mode at earlier times. For a flat \(\Gamma(\omega)\), it reduces to the instantaneous damping term \(-\Gamma\Ph/2\) in Appendix~\ref{sec:waveguide_model_appendix}.

The same memory effect has a spatial interpretation in the waveguide. Fourier transforming the continuum coordinate as in Appendix~\ref{sec:waveguide_model_appendix} gives
\begin{align}
    (\partial_t-\partial_\tau)\Dh(g,\tau;t)
    =-\frac{\Ph(g,t)}{\sqrt{2\pi}}\int\dd{\omega}\,
                         \sqrt{\frac{\Gamma(\omega)}{2\pi}}e^{-i\omega\tau}.
\end{align}
A non-flat coupling spreads the interaction over \(\tau\). The input and output fields must therefore be taken far from this interaction region. We define their frequency components by removing the free evolution at early and late times:
\begin{align}
    \Dh_{\mathrm{in/out}}(g,\omega)
    =\lim_{t_0\to\mp\infty}e^{i\omega t_0}\Dh(g,\omega;t_0),
\end{align}
and the corresponding temporal fields by Fourier transformation with kernel \(e^{-i\omega t}/\sqrt{2\pi}\).
Their projections onto \(u(g)\), denoted by \(\Pinout\), satisfy
\begin{align}
    \Pout(\Omega)=\Pin(\Omega)-\sqrt{\Gamma(\Omega)}\Ph(\Omega).
\end{align}
Here \(\Omega\) is the temporal Fourier frequency. In this representation, the memory integral becomes a frequency-dependent damping and frequency shift:
\begin{align}
    &\qty[\frac{\Gamma(\Omega)}{2}
    -i\qty(\Omega-\Delta(\Omega))]\Ph(\Omega)\nonumber\\
    &\qquad=-iG\ah(\Omega)+\sqrt{\Gamma(\Omega)}\Pin(\Omega),
\end{align}
where the continuum-induced frequency shift is
\begin{align}
    \Delta(\Omega)=\mathcal P\int\frac{\dd{\omega'}}{2\pi}\,
                    \frac{\Gamma(\omega')}{\Omega-\omega'}.
\end{align}
Near the fully excited state, the same decomposition uses \(\Pht(g)\) and \(\Dht(g,\omega)\), with the free detuning terms changing sign as in Appendix~\ref{sec:waveguide_model_appendix}; the resonator interaction becomes two-mode squeezing.

\subsection{Lorentzian and Gaussian examples}

For the Lorentzian profile in Appendix~\ref{sec:waveguide_model_appendix}, Eq.~\eqref{eq:n_to_Gamma} gives \(\Gamma(\omega)=\Gamma\), recovering the frequency-independent coupling.

A Gaussian profile gives a simple example of frequency-dependent coupling. For
\begin{align}
    n(\omega)=\frac{1}{\sqrt{2\pi}\sigma}
    \exp[-\frac{\omega^2}{2\sigma^2}],\label{eq:gaussian_broadening}
\end{align}
the principal-value integral in Eq.~\eqref{eq:n_to_Gamma} can be written using \(x=\omega/(\sqrt{2}\sigma)\) as
\begin{align}
    \mathcal P\int\dd{\omega'}\,\frac{n(\omega')}{\omega-\omega'}
    =\frac{\sqrt{2}}{\sigma}D\qty(x),
\end{align}
where
\begin{align}
    D(x)=e^{-x^2}\int_0^x \dd{t}e^{t^2}
\end{align}
is the Dawson function. Substituting into Eq.~\eqref{eq:n_to_Gamma} gives
\begin{align}
    \Gamma(\omega)=\frac{\sqrt{2\pi}\sigma e^{-x^2}}{2D^2(x)+\frac{\pi}{2}e^{-2x^2}}.
    \label{eq:gaussian_coupling}
\end{align}

The resulting coupling is strongest near the center and falls rapidly in the wings [Fig.~\ref{fig:nonlorentzian_appendix}(b)]. For \(|x|\gg1\), \(D(x)\simeq1/(2x)\), and Eq.~\eqref{eq:gaussian_coupling} gives
\begin{align}
    \Gamma(\omega)\simeq 2\sqrt{2\pi}\sigma x^2 e^{-x^2}=\sqrt{2\pi}\frac{\omega^2}{\sigma}e^{-\omega^2/(2\sigma^2)}.
\end{align}
At the center of any symmetric profile, \(n(\omega)=n(-\omega)\), the principal-value integral vanishes, so Eq.~\eqref{eq:n_to_Gamma} reduces to
\begin{align}
    \Gamma(0)=\frac{2}{\pi n(0)}.
\end{align}
For the Gaussian, this gives \(\Gamma(0)=2\sqrt{2\pi}\sigma/\pi\). More generally, the resonant cooperativity therefore depends only on the local TLS density at the center of a symmetric distribution.

\subsection{Cavity protection}
The weak coupling in the Gaussian wings explains cavity protection~\cite{dinizStronglyCoupling,kuruczSpectroscopicPropertiesa}. Strong resonator--ensemble coupling moves the polariton resonances away from the distribution center, where \(\Gamma(\omega)\) is small. The polaritons then lose less excitation into the dark waveguide and have narrower linewidths.

To see this explicitly, consider a resonant cavity and a centered Gaussian ensemble, with \(G\gg\sigma,\kappa\). The frequency shift \(\Delta(\omega)\) then gives only a small correction, so we approximate each polariton as having equal cavity and ensemble components.
The resonance frequencies are approximately
\begin{align}
    \omega_\pm\simeq\pm G.
\end{align}
If the coupling and frequency shift vary little over each linewidth, we may use the local rate
\begin{align}
    \Gamma(\omega)\simeq\Gamma(\omega_\pm)\equiv\Gamma_\pm.
\end{align}
The dynamics near each resonance are then described by
\begin{align}
    \dot{\ah} & = -\frac{\kappa}{2}\ah - iG\Ph + \sqrt{\kappa}\,\ain,\\
    \dot{\Ph} & = -\frac{\Gamma_\pm}{2}\Ph - iG\ah + \sqrt{\Gamma_\pm}\,\Pin .
\end{align}
Introducing the polariton modes
\begin{align}
    \Ph_\pm = \frac{1}{\sqrt{2}}\left(\ah \pm \Ph\right),
\end{align}
or equivalently \(\ah=(\Ph_+ + \Ph_-)/\sqrt{2}\) and \(\Ph=(\Ph_+ - \Ph_-)/\sqrt{2}\), gives
\begin{align}
    \begin{split}
    \dot{\Ph}_+ & = -iG\Ph_+ - \frac{\kappa+\Gamma_\pm}{4}\Ph_+ - \frac{\kappa-\Gamma_\pm}{4}\Ph_- \\
    &\quad+ \frac{1}{\sqrt{2}}\qty(\sqrt{\kappa}\ain+\sqrt{\Gamma_\pm}\Pin ),
    \end{split}\\
    \begin{split}
    \dot{\Ph}_- &= iG\Ph_- - \frac{\kappa+\Gamma_\pm}{4}\Ph_- - \frac{\kappa-\Gamma_\pm}{4}\Ph_+ \\
    &\quad+ \frac{1}{\sqrt{2}}\qty(\sqrt{\kappa}\ain-\sqrt{\Gamma_\pm}\Pin ).
    \end{split}
\end{align}
The equal cavity and ensemble weights give each polariton an amplitude damping term \((\kappa+\Gamma_\pm)/4\). The remaining off-diagonal terms mix the two polariton branches with strength \((\kappa-\Gamma_\pm)/4\).

When
\begin{align}
    2G \gg \kappa,\Gamma_\pm ,
\end{align}
the off-diagonal terms can be neglected under a rotating-wave approximation.
Each polariton then evolves independently as
\begin{align}
    \dot{\Ph}_\pm ={}& \mp iG\Ph_\pm - \frac{\kappa+\Gamma_\pm}{4}\Ph_\pm \nonumber\\
    &+ \frac{1}{\sqrt{2}}\left(\sqrt{\kappa}\,\ain \pm \sqrt{\Gamma_\pm}\,\Pin\right).
\end{align}
The amplitude decay rate is therefore \((\kappa+\Gamma_\pm)/4\), corresponding to the linewidth
\begin{align}
    \gamma_\pm = \frac{\kappa+\Gamma_\pm}{2}.
\end{align}
Evaluating the Gaussian tail at \(\omega_\pm\simeq\pm G\) shows that \(\Gamma_\pm\), and hence the contribution from decay into the dark waveguide, is exponentially suppressed as \(G\) increases. The linewidth approaches the cavity contribution \(\kappa/2\) within this approximation, consistently with Refs.~\cite{dinizStronglyCoupling,kuruczSpectroscopicPropertiesa}.

\section{Applicability of the waveguide model}\label{sec:waveguide_applicability_appendix}

The waveguide model introduced in the main text applies when the ensemble remains close to either the ground state or the fully excited state during the dynamics of interest, so that the Holstein--Primakoff approximation can be used.
In this appendix, we summarize the conditions required for this description to remain valid.

\subsection{Constraint on collective population}
The basic requirement is that the Holstein--Primakoff expansion remains valid around the relevant reference state.
From the original interaction Hamiltonian in Eq.~\eqref{eq:main_interaction_hamiltonian}, the interaction between the bright mode and the resonator is characterized by
\begin{align}
    \dot{\Ph} &= i \bar{g} \frac{\Sh^z}{\sqrt{N}}\ah+\cdots,\\
    \dot{\ah} &= -i \bar{g}\sqrt{N}\Ph+\cdots .
\end{align}
where \(\Sh^z=\bar{g}^{-2}\sum_j |g_j|^2\sigmah^z_j\) is the coupling-weighted population.

Thus, the population \(\Sh^z\) determines the effective strength of the collective resonator--ensemble interaction.
Starting from \(\Sh^z=\pm N\), the population deviation can be expressed in terms of the bright-mode operator as
\begin{align}
    |\delta \Sh^z| \simeq 2\Ph^\dagger\Ph .
\end{align}
Therefore, if one requires the relative deviation of the effective coupling strength to be smaller than \(\epsilon\), the bright-mode population should satisfy
\begin{align}
    \Ph^\dagger\Ph < \frac{\epsilon}{2}N .
    \label{eq:population_condition}
\end{align}
This condition is the basic applicability condition of the waveguide model.

\subsection{Weak pulses}
Equation~\eqref{eq:population_condition} constrains the strength of the pulses used in the protocol.
The pulses should either be sufficiently weak signal pulses, which generate only a small number of collective excitations, or \pipulse{}s, which switch the ensemble between the near-ground-state and near-excited-state regimes.
Here we first consider weak signal pulses.

Near the ground state, in the steady-state limit, the bright-mode population scales as
\begin{align}
    \Ph^\dagger\Ph\sim\frac{4G^2}{\Gamma^2}\ah^\dagger\ah=C\frac{\kappa}{\Gamma}\ah^\dagger\ah,
\end{align}
where \(C\) is the cooperativity.
Therefore, the intracavity photon number should satisfy
\begin{align}
    \ah^\dagger\ah<\frac{\epsilon N}{2}\frac{\Gamma}{C\kappa}.\label{eq:weak_pulse_ground_condition}
\end{align}

Near the fully excited state, the interaction becomes a two-mode-squeezing interaction.
Because this regime provides gain, the same input pulse can generate a larger bright-mode population.
Below the instability threshold, \(C<1\), the resonant response is enhanced by the gain factor, giving the estimate
\begin{align}
    \Ph^\dagger\Ph\sim\frac{C\kappa}{\Gamma}\frac{1}{(1-C)^2}\ah^\dagger\ah+N_{\mathrm{ASE}},
\end{align}
where \(N_{\mathrm{ASE}}\) denotes excitations generated by amplified spontaneous emission, which is discussed in the next subsection.
Thus, the weak-pulse condition becomes
\begin{align}
    \ah^\dagger\ah<\frac{\epsilon N}{2}\frac{\Gamma}{C\kappa}(1-C)^2.\label{eq:weak_pulse_excited_condition}
\end{align}
This condition becomes increasingly restrictive as \(C\) approaches unity.

\subsection{Purcell effect and amplified spontaneous emission}
Even in the absence of weak input pulses, the ensemble can deviate from the reference state because of spontaneous emission through the resonator.
For a single TLS resonant with the resonator, the Purcell decay rate is approximately
\begin{align}
    \gamma \sim \frac{4\bar{g}^2}{\kappa}.
\end{align}
Therefore, for an ensemble initially close to the excited state, the waveguide model applies only on timescales sufficiently short compared with \(\gamma^{-1}\).
In practice, additional relaxation channels also contribute and define a finite \(T_1\).

This effect is enhanced for an inverted ensemble because spontaneous emission into the collective mode is amplified (ASE).
Below the instability threshold \(C<1\), the ASE rate is enhanced by the gain factor and can be estimated as
\begin{align}
    \gamma_\mathrm{ASE}\sim\frac{1}{1-C}\gamma.
\end{align}
Thus, for higher cooperativity, the waveguide model applies only on shorter timescales.

In the numerical simulations in this work, we use \(\bar{g}/(2\pi)=\SI{100}{Hz}\) and \(\kappa/(2\pi)=\Gamma/(2\pi)=\SI{1}{MHz}\), while \(N\) is varied to obtain different cooperativities \(C\).
This is a typical parameter regime for microwave-coupled spin ensembles~\cite{osullivanRandomAccessQuantum}.
For these parameters, the single-TLS Purcell rate is
\begin{align}
    \gamma \sim \frac{4\bar{g}^2}{\kappa} \sim 2\pi\times\SI{4e-2}{Hz},
\end{align}
corresponding to a Purcell-limited timescale of order \(\gamma^{-1}\sim \SI{4}{s}\).
Thus, Purcell-induced depletion is negligible on the timescales considered in our simulations.

The ASE-enhanced rate is
\begin{align}
    \gamma_\mathrm{ASE}\sim\frac{\gamma}{1-C}.
\end{align}
To obtain a depletion of order \(1\%\) within \(\SI{100}{\micro\second}\), one would require
\begin{align}
    1-C \sim  2.5\times 10^{-3},
\end{align}
or equivalently \(C\sim 0.9975\).
Therefore, ASE-induced depletion becomes appreciable only when the system is tuned very close to the instability threshold \(C=1\), indicating that this effect is negligible in the parameter regime considered here.

\subsection{Adiabaticity of ARP \pipulse{}s}

The \pipulse{}s can also introduce deviations from the Holstein--Primakoff regime if the inversion is imperfect.
Although \pipulse{}s implemented by adiabatic rapid passage (ARP) are robust against frequency and coupling inhomogeneity, high-fidelity inversion requires the adiabatic condition
\begin{align}
    Q=\frac{(2A g)^2}{k} \gg 1 ,\label{eq:arp_adiabatic_condition}
\end{align}
where \(A\) is the drive amplitude and \(k\) is the chirp rate.
When this condition is not satisfied, the inversion error is approximately given by the Landau--Zener formula \cite{LandauEnergy,zenerNonadiabaticCrossing} as
\begin{align}
    |\delta \Sh^z|\sim 2N\exp[-\frac{\pi Q}{2}].
\end{align}
Therefore, the ARP pulse must be sufficiently strong and slow that this error remains much smaller than unity.

For an ensemble with inhomogeneous coupling, the adiabatic condition in Eq.~\eqref{eq:arp_adiabatic_condition} is most restrictive for the weakly coupled TLSs.
However, the relevance of these weakly coupled TLSs to the collective dynamics is weighted by their overlap with the bright mode.
As discussed in Appendix~\ref{sec:silencing_factor_appendix}, the bright mode in coupling space is
\begin{align}
    u(g)=\frac{1}{\bar{g}} g\sqrt{\rhog}.
\end{align}
Thus, each TLS contributes to the stored collective excitation according to its weight in \(P(g)\).
It is therefore important that the coupling distribution does not have a strong divergence near \(g=0\), where the adiabaticity condition is hardest to satisfy.
If the low-coupling behavior is described by a power law,
\begin{align}
    \rhog \propto g^{\alpha},
\end{align}
then
\begin{align}
    u(g)\propto g^{1+\alpha/2}.
\end{align}
Therefore, requiring the bright-mode profile not to diverge at \(g=0\) gives
\begin{align}
    \alpha \geq -2 .
\end{align}
When this condition is satisfied, high-fidelity inversion of the collectively relevant modes can be achieved with sufficiently large drive amplitude and sufficiently slow chirp rate.

This consideration is especially relevant for far-field microwave coupling~\cite{ranjanPulsedElectron}, where the low-\(g\) tail of the coupling distribution can follow a power-law behavior.
In contrast, for evanescently coupled optical platforms~\cite{yangPhotonicIntegration}, the exponential spatial dependence of the field can more strongly suppress weakly coupled emitters, effectively producing a sharper low-\(g\) cutoff and thereby relaxing the adiabaticity requirement.
\section{Experimental considerations}
\label{sec:experiment_appendix}

The microwave and optical experiments of Refs.~\cite{osullivanRandomAccessQuantum,kamelMultimodeRandomAccess} both observe suppression of the first echo in ROSE protocols using ARP pulses. Here we describe the model parameters and procedure used to compare with the microwave experimental data in Fig.~\ref{fig:rose}(e), then estimate how much coupling inhomogeneity is required to explain echo silencing in each experiment.

\subsection{Direct comparison with the microwave experiment}
\label{sec:microwave_comparison}
To illustrate how coupling inhomogeneity explains the observed echo suppression, we compare analytical and numerical predictions with the measured signals from the bismuth-donor spin ensemble of Ref.~\cite{osullivanRandomAccessQuantum} [Fig.~\ref{fig:rose}(e)]. We reconstruct the distribution of coupling strengths from the superconducting resonator's magnetic field, then calculate the pulse sequence both with this distribution and with homogeneous coupling at the same collective coupling strength.

\subsubsection{Reconstruction of the coupling distribution}
To obtain \(\rhog\), we require the spatial distribution of the resonator's zero-point magnetic field. This distribution is shown in Fig.~3(c) of Ref.~\cite{osullivanRandomAccessQuantum}, but the underlying numerical field-map data are not provided. We therefore calculate the magnetic field using simple approximations to reproduce that figure and use the resulting field map to estimate the distribution of coupling strengths. We model a two-dimensional cross-section of two parallel niobium strips carrying opposite currents. We estimate the strip width \(w=\SI{1.15}{\micro\meter}\) and gap \(d=\SI{1.20}{\micro\meter}\) from Fig.~3 of Ref.~\cite{osullivanRandomAccessQuantum}, and take the film thickness \(t_{\mathrm{Nb}}=\SI{100}{nm}\) from its Appendix~A. Following the coordinate convention of Ref.~\cite{osullivanRandomAccessQuantum}, we take silicon to occupy \(X>0\) and the film to occupy \(-t_{\mathrm{Nb}}<X<0\). The strip centers lie at \(Z=\pm(w+d)/2\), and the model assumes translational invariance along the strips.

We use the zero-point current \(I_{\mathrm{zpf}}=\omega_r\sqrt{\hbar/(2Z_r)}\), taking \(\omega_r/(2\pi)=\SI{7.093}{GHz}\) from Ref.~\cite{osullivanRandomAccessQuantum} and \(Z_r=\SI{100}{\ohm}\) from the estimate in Supplemental Sec.~E of the device paper, Ref.~\cite{osullivanSpinResonanceLinewidthsb}. We calculate the magnetic field using the Biot--Savart law, with the transverse current distribution in each strip given in the thin-film approximation by~\cite{rhoderickCurrentDistribution}
\begin{align}
    I(Z)=\frac{I_{\mathrm{strip}}}{\pi\sqrt{(w/2)^2-(Z-Z_c)^2}},
    \qquad |Z-Z_c|<w/2,
\end{align}
where \(I(Z)\) is the current per unit width, \(Z_c\) is the strip center, and \(I_{\mathrm{strip}}=\pm I_{\mathrm{zpf}}\) for the two strips. The current is confined to the film midplane. We obtain the local coupling from \(g(X,Z)/(2\pi)=\gamma_e|B_X(X,Z)|\), using the bound-electron gyromagnetic ratio \(\gamma_e=\SI{27.997}{GHz/T}\) given in Ref.~\cite{osullivanSpinResonanceLinewidthsb}. Visual comparison confirms that the calculated field distribution reproduces well the spatial pattern and field magnitude shown in Fig.~3(c) of Ref.~\cite{osullivanRandomAccessQuantum}.

We take the implanted-layer depth of \(\SI{1}{\micro\meter}\) from Ref.~\cite{osullivanRandomAccessQuantum} and assume a uniform donor density throughout this layer. For the reconstruction, we choose a uniform \(500\times10000\) grid over \(0\leq X\leq\SI{1}{\micro\meter}\) and \(|Z|\leq\SI{30}{\micro\meter}\), and obtain \(\rhog\) from a normalized histogram with 201 equal-width coupling bins, then linearly interpolate and renormalize it on the finer coupling grid used in the waveguide model.

This procedure determines the coupling distribution from the assumed resonator geometry and spatial distribution of donors, without fitting it to the echo amplitude. The density \(\rhog\) describes the fraction of spins with a given coupling strength, whereas the silencing factor depends on their contribution to the bright mode, weighted by \(|u(g)|^2=g^2\rhog/\bar g^2\).

\subsubsection{Pulse sequence, calculation parameters, and comparison with experiment}
The experimental parameters and pulse specifications used below are taken from Ref.~\cite{osullivanRandomAccessQuantum}.
Both calculations use \(\kappa/(2\pi)=\SI{0.4}{MHz}\), \(\Gamma/(2\pi)=\SI{2.4}{MHz}\), zero cavity detuning, and \(C=0.07\). The reference coupling \(g_{\mathrm{ref}}/(2\pi)=\SI{80}{Hz}\) sets the control-pulse amplitude and the coupling in the homogeneous case.

The input is Gaussian with amplitude width \(\sigma_t=\SI{1.07}{\micro\second}\). Two identical WURST pulses have duration \(T_{\mathrm W}=\SI{200}{\micro\second}\), envelope order 20, and positive chirp rate \(k/(2\pi)=\SI{0.02}{MHz^2}\), corresponding to a \(\SI{4}{MHz}\) sweep.
The Rabi frequency used in Fig.~1 of Ref.~\cite{osullivanRandomAccessQuantum} is not explicitly specified. We therefore adopt a representative value of \(\Omega_{\mathrm{ref}}/(2\pi)=\SI{1.7}{MHz}\) at \(g_{\mathrm{ref}}\), based on the WURST pulse parameters marked by triangles in Fig.~6 of the same reference.

For the analytical prediction, we obtain the intracavity amplitude using the steady-state cavity response described in Appendix~\ref{sec:coherent_drive}, with \(A_0=\Omega_{\mathrm{ref}}/(2g_{\mathrm{ref}})\). We integrate Eq.~\eqref{eq:main_arp_phase_integral} to obtain the phase and evaluate the silencing factor using the reconstructed coupling distribution. The intermediate and retrieved echoes follow from Eq.~\eqref{eq:silencing_effect} and the narrowband ROSE result in Appendix~\ref{sec:rose_appendix}.

For the numerical calculation, we use a first-order mean-field simulation with 6000 frequency bins and 201 coupling bins, or one coupling bin at \(g_{\mathrm{ref}}\) for homogeneous coupling. The 200-photon input is centered at \(\SI{50}{\micro\second}\), and the WURST pulses at \(\SI{200}{\micro\second}\) and \(\SI{650}{\micro\second}\), with incident amplitude \(E_0=\sqrt{\kappa}A_0/2\). We evolve to \(\SI{1100}{\micro\second}\), saving 5001 points with relative and absolute tolerances of \(10^{-8}\). The detailed numerical procedure is described in Appendix~\ref{sec:numerical_appendix}. For Fig.~\ref{fig:rose}(e), each dataset is normalized by its input-pulse peak, with the same factor applied to both echoes. The phase and local time origin are aligned within each displayed window; neither echo amplitude is independently fitted.

To quantify the comparison reported in Sec.~\ref{sec:rose}, we compare the measured and predicted amplitudes in the temporal mode of each simulated first echo. Writing the predicted field as \(\alpha_{\mathrm{th}}(t_j)=A_{\mathrm{th}}f_j\), with \(A_{\mathrm{th}}^2=\sum_j|\alpha_{\mathrm{th}}(t_j)|^2\) and \(\sum_j|f_j|^2=1\), we obtain the measured amplitude by projection:
\begin{align}
    A_{\mathrm{exp}}=\left|\sum_j f_j^*\alpha_{\mathrm{exp}}(t_j)\right|.
    \label{eq:echo_mode_test}
\end{align}
Here \(\alpha_{\mathrm{exp}}\) is the background-subtracted experimental field. Taking the magnitude removes the dependence on a constant phase offset, and no amplitude adjustment is made.

To estimate the noise in this mode, we Fourier transform a \SI{13}{\micro\second} signal-free interval before the input pulse, after subtracting its mean. The squared magnitude gives the noise spectrum, which we weight by the squared magnitude of the Fourier transform of the normalized echo mode and integrate to obtain the projected noise variance. We assume that the noise has the same statistical properties in the background and echo windows and is Gaussian with equal, uncorrelated quadratures. The projected complex noise has an in-phase and a quadrature component, and its variance is the sum of their variances. Under the equal-variance assumption, each component therefore has half the projected variance, and \(\sigma\) is the square root of this quadrature variance.

With coupling inhomogeneity, the amplitude differences are \(|A_{\mathrm{exp}}-A_{\mathrm{th}}|=1.37\sigma\) (numerical) and \(1.58\sigma\) (analytical). For homogeneous coupling, the corresponding differences are \(35.9\sigma\) and \(31.6\sigma\). Thus, including coupling inhomogeneity brings the predicted echo amplitude close to the measured value, whereas homogeneous coupling predicts a much larger echo.

\subsection{Adiabaticity estimates for the two experiments}

The preceding comparison shows how the coupling distribution reconstructed from the resonator field accounts for the observed echo suppression in the microwave experiment. To understand the scale of inhomogeneity needed for this effect and assess the same mechanism in the optical experiment, we now use the simpler estimate in Eq.~\eqref{eq:silencing_adiabaticity}. For a Gaussian distribution of coupling strengths, this expression relates the silencing factor to the relative width \(\sigma_g/g_0\) and the ARP adiabaticity factor \(Q=\Omega_0^2/k\). It therefore gives an estimate of the coupling variation required for strong suppression from the pulse parameters alone, without reconstructing the full spatial field distribution. We apply this estimate to both experiments below; the Gaussian assumption is used only for these estimates, while the direct comparison above retains the reconstructed coupling distribution.

\subsubsection{Microwave spin ensemble}
In Ref.~\cite{osullivanRandomAccessQuantum}, the spatially varying resonator field produces coupling inhomogeneity, as also indicated by the strongly damped Rabi oscillations.

For this microwave experiment, the WURST pulse parameters are approximately
\begin{align}
    \Omega_0 \simeq 2\pi\times 1.7~\mathrm{MHz},
    \qquad
    k \simeq 2\pi\times \SI{0.02}{MHz^2},
\end{align}
yielding
\begin{align}
    Q=\frac{\Omega_0^2}{k}\simeq 9.1\times 10^2.
\end{align}
Therefore, a relative coupling inhomogeneity
\begin{align}
    \frac{\sigma_g}{g_0}\gtrsim 4.1\times 10^{-3}
\end{align}
is sufficient to produce strong silencing, \(F<10^{-3}\).
This required variation is only at the sub-percent level.
Thus, even a small fraction of the coupling inhomogeneity naturally present in a planar microwave resonator geometry is enough to suppress the intermediate echo.

\subsubsection{Optical rare-earth-ion ensemble}
Ref.~\cite{kamelMultimodeRandomAccess} uses an optical rare-earth-ion platform based on the optical transition of \(^{171}\mathrm{Yb}^{3+}\) ions in a \(\mathrm{Y}_2\mathrm{SiO}_5\) crystal.
In contrast to the microwave resonator case, the control field is an optical beam propagating through a bulk crystal.
Spatial variation of the optical intensity across the ensemble therefore produces an inhomogeneous Rabi frequency, which is equivalent to coupling inhomogeneity in the present model.
Such variation can arise from the transverse mode profile of the optical beam.

For this experiment, the ARP pulse parameters are
\begin{align}
    \Omega_0 = 2\pi\times 0.35~\mathrm{MHz},
    \qquad
    k = 2\pi\times \SI{0.03}{MHz^2},
\end{align}
yielding
\begin{align}
    Q=\frac{\Omega_0^2}{k}\simeq 25.7.
\end{align}
In this case,
\begin{align}
    \frac{\sigma_g}{g_0}\gtrsim 1.4\times 10^{-1}
\end{align}
is sufficient to achieve \(F<10^{-3}\).
Compared with the microwave experiment, the smaller value of \(Q\) means that a larger relative Rabi-frequency inhomogeneity is required for coupling-induced silencing.
This level of inhomogeneity is still plausible for an optical ensemble.

In both cases, the required relative inhomogeneity is experimentally plausible. These estimates therefore show that coupling-dependent ARP phases provide a consistent and physically realistic explanation for the observed echo silencing in both experiments.
\end{document}